\documentclass[draft]{article}
\usepackage{amsmath,amsfonts,latexsym, amssymb}

\usepackage{color}

\newcommand{\non}{\nonumber}
\newcommand{\bla}{\mbox{\boldmath $\lambda$}}

\newcommand{\fn}{{\frak n}}
\newcommand{\fa}{{\frak a}}

\newcommand{\bS}{{\bf S}}
\newcommand{\bO}{{\bf O}}

\newcommand{\ttau}{\vartheta}

\newcommand{\ov}{\overline}

\newcommand{\e}{\varepsilon}

\newcommand{\pt}{\partial}
\newcommand{\rd}{{\rm d}}
\newcommand{\bR}{{\mathbb R}}
\newcommand{\bbZ}{{\mathbb Z}}

\newcommand{\ba}{{\bf{a}}}

\newcommand{\bx}{{\bf{x}}}
\newcommand{\by}{{\bf{y}}}

\newcommand{\bu}{{\bf{u}}}
\newcommand{\bv}{{\bf{v}}}

\newcommand{\bm}{{\bf{m}}}
\newcommand{\bw}{{\bf{w}}}

\newcommand{\wh}{\widehat}

\newcommand{\bG}{{\bf G}}

\newcommand{\wH}{{K}}

\newcommand{\bZ}{{\bf{Z}}}

\newcommand{\al}{\alpha}

\newcommand{\be}{\begin{equation}}
\newcommand{\ee}{\end{equation}}

\newcommand{\ga}{{\gamma}}

\newcommand{\la}{\lambda}
\newcommand{\Om}{{\Omega}}
\newcommand{\om}{{\omega}}

\newcommand{\cG}{{\cal G}}
\newcommand{\cA}{{\cal A}}

\newcommand{\cN}{{\cal N}}
\newcommand{\cP}{{\cal P}}

\newcommand{\cK}{{\cal K}}
\newcommand{\cF}{{\cal F}}
\newcommand{\cI}{{\cal I}}

\newcommand{\cH}{{\cal H}}

\newcommand{\im}{{\mathfrak{Im}\, }}
\newcommand{\re}{{\mathfrak{Re}\, }}

\newcommand{\E}{{\mathbb E }}
\newcommand{\R}{{\mathbb R }}
\newcommand{\N}{{\mathbb N}}
\newcommand{\Z}{{\mathbb Z}}
\renewcommand{\P}{{\mathbb P}}
\newcommand{\bC}{{\mathbb C}}
\newcommand{\C}{{\mathbb C}}

\newcommand{\wkto}{\rightharpoonup}

\newcommand{\wt}{\widetilde}

\newcommand{\barv}{[v]}
\newcommand{\barZ}{[Z]}

\newcommand{\tr}{\mbox{Tr}\, }

\newtheorem{theorem}{Theorem}

\newtheorem{lemma}[theorem]{Lemma}

\newtheorem{definition}{Definition}
\newcommand{\qed}{\hfill\fbox{}\par\vspace{0.3mm}}

\numberwithin{equation}{section}
\numberwithin{theorem}{section}
\numberwithin{definition}{section}

\title{Universality of Wigner random matrices: a survey of recent results}

\author{L\'aszl\'o Erd\H os${}^1$\thanks{Partially supported
by SFB-TR 12 Grant of the German Research Council}
  \\ \\
Institute of Mathematics, University of Munich, \\
Theresienstr. 39, D-80333 Munich, Germany}
\begin{document}
\date{Sep 1, 2010}

\maketitle

\begin{abstract}

We study the universality of spectral statistics of large random matrices.
We consider  $N\times N$ symmetric, hermitian or quaternion
self-dual random matrices with independent, identically distributed
 entries (Wigner matrices) where the probability distribution  of each matrix element
 is given by 
a measure $\nu$ with zero expectation and with a subexponential decay. 
Our  main result is that the correlation functions of the
 local eigenvalue statistics in the bulk of the spectrum 
 coincide with those of the Gaussian Orthogonal Ensemble (GOE),
the Gaussian Unitary Ensemble (GUE) and the Gaussian Symplectic
Ensemble (GSE), respectively, in the limit $N\to \infty$.
Our approach is based on the study 
of the Dyson Brownian motion via a related new dynamics, the local relaxation flow.

As a main input, we establish that the density of eigenvalues
converges to the Wigner semicircle law and this holds even 
down to the smallest possible scale, and, moreover,  we show that eigenvectors
are fully delocalized. These results hold even without the
condition that the matrix elements are identically distributed,
only independence is used. 
In fact, we give strong estimates 
on the matrix elements of the Green function as well
that imply that the local statistics of
any two ensembles in the bulk are identical if the first
four moments of the matrix elements match. Universality 
at the spectral edges requires matching only two moments.
 We also prove a Wegner type estimate
and that the eigenvalues repel each other on arbitrarily small
scales.

\end{abstract}


\medskip


\newpage
\setcounter{tocdepth}{4}

\tableofcontents

\newpage

\section{Introduction}

This survey is based upon the lecture notes that
the author has prepared for the participants
at the Arizona School of Analysis and Applications, Tucson, AZ
in 2010. The style of the presentation is closer
to the informal style of a lecture than to a formal research article.
For the details and sometimes even for the precise formulation
we refer to the original papers.

In the first introductory section we give an overview about universality
of random matrices, including previous results, history
and motivations. We introduce some basic
concepts such as Wigner matrix, Wigner semicircle law,
Stieltjes transform, moment method, sine-kernel, gap distribution,
level repulsion, bulk and edge universality, invariant
ensembles, Green function comparison theorem, four moment
theorem, local relaxation flow and reverse heat flow.
Some of these concepts will not be used for our main
results, but we included them to help the orientation of the reader.
The selection of the material presented in the
first section is admittedly 
reflects a personal bias of the author
and it is not meant to be comprehensive. 
It is focused on the background material
for the later sections where we
present  our recent results on universality of random matrices.

There are several very active research directions connected 
with random matrices that are not mentioned in this survey at all,
e.g. supersymmetric methods or connection with free probability.
Some other very rich topics, e.g. edge universality 
with the moment method or connections with orthogonal polynomials,
are mentioned only  superficially. We refer the reader to more
comprehensive surveys on random matrices, especially the classical book
of Mehta \cite{M}, the survey of the Riemann-Hilbert approach of Deift
\cite{D}, the recent book of Anderson, Guionnet and Zeitouni \cite{AGZ}
and the forthcoming book of Forrester \cite{F}.
An excellent short summary about the latest developments
is by Guionnet \cite{Gfr}.

\bigskip

Starting from Section \ref{sec:sc}, we present our recent
results that gives 
the shortest and up to now the most powerful 
approach to the bulk universality for $N\times N$ Wigner matrices. 
One of the main results is formulated in Theorem~\ref{mainres}.
These results  were obtained in
collaboration with  J. Ramirez, S. P\'ech\'e, B. Schlein, H.-T. Yau and J. Yin;
see the bibliography for precise references.
In this part we strive for mathematical rigor, but several details 
will be referred to the original papers.  The argument has
three distinct steps:
\begin{itemize}
\item[1.] Local semicircle law (Section \ref{sec:sc});
\item[2.] Universality for Gaussian convolutions via the
 local relaxation flow (Section \ref{sec:sine});
\item[3.] Green function comparison theorem (Section \ref{sec:4mom}).
\end{itemize}
Finally, in Section \ref{sec:put}, we put together the proof from these ingredients.
The main result on universality of local statistics in the bulk
for Wigner matrices is formulated in Theorem~\ref{mainthm}.
Some technical lemmas are collected in the Appendices that can be neglected at
first reading.

\medskip

{\it Convention:} Throughout the paper the letters $C$ and $c$ denote positive
constants whose values may change from line to line and they
are independent of the relevant parameters. Since we will always take the $N\to \infty$
limit at the end, all estimates are understood for sufficiently large $N$.
In informal explanations we will often neglect logarithmic factors, by introducing the
notation $\lesssim$ and $\ll$ to indicate inequality ``up to some $\log N$ factor''.
More precisely, $A\lesssim B$ means $A\le(\log N)^C B$ with some non-negative constant $C$,
and $A\ll B$ means $A\le (\log N)^{-C} B$ with some positive constant $C$.

\bigskip

{\it Acknowledgement.} The author thanks H.-T. Yau for  suggestions to improve
the presentation of this overview.

\subsection{Summary of the main results: an orientation for the reader}\label{sec:mainresult}

We will consider $N$ by $N$ matrices $H=(h_{ij})_{i,j=1}^N$ whose entries are 
real or complex random
variables.
 In most cases we assume that $H$ is hermitian or symmetric,
but our method applies to other ensembles as well
(our results for matrices with quaternion entries will not
be discussed here, see \cite{ESYY}).
We assume that the entries are independent up to 
the symmetry constraint, $h_{ij}= \ov h_{ji}$, they are centered, $\E h_{ij}=0$,
 and their tail probability has a uniform subexponential
decay (see \eqref{subexpuj} later).
 We do {\it not} assume that the matrix elements are identically distributed
but we assume that the variances, $\sigma_{ij}^2:=\E |h_{ij}|^2$ satisfy the
normalization condition
\be
  \sum_{j=1}^N \sigma_{ij}^2 =1, \qquad i=1,2,\ldots, N.
\label{normal}
\ee
i.e., the deterministic
$N\times N$  matrix of variances, $\Sigma= (\sigma_{ij}^2)$,
 is  symmetric and  doubly stochastic.
These conditions guarantee that $-1\le \Sigma \le 1$. We will always assume that $1$ is a simple
eigenvalue of $\Sigma$ and there is a positive number $\delta_->0$ such that
$-1+\delta_-\le \Sigma$. This assumption is satisfied for
practically any random matrix ensembles. Sometimes we will need
a uniform gap condition, i.e. that there exists a positive $\delta_+>0$
such that 
$$
  \mbox{Spec} \; \Sigma \subset [-1+\delta_-, 1-\delta_+]\cup \{ 1\}.
$$
For example,  for the standard Wigner matrix $\sigma_{ij}^2=N^{-1}$ and $\delta_-
=\delta_+=1$. For random band matrices (see \eqref{BM} for the precise
definition) with band width $W$ satisfying  $1\ll W\ll N$, 
the gap $\delta_+$ goes to zero as the size of the matrix increases.

The normalization \eqref{normal} ensures that the bulk of the
spectrum of $H$ lies in the interval $[-2,2]$ and the density
of eigenvalues $\la_1\le \la_2\le \ldots \le \la_N$
is given by the Wigner semicircle law as $N\to \infty$.
Apart from the vicinity of the edges $\pm 2$, the typical
spacing of neighboring eigenvalues is of order $1/N$.
We are interested in the statistics of the eigenvalues in the $N\to\infty$ limit.

\subsubsection{Summary of Section \ref{sec:sc}:
Main results on the local semicircle law}\label{sec:mainlsc}

 In Section \ref{sec:sc}
we prove that the density of eigenvalues
follows the semicircle law {\it down to the smallest possible scale}, i.e.,
to scales only a slightly larger than $1/N$. We will call it {\it local
semicircle law}. 
  The local semicircle law is identified via
the Stieltjes transform of the empirical density of the eigenvalues,
$$
   m(z): =m_N(z)= \frac{1}{N}\sum_{j=1}^N \frac{1}{\lambda_j - z},
 \qquad z= E+i\eta, \;\; E\in\R,
 \; \eta>0,
$$
and we show that $m_N(z)$ converges to the Stieltjes transform of the
semicircle density
$$
   m_{sc}(z):  =\int_\R\frac{\varrho_{sc}(x)\rd x}{x-z}, \qquad  \varrho_{sc}(x):
  =\frac{1}{2\pi}\sqrt{(4-x^2)_+},
$$
in the limit $N\to\infty$. The imaginary part $\eta=\im z$ may depend on $N$
and it corresponds to the local scale on which the density is identified. 
The precision of our approximation is of order $(N\eta)^{-1}$.
Our best result in this direction is Theorem 2.1 of \cite{EYY3},
which we will call the {\it strong local semicircle law}:
\be
  |m(z) - m_{sc}(z)|\le \frac{C(\log N)^L}{N\eta},
\label{mmmbest}
\ee
for some sufficiently large $L$ and
with a very high probability (see Section \ref{sec:best}).
This result holds  even for a more general class Wigner matrices whose
variances are comparable (see \eqref{VV} for precise definition),
the key input is that in this case we have $\delta_+>0$.

\medskip

For even more general Wigner matrices  (they will be called
universal Wigner matrices, see Definition~\ref{def:gen} later),
the key quantity that measures
the precision is the {\it spread of the matrix}, defined by
\be
  M: =\frac{1}{\max_{ij}\sigma_{ij}^2}.
\label{def:spread}
\ee
For typical random band matrices (see \eqref{BM} for the precise definition),
 $M$ is comparable with the band width $W$.
If $M\ll N$, then
the precision of our estimates is determined by $M$ instead of $N$,
for example, in \cite{EYY2} we obtain
\be
  |m(z) - m_{sc}(z)|\le \frac{CN^\e}{M\eta\kappa^2}, \qquad  \kappa := \big| |E|-2\big|,
\label{mmm3}
\ee
for any $\e>0$,
with a very high probability
(see Theorem \ref{lsc} later), or
\be
  |m(z) - m_{sc}(z)|\le\frac{C(\log N)^L}{\sqrt{M\eta}\kappa} 
\label{mmm4}
\ee
was proven in Theorem 2.1 of \cite{EYY}. Note that these estimates
deteriorate near the spectral edges.

It is well known that the identification of the Stieltjes transform
of a measure for the complex parameters $z=E+i\eta$, $E\in \R$,  is equivalent
to knowing the density down to scales essentially $O(\eta)$, thus
we obtain the control on the density down to scales essentially of order $\eta\sim 1/M$.

\bigskip

The Stieltjes transform $m(z)$ can also be viewed as the normalized
trace of the resolvent,
$$
  m(z) = \frac{1}{N} G(z) = \frac{1}{N}\sum_{i=1}^N G_{ii}(z), \qquad G(z): = \frac{1}{H-z}.
$$
In addition to \eqref{mmmbest}, we are able to prove 
 that not only the sum,
but each diagonal element $G_{ii}(z)$ is given by the semicircle law, but 
the precision is weaker:
\be
  \max_i |G_{ii}(z)- m_{sc}(z)|\lesssim \frac{C}{\sqrt{N\eta}}, \qquad z= E+i\eta.
\label{GIIM}
\ee
Finally, we can also show that the off-diagonal resolvent elements are small:
\be
  \max_{ij} |G_{ij}(z)|\lesssim\frac{C}{\sqrt{N\eta}}
\label{GIJ}
\ee
with logarithmic corrections \cite{EYY3}  (see Theorem \ref{45-1}
in Section \ref{sec:best}).
In our previous papers, \cite{EYY, EYY2},
the constant $C$ in \eqref{GIIM} and \eqref{GIJ} 
depended on $\kappa$, i.e. the estimates
deterioriated near the
spectral edge as an inverse power of $\kappa$; the
exponent  depends on whether a positive  uniform lower bound $\delta_+>0$ is available
or not. For more general Wigner matrices, e.g.
for band matrices, we obtain the same estimates
but $M$ replaces $N$ on the right hand sides of \eqref{GIIM} and \eqref{GIJ}
and $C$ depends on $\kappa$.
The precise statements are given in Theorem \ref{lsc}.

The asymptotics of the Stieltjes transform can be translated into the asymptotics
of the counting function (e.g. Theorem \ref{prop:count}) or into a result on
 the location of the eigenvalues (Theorem \ref{prop:lambdagamma}).
Moreover, the local semicircle law easily implies that the
eigenvectors are fully {\it delocalized}
 (see Section~\ref{sec:deloc}).

\subsubsection{Summary of Section \ref{sec:sine}: 
Main results on the bulk universality with Gaussian component}
\label{sec:maingauss}

Bulk universality refers to the fact that local eigenvalue statistics,
i.e., correlation functions of eigenvalues rescaled by a factor $N$,
or the distribution of the gap between consecutive eigenvalues
exhibit universal behavior which is solely determined
by the symmetry class of the ensemble. 

Bulk universality has first been proved for Gaussian Wigner ensembles, i.e., when
 the matrix elements $h_{ij}$  are i.i.d. Gaussian
random variables by Dyson \cite{Dys1} and Mehta  \cite{M2}.
The Gaussian character makes explicit calculations easier that
are needed to identify the limiting correlation functions
(e.g. the celebrated {\it sine kernel} for the hermitian case).
The key fact is that the joint distribution function for the eigenvalues of
such ensembles is explicit and it contains  a Vandermonde determinant structure
from which the local universality can be deduced, see \eqref{expli2}.

It is a natural idea to consider a broader class of matrices that
still have some Gaussian character; the useful concept is
the {\it Gaussian divisible} ensembles, i.e., where the 
probability law of each matrix elements contains a Gaussian
component (Gaussian convolution). 

One approach with Gaussian convolutions is to push the explicit calculations further
by finding a similar Vandermonde structure. 
Based upon an earlier paper of Br\'ezin and Hikami \cite{BH},
Johansson \cite{J} has found
a representation formula for correlation functions
and he was able to prove bulk universality 
 for Gaussian divisible matrices.
For an algebraic reason, this method
is available for the hermitian case only.

The size of the Gaussian component in \cite{J} was substantial;
it was of the same order as the non-Gaussian part.
Using our local semicircle law and a slightly modified version of
an explicit representation formula of Johansson \cite{J} we 
were able to prove  bulk universality for hermitian
Wigner matrices with a tiny Gaussian component of variance
$O(N^{-1+\e})$ with an improved formula in \cite{EPRSY} (Section \ref{sec:conv}).

The second approach  (sketched in Section~\ref{sec:lrf}
and elaborated in Section~\ref{sec:sine})
is to embed the Gaussian divisible ensemble
into a stochastic flow of matrices, and use the key
observation of Dyson \cite{Dy} that under this flow the
eigenvalues perform a specific stochastic dynamics 
with a logarithmic interaction, the celebrated {\it Dyson Brownian Motion}.
Eventually the dynamics relaxes to equilibrium, which is the well known
Gaussian model (GUE, GOE or GSE). The main idea is that
the {\it local}  relaxation is much faster, i.e., the local statistics of eigenvalues
already reach their equilibrium within a very short $t=N^{-\e}$
time (with an explicit $\e>0$). In fact, Dyson \cite{Dy} has
predicted that the time scale to local equilibrium is of order $N^{-1}$,
which we eventually proved in \cite{EYY3}.
Our main result states that the local correlation functions of Gaussian
divisible matrices with a small Gaussian component 
coincide with the correlation functions
of the purely Gaussian ensembles.

This result can be formulated in a general setup and
viewed as a strong local ergodicity of the Dyson Brownian motion
or, in fact, of  any one dimensional stochastic particle
dynamics with  logarithmic interaction. This general formulation appeared first
in \cite{ESYY} and it will be given in Theorem \ref{thmM},
but most of the key ideas were invented  in \cite{ESY4}.
 For the application of this general principle
to random matrices one needs certain apriori information
about the location of the eigenvalues, which we obtain from the local
semicircle law. In particular, using this idea,
the bulk universality for {\it symmetric} Wigner matrices was first proved
in \cite{ESY4}. The cases of quaternion self-dual and sample covariance
matrices were treated in \cite{ESYY}.

\subsubsection{Summary of Section \ref{sec:4mom}:
Main results on the removal of the Gaussian component}

To prove the universality of any Wigner ensemble, we need to
compare it with a Gaussian divisible ensemble for which universality has 
already been proven. Such comparison principle is plausible if
 the Gaussian component is small and indeed a perturbation
argument can be applied. It is essentially a density
argument, stating that Gaussian divisible ensembles
are sufficiently ``dense'' in the space of all Wigner ensembles.

The first result of this type used a {\it reversed  heat flow}
argument \cite{EPRSY}, where we showed that any smooth distribution
can be approximated  with a very high precision
by a Gaussian divisible distribution. Combining this method
with the bulk universality for hermitian Wigner matrices
with a Gaussian component of variance
of order $O(N^{-1+\e})$, we were able
to prove bulk universality for any Wigner ensemble under 
the condition that the distribution of the matrix elements is smooth.

A more robust approach is the Green function comparison theorem
from \cite{EYY}, which states that for two matrix
ensembles, the joint distribution of
the Green  functions coincides, provided that the first four
moments of the probability law of the matrix elements are
identical or very close. The spectral parameter $z$ can
have a very small imaginary part $\im z\sim N^{-1-\e}$,
i.e., these Green functions can detect individual eigenvalues.
The precise statement is given in Theorem \ref{comparison}.
The key input is the local semicircle law involving 
individual matrix elements of the resolvent, \eqref{GIIM}--\eqref{GIJ}.

The combination of the results of Section \ref{sec:maingauss}
on the bulk universality of Gaussian divisible matrices
and the Green function comparison theorem gives the bulk universality
for any Wigner ensemble by a simple matching argument \cite{EYY2}.
 The method applies even to
matrices with comparable variances \eqref{VV}.
 The only condition in the approach is a subexponential
decay for the tail of the probability law of the matrix elements
\eqref{subexpuj}. In fact, this condition can be relaxed to 
a sufficiently fast polynomial decay, but for simplicity we will
not pursue this direction.

The {\it  four moment condition} was first 
observed by Tao and Vu \cite{TV} in the four moment
theorem for eigenvalues (Theorem \ref{thm:TV}). Their key technical
input is also the local semicircle law and its corollary on delocalization of
eigenvectors.
They used  this result  to prove the universality for hermitian Wigner matrices
without smoothness condition but under some moment and
support condition, that especially excluded the Bernoulli distribution.
 The bulk universality for  
hermitian Wigner matrices including the Bernoulli case the was first proved in
\cite{ERSTVY} after combining the
results of \cite{EPRSY} and \cite{TV}.

\bigskip

Finally, in Section~\ref{sec:put} we state the main result (Theorem~\ref{mainthm})
on bulk universality and we summarize how its proof follows
from the previous sections.
Currently the local relaxation flow method combined with the Green function
comparison theorem gives the most general approach to
bulk universality. This path not only proves
bulk universality for general Wigner ensembles, but
it also offers a conceptual understanding how 
universality emerges from simple principles.

\bigskip

In Sections \ref{sec:wig}--\ref{sec:wegner}  we review several facts,
results and methods in connection with  Wigner random matrices
and some related ensembles. 
These sections are meant to provide a general background information.
 In Section \ref{sec:locnew} we also explain the key new ideas listed above
in more details and give a summary of various results
on bulk universality. 
A reader wishing to focus only on the most recent
developments  can skip Sections \ref{sec:wig}--\ref{sec:wegner}
and jump to Section \ref{sec:sc}.

\subsection{Wigner matrix ensemble}\label{sec:wig}

A central question in probability theory is the universality of cumulative
statistics of a large set of independent data. Given an array of
$N$ independent random variables
\be
   (X_1, X_2, \ldots, X_N)
\label{array}
\ee
one forms linear statistics like the mean or the fluctation
\be
   \ov X^{(N)}: = \frac{1}{N}\sum_{j=1}^N X_j, \qquad
   S^{(N)}: = \frac{1}{\sqrt{N}} \sum_{j=1}^N (X_j-\E X_j).
\label{XS}\ee
Under very general conditions,  a universal pattern emerges
as $N\to\infty$: the mean converges
to  its expectation, in particular, it becomes deterministic,
$$
    \ov X^{(N)}\to  \lim_{N\to\infty}\E \ov X^{(N)},
$$
assuming that the latter limit
exists (law of large numbers). 
Moreover, 
 the fluctation $S^{(N)}$ converges to a centered normal Gaussian random variable $\xi$
\be
  S^{(N)} \to \xi   \qquad \mbox{(in distribution)}
\label{SN}
\ee
(central limit theorem),
i.e., the density function of $\xi$ is given by
$f(x) = (\sqrt{2\pi}\sigma)^{-1}\exp{(-x^2/2\sigma^2)}$.
The variance $\sigma^2$ of $\xi$ is given by the average of the
variances of $X_j$,
$$
  \sigma^2 := \lim_{N\to\infty}
 \frac{1}{N}\sum_{j=1}^N \sigma_j^2, \qquad \sigma_j^2:= \E \big[ X_j - \E X_j\big]^2.
$$
 In particular, for independent, identically distributed (i.i.d.)
random variables, 
$\ov X^{(N)}$ converges to the common expectation value of  $X_j$'s, 
and $S^{(N)}$ converges to the centered normal distribution with the 
common variance of $X_j$'s.

The emergence of a single universal distribution, the Gaussian, is a
remarkable fact of Nature. It shows that large systems with many independent components
in a certain sense behave identically, irrespective of the
details of the distributions of the components.

\bigskip

It is natural to generalize this question of universality from arrays \eqref{array} to 
double arrays, i.e., to matrices:
\be\label{XMN}
X^{(N,M)} = 
\begin{pmatrix} X_{11} &  X_{12} & \ldots & X_{1N} \\ 
 X_{21} &  X_{22} & \ldots & X_{2N} \\
\vdots & \vdots &  & \vdots\\
 X_{M1} &  X_{M2} & \ldots & X_{MN} \\
\end{pmatrix}
\ee
with independent entries.
The statistics in question should involve a quantity which reflects
the matrix character and is influenced by all entries, for example
the (Euclidean) norm  of $X^{(N,M)}$. Although the norm of each random realization
of $X^{(N,M)}$ may differ, it is known, for example, that in the limit as $N,M\to\infty$,
such that $N/M\to d$, $0<d\le 1$ is fixed, it becomes deterministic, e.g. we have \cite{MP, Wa}
\be
  \frac{1}{\sqrt{M}} \| X^{(N,M)} \| \to \sigma (1+\sqrt{d})
\label{norm}
\ee
assuming that the matrix elements are centered, $\E X_{ij}=0$, and
their average variance is $\sigma^2$. Note that the typical size of 
$ X^{(N,M)}$ is only of order $\sqrt{M}$ despite that the matrix
has dimensions $M\times N$ 
 filled with elements of size $O(1)$.
If the matrix elements were strongly correlated then the norm
could  be of order $M$. For example, in the extreme case, if all elements were the same,
$X_{ij}= X$, then $\| X^{(N,M)} \|\sim M$. Independence of the matrix elements
prevents such conspiracy and it reduces the typical size of the matrix 
by a factor of $\sqrt{M}$, similarly to  the central limit theorem (note
the $\sqrt{N}$ normalization in \eqref{XS}).

\medskip 

Matrices offer  a much richer structure than studying only their norm.  Assuming $M=N$,
the most important characteristics of a square matrix are the 
 eigenvalues and eigenvectors. As \eqref{norm} suggests, it is convenient
to assume zero expectation for the matrix elements and rescale the matrix
by a factor $N^{-1/2}$ to have a norm of order 1. For most of this
presentation, we will therefore consider
large $N\times N$ square matrices of the form
\be\label{Hdef}
H=H^{(N)} = 
\begin{pmatrix} h_{11} &  h_{12} & \ldots & h_{1N} \\ 
 h_{21} &  h_{22} & \ldots & h_{2N} \\
\vdots & \vdots &  & \vdots\\
 h_{N1} &  h_{N2} & \ldots & h_{NN} \\
\end{pmatrix}
\ee
with {\it centered} entries
\be
   \E\, h_{ij} = 0, \qquad i,j =1,2,\ldots, N.
\label{centered}
\ee
As for the  normalization, we assume  that the  matrix of variances
\be
  \Sigma: = \begin{pmatrix} \sigma^2_{11} &  \sigma^2_{12} & \ldots & \sigma^2_{1N} \\ 
 \sigma^2_{21} &  \sigma^2_{22} & \ldots & \sigma^2_{2N} \\
\vdots & \vdots &  & \vdots\\
 \sigma^2_{N1} &  \sigma^2_{N2} & \ldots & \sigma^2_{NN} \\
\end{pmatrix}, \qquad \sigma_{ij}^2: = \E |h_{ij}|^2.
\label{def:Sigmamatrix}
\ee
is doubly stochastic, i.e.,
for every $i=1,2,\ldots , N$ we have
$$
   \sum_{j} \sigma_{ij}^2 =\sum_{j}\sigma_{ji}^2 =1.
$$
The most natural example is the {\bf mean-field model}, when
$$
   \sigma_{ij}^2 = \frac{1}{N}  \qquad i,j=1,2, \ldots, N,
$$
i.e., each matrix element is of size $h_{ij}\sim N^{-1/2}$.
This corresponds to the standard Wigner matrix.
For most of this presentation the reader can restrict
the attention to this case.

Random matrices are typically subject to some symmetry restrictions, e.g. we will consider
symmetric $(h_{ij}= h_{ji}\in \R)$ or hermitian $(h_{ij} =\ov h_{ji}\in \bC)$
random matrices. We will mostly assume that the matrix elements
are {\it independent} up to the symmetry requirement
(i.e. in case of symmetric or hermitian matrices, the variables $\{ h_{ij}\; : \; i\le j\}$
are independent). This leads us to the

\begin{definition}\label{def:gen}
An $N\times N$ symmetric or hermitian  random matrix \eqref{Hdef} is
called {\bf universal Wigner matrix (ensemble)} if the entries are centered
\eqref{centered}, their variances $\sigma_{ij}^2= \E |h_{ij}|^2$ satisfy
\be
\sum_{j} \sigma_{ij}^2=1, \qquad i=1,2,\ldots, N
\label{sum}
\ee
and $\{ h_{ij}\; : \; i\le j\}$ are independent.
An important subclass of universal Wigner ensembles is
called {\bf generalized Wigner matrices (ensembles)}
if, additionally, the variances are comparable, i.e. 
\be
 0< C_{inf}\le   N\sigma_{ij}^2  \le C_{sup} < \infty, \qquad i,j=1,2, \ldots, N,
\label{VV}
\ee
holds with some fixed positive constants $C_{inf}$, $C_{sup}$. 
In the special case  $\sigma_{ij}^2=1/N$, we recover the
original definition of the {\bf Wigner matrices} or {\bf Wigner ensemble}
\cite{W}.
\end{definition}

The most prominent Wigner ensembles  are the {\it Gaussian Orthogonal Ensemble} (GOE)
and the {\it Gaussian Unitary Ensemble} (GUE); i.e., symmetric and hermitian
Wigner matrices with rescaled matrix elements $\sqrt{N} h_{ij}$ being  standard Gaussian variables
(in the hermitian case, $\sqrt{N} h_{ij}$ is a standard complex Gaussian variable, i.e.
 $\E |\sqrt{N} h_{ij}|^2 =1$). 

For simplicity of the presentation, in case of the Wigner ensembles, we will
assume that $h_{ij}$, $i<j$,  are identically distributed (i.e. not only
their variances are the same). In this case we fix a distribution $\nu$
and we assume that the rescaled matrix elements $\sqrt{N}h_{ij}$ are
distributed according to $\nu$. 
Depending on the symmetry type, the
diagonal elements may have a slightly different distribution,
but we will omit this subtlety from the discussion.
The distribution $\nu$ will be called the {\it single entry distribution of $H$}.

We will sometimes mention a special class of universal Wigner matrices that have a
band structure; they will be called {\it random band matrices}. The variances are given by 
\be\label{BM}
   \sigma^2_{ij} = W^{-1} f\Big(\frac{ [i-j]_N}{W}\Big),
\ee
where $W\gg 1$, $f:\bR\to \bR_+$ is a bounded nonnegative symmetric function with 
$\int f(x)\rd x =1$ 
and we defined $[i-j]_N\in \bZ$ by the property
that  $[i-j]_N\equiv i-j \; \mbox{mod} N$
and $-\frac{1}{2}N < [i-j]_N \le\frac{1}{2}N$.  
 Note that 
the relation \eqref{sum} holds only asymptotically as $W\to \infty$
but this can be remedied by an irrelevant rescaling.
One can even consider {\it $d$-dimensional band matrices}, where the rows and
columns are labelled by a finite lattice $\Lambda\subset \bbZ^d$ and
$\sigma_{ij}^2$
 depends on the difference $i-j$ for any $i,j\in \Lambda$.

Another class of random matrices, that even predate Wigner, 
 are the {\it random covariance matrices}.
These are matrices of the form 
\be
   H = X^*X,
\label{X*X}
\ee
where $X$ is a rectangular $M\times N$ matrix of the form  \eqref{XMN}
with centered i.i.d. entries with variance $\E |X_{ij}|^2= M^{-1}$.
Note that the matrix elements of $H$ are not independent,
but they are generated from the independent matrix elements of $X$ in
a straightforward way. These matrices appear in statistical samples
and were first considered by Wishart \cite{Wish}.
In the case when  $X_{ij}$ are centered Gaussian, the random
covariance matrices are called {\it Wishart matrices or ensemble.}

\subsection{Motivations: from Schr\"odinger operators to the $\zeta$-function}\label{sec:mot}

We will primarily study the eigenvalue statistics of large random matrices
and some results about eigenvectors will also be mentioned.
The main physical motivation is that a random matrix can model
the Hamilton operator of a disordered quantum system. 
The symmetry properties of $H$ stem from this consideration:
symmetric matrices  represent Hamiltonians of systems 
with time reversal invariance (e.g. no magnetic field),
hermitian matrices correspond to systems without time reversal symmetry.
(There is a third class of matrices, the quaternion self-dual matrices,
most prominently modelled by the {\it Gaussian Symplectic Ensemble (GSE)},
that describe systems with odd-spin and no rotational symmetry, but we
will not discuss it in detail.)

E. Wigner has originally invented random matrices to mimic the eigenvalues
of the  unknown Hamiltonian of heavy nuclei; lacking any
information, he assumed that the matrix elements are i.i.d.
random variables subject to the hermitian condition. 
His very bold vision was that, although such a crude
approximation cannot predict individual energy levels (eigenvalues)
of the nucleus, their {\it statistical properties} may be 
characteristic to some global feature shared by any nucleus.
By comparing measured data of energy levels of nuclei with
numerical calculations of eigenvalues of certain random matrices,
he found that the {\it level statistics}, i.e., the
distribution of the energy gaps between neighboring energy levels (eigenvalues),
show remarkable coincidence and robustness. In particular, he observed
that energy levels tend to repel each other, a significant
difference from the level statistics of fully uncorrelated
 random points (Poisson point process).  Similar feature
was found for random matrices:
even Wigner matrices that are ``as stochastic as possible''
delivered plots of strongly correlated (repelling) eigenvalues. This
correlation is due to the underlying fundamental symmetry
of the matrix ensemble, in particular symmetric and hermitian
matrices were found to have a different strength of level repulsion,
but within a fixed symmetry class a universal pattern emerged.
For more details on the history of this remarkable
discovery, see \cite{M}.

Universality of local eigenvalue statistics is believed to hold
for a much broader class of matrix ensembles than we have introduced.
There is  no reason to believe that the matrix
elements of the Hamiltonian of the heavy nuclei are indeed 
i.i.d. random variables. Conceivably, the matrix
elements need not be fully independent or identically distributed for universality.
There is little known  about matrices with 
correlated entries, apart from the unitary invariant ensembles (Section \ref{sec:inv})
that represent a very specific 
correlation. In case of a certain class of Wigner matrices with weakly correlated
entries, the semicircle law and its Gaussian fluctuation
have been proven \cite{SSh1, SSh2}.

Much more studied are
various classes of random matrices with independent but not identically distributed entries.
The most prominent example is the tight binding Anderson model 
\cite{A}, i.e., a Schr\"odinger operator,
$-\Delta +\lambda V$,  on a regular square lattice $\Z^d$ with a random on-site
potential $V$ and disorder strength $\lambda$. This model describes
electron propagation (conductance) in an ionic lattice with  a disordered environment.
 Restricted to a finite box,
it can be represented by a matrix whose diagonal elements are i.i.d. random variables;
the deterministic off-diagonal elements are  given by the Laplacian.

\medskip

The general formulation of the {\it universality conjecture for random Schr\"odinger
operators}
states that there are two distinctive  regimes depending on the energy and
the disorder strength. In the strong disorder regime, the eigenfunctions
are localized and the local spectral statistics are Poisson.
In the weak disorder regime, the eigenfunctions are delocalized
and the local statistics coincide with those of a Gaussian 
matrix ensemble. 
Separate conjectures, that will not be discussed here, relate these two regimes to chaotic
vs. integrable behavior of the underlying classical dynamical system. According to
the Berry-Tabor conjecture \cite{BT},
Poisson statistics of eigenvalues should emerge from quantizations
of integrable  classical dynamics, while random matrix theory
stems from quantization of chaotic classical dynamics (Bohigas, Giannoni, Schmit \cite{BGS})

\medskip

Returning to the more concrete Anderson model,
 in space dimensions three or higher and for weak randomness,
the model is conjectured to exhibit metal-insulator transition,
 i.e.,\ in $d\geq 3$ dimensions the eigenfunctions of  $-\Delta+\lambda V$
are delocalized for small  $\lambda$, while they are localized for large $\la$.
It is a  fundamental open mathematical question  to establish this transition.

 The localization regime at large disorder or near the spectral
edges has been well  understood by Fr\"ohlich and Spencer
with the multiscale technique \cite{FS, FMSS}, and later by Aizenman and Molchanov
by the  fractional moment method \cite{AM}; many other
works have since contributed to this field. In particular, it
has been established that the local eigenvalue statistics are Poisson \cite{Mi} and
that the eigenfunctions are exponentially localized
with an upper bound on the localization length that diverges
 as the energy parameter approaches the presumed phase transition 
point \cite{Spen, Elg}.

The progress in the delocalization regime has been much slower. For the Bethe lattice, 
corresponding to the infinite-dimensional case, delocalization has been established in 
\cite{Kl, ASW, FHS} (in an apparent controversy to the general conjectures,
the eigenvalue statistics, however, are Poisson but for a well understood specific
reason \cite{AW}).
In finite dimensions only partial results are available. The existence of an absolutely 
continuous spectrum (i.e.,\ extended states) has been shown for a rapidly decaying potential, 
corresponding to a scattering regime \cite{RS, B, Den}.  Diffusion
has been established  for a heavy quantum particle immersed in a
phonon field in $d\geq 4$ dimensions \cite{FdeR}.
For the original Anderson Hamiltonian  with a small coupling constant $\lambda$,
the eigenfunctions have a
localization length of at least $\lambda^{-2}$ 
\cite{Ch}. The time and space scale 
$\lambda^{-2}$ corresponds to
the kinetic regime where the quantum evolution can be
modelled by a linear Boltzmann equation \cite{Sp1, EY}. Beyond this
time scale the dynamics is diffusive. This has been established
in the scaling limit $\lambda\to0$
up to time scales $t\sim \lambda^{-2-\kappa}$ with an explicit $\kappa>0$
in \cite{ESY}.

There are no rigorous results on the local spectral statistics of the Anderson model
in the disordered regime, but
it is conjectured -- and supported by numerous arguments in
the physics literature, especially by supersymmetric methods
 (see \cite{Efe}) -- that the local 
correlation functions of the eigenvalues of the finite volume Anderson
model follow the GOE statistics in the thermodynamic limit.
GUE statistics are expected if an additional magnetic field breaks the time-reversal symmetry
of the Anderson Hamiltonian.
Based upon this conjecture, the local eigenvalue statistics are
used to compute the phase diagram numerically.
It is very remarkable that the random Schr\"odinger operator,
represented by a very sparse random matrix, exhibits
the same universality class as the full Wigner matrix, at least in
a certain energy range.

\bigskip

Due to their mean-field character, Wigner matrices are simpler to study
than the Anderson model  and  they are always in the delocalization regime.
In this survey we mainly focus on Wigner matrices, but
we keep in mind the original motivation from general disordered systems.
In particular, we will study not only eigenvalue statistics
but also eigenvectors that are shown to be completely delocalized \cite{ESY4}.
The local
spectral statistics in the bulk are universal, i.e.,\ it follows the statistics
of the corresponding Gaussian ensemble (GOE, GUE, GSE), depending
on the symmetry type of the matrix. This topic will be the main
goal of this presentation.

\bigskip

To close this section, we mention some other possible research
directions  that we will not pursue here further. The list is incomplete.

A natural intermediate class of
ensembles between the fully stochastic Wigner matrices and the
Anderson model with diagonal randomness  is the family of {\it random band matrices}.
These are hermitian or symmetric random matrices $H$ with 
independent but not identically distributed entries. The variance
of $h_{ij}$ depends only on $|i-j|$ and it becomes negligible
if $|i-j|$ exceeds a given parameter $W$, the band-width; for example,
$\sigma_{ij}^2=\E |h_{ij}|^2 \sim \exp(-|i-j|/W)$. It is conjectured \cite{Fy}
that the system is completely delocalized if $W\gg \sqrt{N}$, otherwise
the localization length is $W^2$. Moreover,
 for narrow bands, $W\ll \sqrt{N}$, the local eigenvalue
statistics are expected to be Poisson, while for broad bands, $W\gg \sqrt{N}$
they should be given by GUE or GOE, depending on the symmetry class.
Localization properties of $H$ for $W\ll N^{1/8}$ and an $O(W^8)$ upper bound 
on the localization length 
have been  shown by J. Schenker \cite{Sch} but not local statistics.
{F}rom the delocalization side, with A. Knowles we recently proved \cite{EK, EK2}
diffusion up to time scale $t\ll W^{1/3}$ which implies that
the localization length is at least $W^{1+1/6}$.

We mention that universality of local eigenvalue statistics is often
investigated by supersymmetric techniques in the physics literature.
 These methods are extremely powerful
to extract the results by saddle point computations, 
but the analysis justifying the saddle point
approximation still lacks mathematical rigor.
So far only the density
of states has been investigated 
rigorously by using this technique  \cite{DPS}.
Quantum diffusion can also be studied by supersymmetry
and certain intermediate models can be rigorously analyzed \cite{DS, DSZ}.

Finally, we point out  that we focused on the physical motivations coming from 
disordered quantum systems, but random matrices appear in
many other branches in physics and mathematics. It is a fundamental
object of nature with an extremely rich structure. The most remarkable
connection is with the $\zeta$-function. It is conjectured
that the roots of the Riemann $\zeta$-function, $\zeta(s):= \sum_{n=1}^\infty n^{-s}$,
lying on the vertical line $\re s = \frac{1}{2}$, have the same 
local statistics as the GUE (after appropriate rescaling). 
A review and further references
to  many numerical evidences is found in the classical book of Mehta \cite{M}.

\subsection{Eigenvalue density and delocalization}\label{sec:density}

For symmetric or hermitian matrix $H$,
let $\lambda_1\le \lambda_2\le \ldots \le \lambda_N$ denote the
eigenvalues. They form a random point process on the real line
with a distribution generated from the joint probability law 
of the matrix elements.  Since the functional relation between
matrix elements and eigenvalues is highly nontrivial, the product measure
on the entries turns out to generate a complicated and highly correlated
measure for the eigenvalues.
Our main goal is to understand this induced measure.

Under the chosen normalization \eqref{sum},
the typical size of the eigenvalues is of order one. We will prove a much more precise
statement later, but it is instructive to have a rough feeling about the size
via computing $\tr H^2$ in two ways:
$$
   \sum_{i=1}^N\lambda_i^2 = \tr H^2 = \sum_{i,j=1}^N |h_{ij}|^2.
$$
Taking expectation and using \eqref{sum} we have
$$
  \frac{1}{N}\sum_i \E \lambda_i^2  = \frac{1}{N}\sum_{ij} \sigma_{ij}^2 = 1
$$
i.e., in an average sense $\E \lambda_i^2=1$.

\subsubsection{Wigner semicircle law and other canonical densities}

The empirical distribution of eigenvalues follows a universal pattern,
the {\it Wigner semicircle law}. To formulate it more precisely, note that
the typical spacing between neighboring eigenvalues is of order $1/N$, so
in a fixed interval $[a,b]\subset \R$, one expects macroscopically many 
(of order $N$) eigenvalues.
More precisely, it can be shown (first proof was given by Wigner \cite{W})
 that  for any fixed $a\le b$ real numbers,
\be
   \lim_{N\to\infty} \frac{1}{N}\#\big\{ i \; : \; \lambda_i\in [a,b]\big\}
  = \int_a^b \varrho_{sc}(x)\rd x, \qquad \varrho_{sc}(x) : = \frac{1}{2\pi}
  \sqrt{(4-x^2)_+},
\label{sc}
\ee
where $(a)_+:=\max\{ a, 0\}$ denotes the positive part of the number $a$.
Note the emergence of the universal density, the semicircle
law, that is independent of the details of the distribution of
the matrix elements.

The semicircle law is characteristic for the universal Wigner matrices
(see Definition \ref{def:gen}).
For random square matrices with independent entries but {\it without symmetry}
(i.e., $h_{ij}$ are independent for all $i,j$) a similar universal pattern
emerges, the {\it circular law}. For example, if $h_{ij}$ are centered i.i.d.
random variables with common variance $\sigma_{ij}^2 = N^{-1}$, then
the empirical density of eigenvalues converges to the uniform measure on
the unit disk in the complex plane \cite{TVK}.
If independence is dropped, one can get many different 
density profiles.

For example,
in case of the random covariance matrices \eqref{X*X}, 
the empirical density of eigenvalues $\lambda_i$ of $H$
converges to the {\it Marchenko-Pastur law} \cite{MP}
in the limit when $M,N\to \infty$
such that $d= N/M$ is fixed $0\le d\le 1$:
\be
   \lim_{N\to\infty} \frac{1}{N}\#\big\{ i \; : \; \lambda_i\in [a,b]\big\}
  = \int_a^b \varrho_{MP}(x)\rd x, \qquad \varrho_{MP}(x) : = \frac{1}{2\pi d}
  \sqrt{\frac{\big[ (\lambda_+-x)(x-\lambda_-)\big]_+}{x^2}}
\label{MP}
\ee
with $\lambda_\pm : = (1\pm \sqrt{d})^2$ being the spectral edges.
Note that in case $M\le N$, the matrix $H$ has macroscopically many
zero eigenvalues, otherwise the spectra of $XX^*$ and $X^*X$ coincide
so the Marchenko-Pastur law can be applied to all nonzero eigenvalues
 with the role of $M$ and $N$ exchanged.

\subsubsection{The moment method}\label{sec:moment}

The eigenvalue density is  commonly approached via the fairly robust 
{\it moment method} (see \cite{AGZ} for an expos\'e)
that was also the original approach of Wigner
to prove the semicircle law \cite{W}. For example, for hermitian Wigner matrices,
it consists of computing traces of high powers of $H$, i.e.,
$$
   \E\, \tr H^{2k}
$$
by expanding the product as 
$$
  \E\, \sum_{i_1, i_2, \ldots i_{2k}} h_{i_1i_2}h_{i_2i_3}\ldots h_{i_{2k}i_1}
$$
and noticing that each factor $h_{xy}$ must be at paired with at least another copy
$h_{yx}= \bar h_{xy}$, otherwise the expectation value is zero. The possible
 index sequences that satisfy this pairing conditions can be classified according
to their complexity, and it turns out that the main contribution comes from
the so-called {\it backtracking paths}.  These are index sequences
$i_1i_2i_3\ldots i_{2k}i_1$, returning to the original index $i_1$, 
that can be successively  generated by a substitution rule
$$
   a\to aba, \qquad b\in \{ 1, 2,\ldots, N\}, \quad b\ne a,
$$
with an arbitrary index $b$. These index sequences satisfy the pairing condition
in an obvious manner and it turns out that they involve the largest possible number ($N^k$)
independent indices.  The number of backtracking paths is explicitly
 given by the Catalan numbers, $C_k = \frac{1}{k+1}{2k\choose k}$, so 
$ \E \tr H^{2k}$ can be computed fairly precisely for each finite $k$:
\be
  \frac{1}{N} \E \tr H^{2k} = \frac{1}{k+1}{2k\choose k} + O_k(N^{-2}).
\label{cat}
\ee
Note that the number of independent labels, $N^k$, exactly cancels the
size of the $k$-fold product of variances, $(\E |h|^2)^k=N^{-k}$.
If the distribution of the matrix elements is symmetric, then
the traces of odd powers all vanish since they can never satisfy the
pairing condition. Without the symmetry condition the traces of odd powers are
non-zero but negligible.

We will compute the trace of the resolvent, or the {\it Stieltjes transform}
of  the {\it empirical density} 
$$
\varrho_N(\rd x): = \frac{1}{N}\sum_1^N \delta(x-\lambda_j)
$$
of the eigenvalues, i.e. we define
\be
   m(z)= m_N(z): = \frac{1}{N}\tr \frac{1}{H-z} = \frac{1}{N}\sum_{j=1}^N
  \frac{1}{\lambda_j-z}=\int_\R \frac{\rd\varrho_N(x)}{x-z}
\label{St}
\ee
for any $z=E+i\eta$, $E\in \R$, $\eta>0$.
For large $z$ one can expand $m_N$ as follows
\be
  m_N(z) =  \frac{1}{N}\tr \frac{1}{H-z} =
 -\frac{1}{Nz}\sum_{m=0}^\infty \Big( \frac{H}{z}\Big)^m,
\label{mexp}
\ee
so after taking the expectation, using \eqref{cat} and neglecting the error terms, we get
\be
  \E\, m_N(z) \approx  - \sum_{m=0}^\infty \frac{1}{k+1}{2k\choose k} z^{-(2k+1)},
\label{anal}
\ee
which, after some calculus, can be identified as the 
power series of $\frac{1}{2}(-z+\sqrt{z^2-4})$.
The approximation becomes exact in the $N\to\infty$ limit. 
Although the expansion \eqref{mexp} is valid
only for large $z$, given that the limit is an analytic function of $z$, one
can extend the relation
$$
   \lim_{N\to\infty}\E m_N(z) = \frac{1}{2}(-z+\sqrt{z^2-4})
$$
by analytic continuation to the whole upper half plane $z=E+i\eta$, $\eta>0$.
It is an easy exercise to see that this is exactly the Stieltjes transform of
the semicircle density, i.e.,
\be
  m_{sc}(z):=\frac{1}{2}(-z+\sqrt{z^2-4}) = \int_\R \frac{\varrho_{sc}(x)\rd x}{x-z}.
\label{msc}
\ee
The square root function is chosen with a branch cut in the segment $[-2, 2]$
so that $\sqrt{z^2-4}\sim z$ at infinity. This guarantees that 
 $\im m_{sc}(z)>0$ for $\im z>0$.
Since the Stieltjes transform identifies the measure uniquely, and pointwise convergence
of Stieltjes transforms implies weak convergence of measures, we obtain 
\be
   \E \,\rd\varrho_N(x) \wkto \varrho_{sc}(x)\rd x.
\label{global:sc}
\ee
With slightly more efforts one can show that 
\be
   \lim_{N\to\infty} m_N(z) = \frac{1}{2}(-z+\sqrt{z^2-4})
\label{limst}
\ee
holds with high probability, i.e.,
the convergence holds
also in probability not only in expectation.
For more details, see \cite{AGZ}.

\subsubsection{The local semicircle law}\label{sec:locsc}

The moment method can typically identify the resolvent for any fixed $z$
and thus give the semicircle law as a weak limit, i.e., 
\eqref{sc} will hold {\it for any fixed} interval $I:=[a,b]$ as $N\to\infty$.
However, a fixed interval $I$ with length $|I|$ typically contains of order $N|I|$ eigenvalues.
It is natural to ask whether the semicircle law holds {\it locally} as well, i.e.,
for intervals whose length may shrink with $N$, but still $N|I|\gg 1$.
Eventually, the semicircle law is a type of law of large numbers
 that should require only that the number of random
objects in consideration goes to infinity. Due to the formula
$$
    \im m_N(z) = \frac{1}{N} \sum_{i=1}^N \frac{\eta}{(\la_i - E)^2 +\eta^2} \sim 
  \frac{\pi}{N}\sum_{i=1}^N \delta_\eta(\la_i-E), \qquad
  z= E+i\eta,
$$
where $\delta_\eta$ denotes an approximate delta function on scale $\eta$,
we see that knowing the Stieltjes transform for some $z\in \C$ with $\im z= \eta$
is essentially equivalent to knowing the local density on scale $\eta$, i.e.,
in an interval of length $|I|\sim \eta$.

In \cite{ESY1, ESY2, ESY3} we proved that the {\it local} semicircle
law holds on the smallest possible scale of $\eta\gg 1/N$, i.e.,
the limit \eqref{sc} holds even if the length of $I=[a,b]$ is essentially
of order $1/N$, hence it typically contains only large but finite 
number of eigenvalues. This will be the key technical input for further investigations on
local spectral statistics. 
There are several versions of the local semicircle law; we will give
three precise statements: Theorem \ref{wegnerlsc} (from \cite{ESY3}), Theorem \ref{lsc}
(from \cite{EYY2}) and Theorem \ref{45-1} (from \cite{EYY3}).

The method of the proof is different
from the moment method, but we still work with the resolvent, or the Stieltjes
transform. The key observation (see also several previous works, e.g.
\cite{BY, MP}) is that the Stieltjes transform $m_{sc}(z)$ of the
semicircle density $\varrho_{sc}$   satisfies the following
simple quadratic equation:
\be
   m_{sc}(z) + \frac{1}{z+ m_{sc}(z) }= 0,
\label{selfcons}
\ee
and among the two possible solutions, $m_{sc}(z)$ is identified
as explained after \eqref{msc}.
The strategy is that expanding the empirical Stieltjes transform $m_N(z)$
\eqref{St} according to minors of $H$, we prove that $m_N$ satisfies
the self-consistent equation \eqref{selfcons} approximately and with a high probability:
$$
   m_N(z) + \frac{1}{z+ m_N(z)} \approx 0.
$$
Then we conclude the proof of $m_N\approx m_{sc}$ 
by invoking the stability of the equation \eqref{selfcons}.
Since the stability deteriorates near the edges, $E=\re z \approx \pm 2$,
the estimate will be weaker there, indicating that the eigenvalue fluctuation
is larger at the edge.

Our best results in this direction are obtained in \cite{EYY3} (which is partly
a streamlined version of \cite{EYY, EYY2}), where not only the
 trace of the Green function \eqref{St} but also individual diagonal
elements were shown to be given by the semicircle law.
The results were already listed informally in Section \ref{sec:mainlsc}
and we pointed out that they hold also for universal Wigner matrices,
see Theorems \ref{lsc} and \ref{45-1}.

  For the universal Wigner ensembles Guionnet \cite{gui}
 and Anderson-Zeitouni  \cite{AZ} already proved that the density of the
 eigenvalues converges to the Wigner semi-circle law on a large scale, our result
improves this to small scales.
For example, for  band matrices (for Definition see \eqref{BM})
with band width $W$ we obtain that the semicircle law holds down to energy scales $1/W$.
The delocalization
length is shown to be at least as large as the band width $W$.
We note that a certain three dimensional version of 
 Gaussian band matrices was also considered by 
Disertori, Pinson and Spencer \cite{DPS} using the supersymmetric method. 
They proved that the expectation of the density of eigenvalues 
is smooth and it  coincides with the Wigner semicircle law
up to a precision determined by the bandwidth.

\subsubsection{Density of eigenvalues for invariant ensembles}

There is another natural way to define  probability distributions on symmetric or hermitian
matrices apart from directly imposing a given probability law $\nu$ on their entries.
They are obtained by defining a density function  directly on the set of matrices:
\be
   \cP(H) \rd H : = Z^{-1} \exp{ (-N\tr V(H)) }\rd H.
\label{ph}
\ee
Here $\rd H = \prod_{i\le j} \rd H_{ij} $ is the flat Lebesgue measure
(in case of hermitian matrices and $i<j$, $\rd H_{ij} $ is the Lebesgue measure
on the complex plane $\bC$). The function $V:\bR\to\bR$ is assumed to
grow mildly at infinity (some logarithmic growth would suffice)
to ensure that the measure defined in \eqref{ph} is finite, and $Z$ is
the normalization factor. Probability distributions of the form \eqref{ph}
are called {\it invariant ensembles} since they are invariant under
the orthogonal or unitary conjugation (in case of symmetric or hermitian
matrices, respectively). For example,  in the hermitian case,
for any fixed unitary matrix $U$, the transformation
$$
    H \to U^* H U
$$
leaves the distribution \eqref{ph} invariant thanks to $\tr V(U^*HU)=\tr V(H)$.

Wigner matrices and  invariant ensembles form two different universes with
quite different mathematical tools  available for their studies. In fact, these two
classes are almost disjoint, the Gaussian ensembles being the
only invariant Wigner matrices. This is the content of the following lemma
(\cite{D} or Theorem 2.6.3 \cite{M}).

\begin{lemma}\label{WH}
Suppose that the symmetric or hermitian matrix ensembles given in \eqref{ph}
have independent entries $h_{ij}$, $i\le j$. Then $V(x)$ is a quadratic
polynomial, $V(x) = a x^2 + bx + c$ with $a>0$. This means that apart from 
a trivial shift and  normalization, the ensemble is  GOE or GUE.
\end{lemma}

The density of eigenvalues of the invariant ensemble \eqref{ph}
is determined by a variational problem \cite{D}. It is given by 
the equilibrium density  of a gas with a logarithmic self-interaction 
and external potential $V$, i.e., as the solution of
$$
   \inf_{\varrho } \Big\{ \int_\R\int_\R \log |s-t|^{-1} \varrho(\rd s) \varrho(\rd t) 
  +\int V(t) \varrho(\rd t)\Big\},
$$
where the infimum is taken over all probability measures $\varrho$. Under some
mild conditions on $V$, the equilibrium measure is absolutely continuous,
$\varrho_{eq}(\rd t)= \varrho_{eq}(t)\rd t$ and it has compact support. If $V$ is a polynomial,
then the support consists of finitely many intervals. The 
empirical density of eigenvalues  converges to  $\varrho_{eq}$
in the sense of \eqref{sc} where $\varrho_{sc}$ is replaced with
the function $\varrho_{eq}$.
It is an easy  exercise to check that
the solution of this variational problem for the Gaussian case, $V(x)= x^2/2$,
is indeed $\varrho_{sc}$.

\subsubsection{Delocalization of eigenvectors}

Apart from the statistics of the eigenvalues, one may also
study the eigenvectors of a random matrix.
In light of the universality conjecture about disordered systems
explained in Section \ref{sec:mot}, it is a challenging
question to test this hypothesis  on the level of eigenvectors
as well.   Wigner matrices are mean-field models and from the physics
intuition they are always in the delocalized regime.  Of course
they are still finite matrices, so they cannot have absolutely continuous
spectrum, a standard signature for delocalization that people 
working in random Schr\"odinger operators are often looking for.
But the delocalization of eigenvectors is a perfectly
meaningful question for large but finite matrices as well. Surprisingly,
this question was largely neglected both by the random matrix
community and the random Schr\"odinger operator
community within mathematics until T. Spencer has raised it
recently.
He pointed out in  a lecture that
in the case of the Gaussian ensembles, a simple invariance
argument proves that the eigenvectors $\bv\in \C^N$ are
fully delocalized in the sense that their $\ell^4$-norm is
$\| \bv\|_4 \sim N^{-1/4}$ (assuming $\|\bv\|_2=1$).
This is a signature of strong delocalization, since,
on the one hand, by Schwarz inequality
$$
  N^{-1/2} \| \bv\|_2=
 \Big( \frac{1}{N}\sum_{i=1}^N |v_i|^2\Big)^{1/2} 
 \le \Big( \frac{1}{N}\sum_{i=1}^N |v_i|^4\Big)^{1/4} 
  = N^{-1/4} \| \bv\|_4,
$$
i.e., $\|\bv\|_4\ge N^{-1/4}$ always holds,  on the
other hand this inequality is essentially saturated if all
coordinates of the eigenvector are of approximately the
same size, $|v_i|\sim N^{-1/2}$. 

The simple invariance argument works
only for the Gaussian case, where the unitary invariance is present,
but  it is a natural question to ask whether eigenvectors of 
Wigner ensembles  are also delocalized and the answer
is affirmative. We have proved   \cite[Corollary 3.2]{ESY3}
that if $\bv$ is an $\ell^2$-normalized eigenvector
of  a Wigner matrix $H$ with eigenvalue $\lambda$
away from the edge
$$
   H\bv=\la \bv,  \qquad \la \in[-2+\kappa, 2-\kappa]
$$
for some $\kappa>0$, then the $\ell^p$ norm of $\bv$,
for any $2<p<\infty$ is bounded by
\be
   \|\bv\|_p\le QN^{-\frac{1}{2}+\frac{1}{p}}
\label{eq:deloc}
\ee
with a very high probability, the set of exceptional events being
subexponentially small in $Q$ for large $Q$.
A similar bound with a logarithmic correction holds for
$p=\infty$ as well.
The precise statement will be given in Theorems \ref{thm:deloc} and \ref{deloc:trad}.
It is essentially a straighforward corollary of the local semicircle law,
Theorem \ref{lsc}, that was informally 
outlined in Section \ref{sec:locsc}.

Note that $\bv$ is an $\ell^2$-normalized eigenvector, then
the size of the $\ell^p$-norm of $\bv$, for $p>2$, gives
information about delocalization.  Complete delocalization occurs
when $\|\bv \|_p \lesssim N^{-1/2+ 1/p}$ since this corresponds 
to the $\ell^p$-norm of the fully delocalized vector
$\bv = (N^{-1/2}, N^{-1/2}, \ldots , N^{-1/2})$.
In contrast, a fully localized eigenvector,
$\bv = (0,0, \ldots, 0, 1, 0, \ldots 0)$ has $\ell^p$ norm one.

\subsection{Local statistics of eigenvalues: previous results.}

A central question concerning random matrices is the universality conjecture 
which states that  {\it local statistics}  of eigenvalues of large 
$N\times N$ square matrices $H$ 
are determined by  the symmetry type of the ensembles 
but are otherwise  independent of the details of  the distributions. 
It turns out that local statistics exhibit even stronger universality
features then the eigenvalue density.

The terminology ``local statistics'' refers to observables that can distinguish among
individual eigenvalues. For all ensembles we presented so far, we used
a normalization such that the typical eigenvalues remain in
a compact set as $N\to\infty$, in other words, the limiting density
function $\varrho$ was compactly supported. In this case, the typical spacing
between neighboring eigenvalues is of order $N^{-1}$. This holds
{\it in the bulk of the spectrum}, i.e., at a positive distance
away the {\it spectral edges}. The spectral edges are characterized by the
points where $\varrho$ goes to zero. For example,  for the Wigner semicircle 
distribution,
$\varrho_{sc}$, they are at $\pm 2$, for the Marchenko-Pastur distribution \eqref{MP}
they are at $\lambda_\pm$, and for certain invariant ensembles the support of the
eigenvalue density 
 might consist of several intervals i.e., it can have more than two
spectral edges.

\subsubsection{Bulk universality: the sine kernel and 
the gap distribution}\label{sec:sinegap}

To see individual eigenvalues and their joint distribution 
in the bulk spectrum, one needs to ``zoom out'' the point process of
the eigenvalues by magnifying it by a factor of $N$. We fix two real numbers,
$\al_1,\al_2$ and an energy $E$ with
$\varrho(E)>0$, and we ask the probability that there is an eigenvalue
at $E+\al_1/[N\varrho(E)]$ {\it and}  simultaneously there is an eigenvalue at 
$E+\al_2/[N\varrho(E)]$ (the normalization is chosen such
that the typical number of eigenvalues between these to points 
is independent of $E$).
It turns out that the answer is independent of the details of the
ensemble and of the energy $E$, it depends only on the symmetry type. For
example, for the hermitian case, it is given by
\be
  \P \Big\{ \mbox{there are eigenvalues $\lambda\in E+\frac{\al_1+\rd \al_2}{N\varrho(E)}$
 and  $\lambda' \in E+\frac{\al_2+\rd\al_2}{N\varrho(E)}$}\Big\} = \Big[ 1-
 \Big(\frac{\sin{\pi(\al_1-\al_2)}}{\pi(\al_1-\al_2)}\Big)^2\Big]\rd\al_1\rd \al_2.
\label{sinesimple}
\ee
The function on the r.h.s. is obtained from the celebrated {\bf sine kernel} and it
 should be viewed
as a two by two determinant of the form
\be
  \det \big( K(\al_i - \al_j)\big)_{i,j=1}^2, \qquad K(x) := \frac{\sin \pi x}{\pi x}.
\label{sine}
\ee
The explicit formula for the $K$ kernel in the symmetric case
is more complicated (see \cite{D}), but it is universal and the correlation function
has the same {\it determinantal structure}.

Note that \eqref{sinesimple} contains a much more delicate information
about the eigenvalues than the semicircle law \eqref{sc}. First, it is a local
information after a magnification to a scale where individual eigenvalues matter.
Second, it expresses a correlation among two eigenvalues. For example, 
due to $\frac{\sin y}{y}\to 1$ as $y\to 0$, we see that the eigenvalues
repel each other.

In general, the $k$-th correlation functions (or $k$-point marginals) give
information about the joint behavior of a $k$-tuple of eigenvalues. Their definition
is as follows:

\begin{definition}
Let $p_N(\lambda_1, \lambda_2, \ldots ,\lambda_N)$ be the joint
symmetrized probability distribution of the eigenvalues. For any $k\ge 1$,
the $k$-point 
correlation function is defined by
\be
  p^{(k)}_N(\la_1, \la_2, \ldots, \la_k): = \int_{\bR^{N-k}} p_N( \la_1, \ldots,\la_k, \la_{k+1},
\ldots \la_N) \rd\la_{k+1} \ldots \rd\la_{N}.
\label{pk}
\ee
\end{definition}

{\it Remark.} We usually label the eigenvalues in increasing order. For the
purpose of this definition, however, we dropped this restriction and
we consider $p_N(\lambda_1, \lambda_2, \ldots ,\lambda_N)$ to be a
symmetric function of $N$ variables, $\bla =(\la_1, \ldots, \la_N)$
on $\R^N$. Alternatively, one could consider
the density $\wt p_N(\bla)= N! p_N(\bla)\cdot {\bf 1}(\bla\in \Xi^{(N)})$,
where
$$
  \Xi^{(N)}:= \{ \lambda_1 < \lambda_2 <\ldots < \la_N\} \subset \R^N.
$$

\bigskip

The significance of the $k$-point correlation functions is that they
give  the expectation value of observables (functions) $O$ depending on $k$-tuples of
 eigenvalues
via the formula
$$
 \frac{(N-k)!}{N!}\, \E
\sum_{i_1, i_2, \ldots, i_k =1}^N  O\big(\la_{i_1}, \la_{i_2}, \ldots, \la_{i_k}\big)
 = \int_{\R^k} O(x_1, x_2,\ldots, x_k) p^{(k)}_N(x_1, x_2,\ldots, x_k)\rd x_1 \ldots \rd x_k,
$$
where the summation is over all {\it distinct} indices $i_1, i_2, \ldots, i_k$
and the prefactor is a normalization of the sum.

For example, the one-point function $p_N^{(1)}$ expresses the density, in particular, by choosing
the observable $O(x) = {\bf 1}(x\in [a,b])$ to be the characteristic function of $[a,b]$, we have
$$
 \frac{1}{N}\#\{ i\;: \; \la_i\in [a,b]\}=  \frac{1}{N} \sum_{i=1}^N O(\la_i) 
  = \int O(x) p_N^{(1)}(x) \rd x = \int_a^b p_N^{(1)}(x) \rd x.
$$
Therefore, the Wigner semicircle law \eqref{sc} states that 
$p_N^{(1)}$ converges weakly to $\varrho_{sc}$ as $N\to\infty$.

The sine kernel universality in the hermitian case
 expresses that the (weak) limit of the rescaled $k$-point correlation function, as
$N\to\infty$,
is given by the determinant of $K(x)$ from \eqref{sine}, i.e.,
\be
  \frac{1}{[\varrho(E)]^k}
 p_N^{(k)}\Big( E+ \frac{\al_1}{N\varrho(E)}, E + \frac{\al_2}{N\varrho(E)},
 \ldots ,E+ \frac{\al_k}{N\varrho(E)}\Big) \rightharpoonup  
\det \big( K(\al_i - \al_j)\big)_{i,j=1}^k
\label{sineres}
\ee
for any fixed $E$, as a weak convergence of functions in the variables $(\al_1, \ldots, \al_k)$.

\medskip

Once the $k$-point correlation functions are identified, it
is easy to derive limit theorems for other quantities related
to individual eigenvalues. The most interesting one is
the {\it gap distribution}, i.e., the distribution of
the difference of neighboring eigenvalues, $\lambda_{j+1}-\lambda_j$.
Note that it apparently involves only two eigenvalues, 
but it is not expressible solely by two point correlation
function, since the two eigenvalues must be consecutive. Nevertheless,
the gap distribution can be expressed in terms of all
correlation functions as follows.

Fix an energy $E$ with
$|E|< 2$. For  $s>0$ and for some $N$-dependent parameter $t$ with $1/N \ll t\ll 1$ let
$$
  \Lambda(s)=\Lambda_N(s):=  \frac{1}{2Nt\varrho_{sc}(E)} \#\Big\{
  1\le j\le N-1\; : \; \la_{j+1}-\la_j \le \frac{s}{N\varrho_{sc}(E)},\; |\la_j-E|\le t\Big\}
$$
i.e., the proportion of rescaled eigenvalue differences below a threshold $s$
in a large but still microscopic vicinity of an energy $E$. Let $\cK_\alpha$ be the 
operator acting 
on $L^2((0, \alpha))$ with integral  kernel $K(x,y):=\frac {\sin \pi(x-y)}
{\pi(x-y)}$.
 Then for any $E$ with $|E|< 2$ and for any $s> 0$ we have
 \be\label{maineq2}
\lim_{N\to \infty}  \E \, \Lambda_N ( s) 
= \int_0^s  p(\alpha)\; \rd \alpha  , \qquad p(\alpha): = 
\frac {\rd^2} {\rd \alpha^2}  \det (1 - \cK_\alpha),
\ee
where $\det$ denotes the Fredholm determinant of the
 operator $1-\cK_\al$ (note that $\cK_\al$ is a compact operator).
 The density function $p(s)$ 
of the nearest neighbor eigenvalue spacing behaves, with
a very good but not exact approximation (called the {\it Wigner surmise}),
as $p(s)\approx \frac{\pi s}{2}e^{-\pi s^2/4}$ for the symmetric
case and $p(s)\approx 32\pi^{-2} s^2 e^{-4s^2/\pi}$ for
the hermitian case \cite{M}.

Note that this behavior is in sharp contrast to the level spacing statistics
of the {\it Poisson point process}, where the corresponding density is $p(s)=e^{-s}$
(after rescaling the process so that the mean distance is one).
In particular, random matrices exhibit {\it level repulsion} whose
strength depends on the symmetry class (note the different behavior of
$p(s)$ near $s\approx 0$).

For the proof of \eqref{maineq2},
 we can use the exclusion-inclusion formula to express
\begin{align}
  \E \,\Lambda(s) =  \frac{1}{2Nt\varrho} \sum_{m=2}^\infty & (-1)^m {N\choose m}
  \int_{-t}^t \rd v_1 \ldots \int_{-t}^t \rd v_m {\bf 1}\Big\{ \max|v_i-v_j|\le 
  \frac{s}{N\varrho}\Big\} \non\\
  & \times p^{(m)}_N( E+ v_1, E+v_2, \ldots, E+v_m),
\end{align}
where $\varrho=\varrho_{sc}(E)$.
After a change of variables,
\be
\begin{split} \label{ex}
\E \; \Lambda ( s)= \; & \; \frac 1 { 2 N t\varrho}
 \sum_{m=2}^\infty (-1)^m   
\int_{-N\varrho t }^{N\varrho t }   \rd z_1 \ldots 
 \int_{-N\varrho t }^{N\varrho t  }
  \rd z_m \\ 
&   \times\,   {N \choose m} \,  \frac{1}{(N\varrho)^m}
   p^{(m)}_{N} \Big(u+ \frac {z_1}{N \rho}  , 
 \ldots, u+ \frac {z_m}{N \rho}\Big) {\bf 1} \Big\{\max |z_i-z_j| 
\le  s\Big\}   \\
= \; & \;
\frac 1 { 2 N t\varrho} \sum_{m=2}^\infty (-1)^m  m 
\int_{-N\varrho t }^{N\varrho t  }   
\rd z_1  \int_0^s \rd a_2 \ldots \int_0^s \rd a_m  \\
&   \times\,   {N \choose m} \,  \frac{1}{(N\varrho)^m}\;
  p^{(m)}_{N} \Big(u+ \frac {z_1}{N \rho}  , 
u+ \frac {z_1+a_2}{N \rho} ,
 \ldots, u+ \frac {z_1+a_m}{N \rho}\Big),
\end{split}
\ee
where the factor $m$ comes from considering the integration
 sector $z_1\le z_j$,
$j\ge 2$.
Taking $N\to \infty$ and using \eqref{sineres}, we get
\be
   \lim_{N\to\infty} \E \; \Lambda (s) 
= \sum_{m=2}^\infty \frac {(-1)^m }{(m-1)!}  
  \int_0^s \rd a_2 \ldots  \int_0^s  \rd a_m   \,   \,  \det 
 \left ( \frac {\sin \pi(a_i-a_j)} {\pi(a_i-a_j)}  \right )_{i,j=1}^m,
\label{fred}
\ee
where in the last determinant term we set $a_1=0$. 
The interchange of the limit and the summation can be
easily justified by an alternating series argument.
We note that
the left hand side of \eqref{fred} is $\int_0^s p(\al)\rd \al$,
where $p(\al)$ is the second derivative of the Fredholm determinant
$\det (1-\cK_\al)$ given in \eqref{maineq2} (see \cite{Simon} or \cite{AGZ} for more details).
We thus have
\be
\label{elambda}
\lim_{N\to\infty} \E \; \Lambda_N ( s ) =\int_0^s p(\al)\rd \al.
\ee

\subsubsection{Edge universality: the Airy kernel}

Near the spectral edges and under a different scaling
another type of universality emerges. It also has a determinantal form,
but the kernel is given by the Airy kernel,
$$
   A(x,y) := \frac{\mbox{Ai}(x) \mbox{Ai}'(y)-\mbox{Ai}'(x)\mbox{Ai}(y)}{x-y}
$$
where $\mbox{Ai}(x)$ is the Airy function, i.e.,
$$
  \mbox{Ai}(x) =\frac{1}{\pi}\int_0^\infty \cos\Big(\frac{1}{3} t^3 + xt\Big)\rd t
$$ 
which is the solution to the second order differential equation, $y'' - xy=0$, with vanishing
boundary condition at $x=\infty$.
The result, that is analogous to \eqref{sineres}, at the upper spectral edge $E=2$
of the hermitian Wigner matrices, is the following weak limit as $N\to\infty$
\be
 p_N^{(k)}\Big( 2+ \frac{\al_1}{N^{2/3}}, 2 + \frac{\al_2}{N^{2/3}},
 \ldots , 2+ \frac{\al_k}{N^{2/3}}\Big) \rightharpoonup  
\det \big( A(\al_i,\al_j)\big)_{i,j=1}^k.
\label{airyres}
\ee
Similar statement holds at the lower spectral edge, $E=-2$.
For Wigner matrices this was first proved by Soshnikov \cite{Sosh}
following the work of Sinai and Soshnikov \cite{SS} and
recently a different proof was given by Tao and Vu \cite{TV2}
and by Erd{\H o}s, Yau and Yin \cite{EYY3}, see Section \ref{sec:newedge}.
Note the different magnification factor $N^{2/3}$ that expresses the
fact that near the edge the typical eigenvalue spacing is $N^{-2/3}$.
Intituively, this spacing is consistent with the
semicircle law, since
$$
   \# \{ \lambda_j \ge 2-\e\}\approx \frac{N}{2\pi}\int_{2-\e}^2 \sqrt{4-x^2}\rd x
  =\frac{2}{3\pi} \e^{3/2}N,
$$
so we expect finitely many eigenvalues at a distance $\e\sim N^{-2/3}$ away
from the edge. Note however, that this argument is not rigorous, since
the semicircle law \eqref{sc} requires the test interval $[a,b]$ to be fixed,
independent of $N$.  Recently we proved a strong form of
the local semicircle law in \cite{EYY3}
(see Theorem \ref{45-1} later) which rigorously justifies this argument.

The largest eigenvalue $\la_N$ may extend above 2, but not more than by $O(N^{-2/3})$.
More precisely, the distribution function of the largest eigenvalue
is given by another universal  function, the {\it Tracy-Widom distribution \cite{TW}}
$$
  \lim_{N\to\infty}\P \Big(\lambda_N \le 2+ \frac{s}{N^{2/3}}\Big) = 
F_{2,1}(s): = \exp\Big(-\int_s^\infty (x-s) \cdot q^2(x) \rd x\Big)
$$
where $q(s)$ is the solution to the Painl\'eve II differential equation
$q''(s)= sq(s) + 2q^3(s)$ with asymptotics $q(s)\sim \mbox{Ai}(s)$
at $s=+\infty$ as a boundary condition. One can prove that 
$$
   F_{2,1}(s) \sim 1- \frac{1}{16\pi s^{3/2}}e^{-\frac{4}{3}s^{3/2}}
$$
 as $s\to\infty$, i.e., eigenvalues beyond the $O(N^{-2/3})$ scale 
are superexponentially  damped.
 Similar formula holds for symmetric matrices as well \cite{TW2}.
Note that, in particular, this result precisely identifies  the limiting 
distribution of the norm of a large Wigner matrix.

\medskip

The edge universality is  commonly approached via the
 moment method  presented in Section \ref{sec:density}.
The error term in \eqref{cat} deteriorates as $k$ increases, but with a more careful
classification and evaluation of the possible pairing structure,
it is possible to determine the moments up to order $k=O(N^{2/3})$,
 see \cite{Sosh}. We just mention the
simpler result 
\be
  \frac{1}{N} \E \tr H^{2k} = \frac{2^{2k}}{\sqrt{\pi k^3}} (1+ o(1))
\label{23}
\ee
as long as $k= o(N^{2/3})$.
Such precision is sufficient to identify the upper spectral edge of
$H$ with a precision almost $N^{-2/3}$ since
$$
   P\big( \lambda_N \ge 2+\e\big)\le \frac{\E\tr H^{2k}}{(2+\e)^{2k}}
   \le \frac{CN}{k^{3/2}(1+\frac{\e}{2})^{2k}} =o(1)
$$
if $\e \ge (\log N)/k \gg N^{-2/3}\log N$.  The computation \eqref{23} can be refined
to include powers of order $k\sim N^{2/3}$ and identify the common distribution
of the largest eigenvalues precisely \cite{Sosh}.
 We remark that the original
work of Soshnikov assumed that the single entry distribution is symmetric
and all its moments are finite, this condition has been
subsequently relaxed \cite{Ruz, P2, TV3}.

\medskip

The moment method does not seem to be applicable beyond
Soshnikov's scale, i.e., for $k$ much larger than $N^{2/3}$. 
On the other hand,  bulk universality would require to compute 
moments up to $k\sim O(N)$ since $1/k$ is essentially the resolution scale for which
knowing the moments of order $k$ precisely still gives some information.
The proof of the  bulk universality requires completely new methods.

\medskip

We mention a useful rule of thumb. There is a strong relation among
controlling   $e^{-itH}$, $H^k$ and $(H-z)^{-1}$.
Modulo some technicalities and logarithmic factors, the following three statements are
roughly equivalent for any $0<\e \ll 1$:
$$
  \mbox{   $e^{-itH} $ can be controlled up to times $|t|\le \e^{-1}$ }
$$
$$
\mbox{   $H^k$ can be controlled up to powers $k\le \e^{-1}$ }
$$
$$
\mbox{   $(H-z)^{-1}$ can be controlled down to  $\im z = \eta\ge \e$ }.
$$
These relations follow from the standard identities
$$
   \frac{1}{H-z} = i\int_0^\infty e^{-it(H-z)} \rd t,  \qquad z=E+i\eta, \;\; \eta>0
$$
$$
  e^{-itH} = \sum_{k=0}^\infty \frac{(-itH)^k}{k!} = 
 \frac{1}{2\pi i} \int_\gamma \frac{e^{-itz}}{H-z} \rd z
$$
(where the contour $\gamma$ encircles the spectrum of $H$).

\subsubsection{Invariant ensembles}\label{sec:inv}

For ensembles that remain invariant under the
transformations $H\to U^*HU$ for any unitary matrix $U$ (or, in case
of symmetric matrices $H$, for any $U$  orthogonal matrix), the joint
probability density function of all the $N$ eigenvalues can be explicitly
computed.  These ensembles are typically given by the
probability density \eqref{ph}.
The eigenvalues are strongly correlated and they
 are distributed according to a Gibbs measure
with a long range logarithmic interaction potential
(this connection was exploited first in \cite{Dy1}).
The joint probability density of the eigenvalues of $H$
can be computed explicitly:
\be
    p_N(\lambda_1, \lambda_2, \ldots , \lambda_N) 
  = \mbox{const.} \prod_{i<j} (\lambda_i-\lambda_j)^\beta \prod_{j=1}^N
   e^{- N\sum_{j=1}^N V(\lambda_j)},
\label{expli}
\ee
where $\beta=1$ for symmetric and $\beta=2$ for hermitian ensembles.
In particular, for the Gaussian case, $V$ is quadratic and thus the
joint distribution of the GOE ($\beta=1$) and GUE ($\beta=2$) eigenvalues is given by
\be
    p_N(\lambda_1, \lambda_2, \ldots , \lambda_N) 
  = \mbox{const.} \prod_{i<j} (\lambda_i-\lambda_j)^\beta \prod_{j=1}^N
   e^{- \frac{1}{4} \beta N \sum_{j=1}^N \lambda_j^2}.
\label{expli2}
\ee
It is often useful to think of this measure as a Gibbs measure of the form
\be
   \mu_N(\rd\bla) = p_N(\bla)\rd \bla = \frac{e^{-N\cH(\bla)}}{Z}, 
\qquad \cH(\bla) :=  
 \sum_{i=1}^N V(\la_i) -  \frac{\beta}{N} \sum_{i< j} 
\log |\la_{j} - \la_{i}| 
\label{H}
\ee
with the confining potential $V(\la)=\frac{\beta}{4}\la^2$.
The proof of \eqref{expli} is a direct (but involved) calculation; it
is based upon a change of variable. We sketch it for the hermitian (unitary
invariant) case.
The key observation is that
invariance of the measure under conjugation implies that the
eigenvalues, organized in a diagonal matrix $D=\mbox{diag}(\la_1, \la_2, \ldots,
\la_N)$, are independent of the eigenvectors, organized in a unitary
 matrix $U$. Writing $H= UDU^*$, one obtains
that $\cP(H)\rd H$ factorizes as
$$
  \cP(H)\rd H = e^{-N\tr V(H)}\rd H=
 \Big[p_N(\lambda_1, \lambda_2, \ldots , \lambda_N) \rd \la_1\ldots \rd\la_N
 \Big] \rd U,
$$
where $\rd U$ denotes the uniform (Haar) measure on the unitary group $U(N)$
and $p_N$ is the induced density function on the diagonal matrices (or its entries). Thus
the computation of the function $p_N$ amounts to computing the Jacobian
of the change of variables from the matrix elements of $H$ to the parametrization
coordinates in terms of  eigenvalues and eigenvectors.
The result is 
\be
  \rd H = \mbox{(const.)} \big[ \Delta_N(\bla)\big]^\beta \rd \bla \rd U, \qquad
\Delta_N(\bla):=\prod_{i<j} (\lambda_i-\lambda_j) ,
\label{VDM}
\ee
where $\beta=1$ is the symmetric and $\beta=2$ is the hermitian case,
 see \cite{AGZ} or Section 3.1--3.3
of \cite{M}  for details.

Especially remarkable is the emerging Vandermonde determinant
in \eqref{expli} which directly comes from integrating out the
Haar measure. Note that the symmetry type of the ensemble appears
through the exponent $\beta$. Only $\beta=1,2$ or 4 cases correspond
to  matrix ensembles of the form \eqref{ph}, namely, to the symmetric, hermitian
and quaternion self-dual  matrices. We will not give the precise
definition of the latter (see, e.g. Chapter 7 of \cite{M} or \cite{ESYY}), just mention
that this is the natural generalization of symmetric or hermitian matrices
to quaternion entries and they have real eigenvalues.

\bigskip

Irrespective of any underlying matrix ensemble, one can nevertheless
study the distribution \eqref{expli} for any $\beta>0$; these are called
the {\it general $\beta$-ensembles.}  In fact, for the Gaussian case, $V(\la)= \la^2/2$,
there are corresponding tridiagonal matrix ensembles for any $\beta>0$, obtained
from successive Householder transformations, whose
eigenvalue distribution is described by \eqref{expli}, see \cite{DE} for an overview.
Using the tridiagonal structure, methods from the theory of Jacobi matrices can
be applied. For example, the universality at the edge eigenvalues
is understood in a sense that they are shown to converge to the lowest eigenvalues
for a one dimensional Schr\"odinger operator with a white noise drift
and, in particular, the $\beta$-analogue of the Tracy-Widom distribution
has been identified in \cite{RRV, RR} following the conjectures of \cite{ES}.
A different method, the {\it Brownian carousel
representation} \cite{VV}, has been used to generalize the tail distribution of
large eigenvalue gaps (``Wigner surmise'') for Gaussian $\beta$-ensembles \cite{VV2}.
More precisely, it has been shown
that the probability under the distribution \eqref{expli} with $V(\la)= \la^2/2$
that there is no point  falling into a fixed interval of length $s$ (after locally rescaling
the spectrum so that the typical distance is $2\pi$) is given by
$$
  q_s = (\kappa_\beta+o(1)) s^{\gamma_\beta} \exp{\Big( -\frac{\beta}{64}s^2 +
\Big(\frac{\beta}{8}-\frac{1}{4}\Big)s\Big)}, \qquad \gamma_\beta:= \frac{1}{4}
 \Big(\frac{\beta}{2}  +\frac{2}{\beta}-3\Big), \qquad \kappa_\beta>0,
$$
as $s\to \infty$ (after $N\to\infty$ limit).

The bulk universality, i.e., the analogue of the 
sine-kernel behavior \eqref{sineres} for general $\beta$-ensembles is unproven, even for
the Gaussian case. 
The main difficulty is that  \eqref{expli} represents an $N$-particle system
with a long range interaction. We can write the joint density as a
Gibbs measure \eqref{H};
we have $N$ particles in a confining potential $V$ that repel each other
with  a potential that has locally a logarithmic repulsion, but also a large
(in fact increasing) long range component. Standard methods from
statistical physics to construct and analyse Gibbs measures do not seem to
apply. Although here we do not attempt to construct the analogue of an
infinite volume Gibbs measure, we only want to compute correlation functions,
but even this is a daunting task with standard methods unless an extra structure
is found.

\subsubsection{Universality of classical invariant ensembles via orthogonal
polynomials}

Much more is known about the classical invariant ensembles, i.e., the $\beta=1,2,4$ cases,
with a general potential $V$. For these specific
values an extra mathematical structure emerges, namely
the orthogonal polynomials with respect to the weight function $e^{-NV(\lambda)}$
on the real line.
This approach was originally applied by Mehta and Gaudin \cite{M, MG}
to compute the gap distribution for the Gaussian case that involved
classical Hermite orthonormal polynomials.  
 Dyson \cite{Dys1} computed the local correlation functions for
a related ensemble (circular ensemble) that was extended to the standard Gaussian
ensembles by Mehta \cite{M2}. Later  a
general method using orthogonal polynomials has been developed to tackle a very general
class of unitary ensembles 
(see, e.g. \cite{BI, D,  DKMVZ1, DKMVZ2, FIK, M, PS} and references therein).

  For simplicity, to illustrate the connection, we will consider
the hermitian case $\beta=2$ with a Gaussian potential $V(\la)=\la^2/2$
(which, by Lemma \ref{WH}, is also a Wigner matrix ensemble, namely the GUE). To simplify the
presentation further, for the purpose of this argument only, we rescale the eigenvalues
$\lambda \to \sqrt{N}\la$, which effectively removes the factor $N$ from the
exponent in \eqref{expli}. (This pure scaling works only in the Gaussian case,
but it is only a technical convenience to simplify formulas.)

Let $P_k(x)$ be the $k$-th orthogonal polynomal with respect to the weight function $e^{-x^2/2}$
with leading coeffient 1.  
Let 
$$
   \psi_k(x): = \frac{e^{-x^2/4}P_k(x)}{ \| e^{-x^2/4}P_k\|}
$$
be the corresponding orthonormal function, i.e.,
\be
   \int \psi_k(x)\psi_\ell(x) \rd x  =\delta_{k,\ell}.
\label{orthog}
\ee
In the particular case of the Gaussian weight function, $P_k$ is  given by the Hermite polynomials
$$
   P_k(x) = H_k(x) := (-1)^k e^{x^2/2} \frac{\rd^k}{\rd x^k} e^{-x^2/2}
$$
and
$$
  \psi_k(x) = \frac{P_k(x)}{(2\pi)^{1/4} (k!)^{1/2}} \; e^{-x^2/4}
$$
but for the following discussion we will not need these explicit formulae.

The key observation is that by simple properties of the Vandermonde determinant, we have
\be
 \Delta_N(\bx)=  \prod_{1\le i<j\le N} (x_j-x_i) = \det\big( x_i^{j-1}\big)_{i,j=1}^N = \det
 \big( P_{j-1}(x_i)\big)_{i,j=1}^N,
\label{vd}
\ee
exploiting that $P_j(x) = x^j + \ldots $ is a polynomial of degree $j$ with
leading coefficient equal one.
Define the kernel
$$
    K_N(x,y):= \sum_{k=0}^{N-1} \psi_k(x)\psi_k(y),
$$
i.e., the projection kernel onto the subspace spanned by the
first $K$ orthonormal functions.
Then \eqref{vd} immediately implies
\begin{align}
  p_N(x_1, \ldots, x_N) = &  C_N \Big[ \det \big( P_{j-1}(x_i)\big)_{i,j=1}^N\Big]^2
  \prod_{i=1}^N e^{-x_i^2/2}\non\\
  =&  C_N'  \Big[ \det \big( \psi_{j-1}(x_i)\big)_{i,j=1}^N\Big]^2
 = C_N' \det \big( K_N(x_i,x_j)\big)_{i,j=1}^N, \non
\end{align}
where in the last step we used that 
the square of the matrix $\big( \psi_{j-1}(x_i)\big)_{i,j=1}^N$ is exactly
$\big( K_N(x_i,x_j)\big)_{i,j=1}^N $ and we did not follow the precise
constants for simplicity.

To compute the correlation functions, we expand the determinant:
\begin{align}
   p_N^{(k)}(x_1,\ldots, x_k) = & C_{k,N} \int_{\R^{N-k}}
   \det \big( K_N(x_i,x_j)\big)_{i,j=1}^N \prod_{i=k+1}^N \rd x_i \non \\
 =&  C_{k,N}\sum_{\sigma, \tau \in S_N} (-1)^{\tau+\sigma}\int_{\R^{N-k}}
  \prod_{j=1}^N \psi_{\sigma(j)-1}(x_j)\psi_{\tau(j)-1}(x_j) \prod_{i=k+1}^N \rd x_i \non \\
  = &  C_{k,N} \sum_{ \al_1< \al_2 < \ldots < \al_k}
 \sum_{\sigma, \tau \in S_k} (-1)^{\tau+\sigma}
  \prod_{j=1}^k \psi_{\al_{\sigma(j)}-1}(x_j)\psi_{\al_{\tau(j)}-1}(x_j) \non\\
  = &  C_{k,N} \sum_{ \al_1< \al_2 < \ldots < \al_k}
 \Big[ \det\big(  \psi_{\al_j-1}(x_j) \big)_{i,j=1}^k\Big]^2, \non
\end{align}
where $S_N$ is the permutation group on $N$ elements and $(-1)^\tau$ is the parity character
of the permutation. In the third line we used \eqref{orthog} to perform 
the integrations that have set $\sigma(j)=\tau(j)$ for all $j\ge k+1$
and we denoted by $\{ \al_1, \al_2, \ldots, \al_k\}$ the ordering 
of the set $\{ \sigma(1), \sigma(2), \ldots ,\sigma(k)\} = \{ \tau(1), \tau(2), 
\ldots , \tau (k)\}$. Finally, using  that the matrix
$ [K_N(x_i,x_j)]_{i,j=1}^k $ can be written as $A^tA$ with $A_{ij} = \psi_{i-1}(x_j)$
and using the Cauchy-Binet expansion formula for the determinant of a product matrix, we get
$$
    \det [K_N(x_i,x_j)]_{i,j=1}^k = \sum_{\al_1< \al_2 < \ldots < \al_k}
\Big[ \det\big(  \psi_{\al_j-1}(x_j) \big)_{i,j=1}^k\Big]^2 .
$$
Apart from the constant, that can be computed, we thus proved that
$$
   p_N^{(k)}(x_1, \ldots , x_k)= \frac{(N-k)!}{N!}  \det [K_N(x_i,x_j)]_{i,j=1}^k,
$$
i.e., the correlation functions have a determinantal structure.

In order to see the sine-kernel \eqref{sine} emerging, we need a basic algebraic
property of the orthogonal polynomials, the Christoffel--Darboux formula:
$$
  K_N(x,y)=  \sum_{j=0}^{N-1} \psi_j (x) \psi_j(y)  = 
  \sqrt{N} \Bigg[ \frac{\psi_N(x) \psi_{N-1}(y) - \psi_N(y) \psi_{N-1}(x)}{x-y} 
  \Bigg].
$$
Furthermore,  orthogonal polynomials of high degree have 
asymptotic behavior as $N\to\infty$
\be
 \psi_{2m} (x)\approx \frac{(-1)^m}{N^{1/4}\sqrt{\pi}} \cos\big( \sqrt{N}x \big)
 + o(N^{-1/4}), \qquad  \psi_{2m+1} (x)\approx 
 \frac{(-1)^m}{N^{1/4}\sqrt{\pi}} \sin\big( \sqrt{N}x \big)
 + o(N^{-1/4}),
\label{asympt}
\ee
for any $m$ such that $|2m-N|\le C$. The approximation is uniform
for $|x|\le CN^{-1/2}$.  These formulas will be useful if we
set $E=0$ in \eqref{sineres}, since we rescaled the eigenvalues
by a factor of $\sqrt{N}$, so the relation between the notation of \eqref{sineres} 
(for $k=2$)
and $x,y$ is
\be 
  x= \sqrt{N}\Big( E+ \frac{\al_1}{N\varrho(E)}\Big), \qquad
  y= \sqrt{N}\Big( E+ \frac{\al_2}{N\varrho(E)}\Big).
\label{xy}
\ee
For different values of $E$ one needs somewhat different asymptotic formulae for
the orthogonal polynomials.

We can thus compute that
\begin{align}
K_N(x,y)
 \approx & \frac{1}{\pi}
  \frac{\sin (\sqrt{N} x) \cos(\sqrt{N} y) - \sin (\sqrt{N}y)
 \cos(\sqrt{N} x)}{x-y} =  \frac{\sin \sqrt{N}(x-y)}{\pi(x-y)}.
\end{align}
Using \eqref{xy} and
that  $\varrho(0)= \pi^{-1}$,  we have
$$
   \frac{1}{\varrho(0)\sqrt{N} } K_N(x,y) \approx 
 \frac{\sin \pi(\al_1-\al_2)}{\pi(\al_1-\al_2)},
$$
which gives \eqref{sineres} for $E=0$ after undoing the $\la\to \sqrt{N}\la$
magnification.

The main technical input is the refined asymptotic formulae \eqref{asympt} for
orthogonal polynomials. In case of the classical orthogonal
polynomials (appearing in the standard Gaussian Wigner and Wishart ensembles)
 they are usually obtained by a Laplace asymptotics
from their integral representation. For a general potential $V$ the
corresponding analysis is quite involved and depends on the regularity
properties of $V$. One successful approach was initiated by Fokas, Its and Kitaev \cite{FIK}
and by P. Deift  and collaborators 
via the Riemann-Hilbert method, see \cite{D} and references therein.
An alternative method was presented in \cite{L, LL} using more
direct methods from orthogonal polynomials. 

There have been many refinements and improvements in this very active
research area related to invariant ensembles as it reveals fruitful connections
between random matrices, orthogonal polynomials, complex analysis and
even combinatorics (see \cite{D}). One common input, however,
is the explicit formula \eqref{expli} for the joint probability density
that allows one to bring in orthogonal polynomials. We now depart from this
topic and we will focus on ensembles when such explicit formula is not
available; the most
prominent example is the Wigner matrix. Apart from the Gaussian case,
no explicit formula is available for the joint eigenvalue
distribution. Thus the basic algebraic connection between
eigenvalue ensembles and orthogonal polynomials is lacking
and completely new methods needed to be developed. 
In the next section we summarize recent results in this direction.

\subsection{Local statistics of eigenvalues: new results}\label{sec:locnew}

\subsubsection{Hermitian matrices with 
Gaussian convolutions}\label{sec:conv}

The first rigorous partial result for bulk universality in the non-unitary case
was given by  Johansson  \cite{J}, see also Ben Arous and P\'ech\'e \cite{BP}
for extending \cite{J} to the full bulk spectrum 
and the recent improvement \cite{J1} on weakening moment conditions.
The main result states that the bulk universality holds for Gaussian divisible 
{\it hermitian} ensembles, i.e., hermitian ensembles   of the form
\be
H = \sqrt{1-\e}\wh H+ \sqrt{\e} V,
\label{HaV1}
\ee
where $\wh H$  is a hermitian Wigner matrix, 
$V$ is an independent standard GUE matrix
and $\e$ is a positive constant of order one, independent of $N$.

We will often use the parametrization
\be
H = e^{-t/2}\wh H+ (1- e^{-t})^{1/2} V.
\label{HaV}
\ee
If embedded in a flow, then
 $t$ can be interpreted as time of an {\it Ornstein--Uhlenbeck (OU) process}.
This formalism incorporates the idea that matrices with Gaussian convolutions
can be obtained as a matrix valued stochastic process, namely as the
solution of the following stochastic differential equation:
\be
  \rd H_t = \frac{1}{\sqrt{N}}\, \rd \beta_t - \frac{1}{2} H_t \rd t, \qquad H_0 =\wh H,
\label{process}
\ee
where $\beta_t$ is a hermitian matrix valued process
whose diagonal matrix elements are standard real Brownian motions
and whose off-diagonal matrix elements are standard complex Brownian motions.
The distibution of the solution to \eqref{process} for any fixed $t$
coincides with the distribution of \eqref{HaV}.
Note that infinite time, $t=\infty$, corresponds to the GUE ensemble, so
the matrices \eqref{HaV} {\it interpolate} between the Wigner matrix $\wh H$
and the GUE. This point of view will be extremely useful in the sequel
as it allows us to compare Wigner matrices with Gaussian ones if the effect of the time
evolution is under control.

Alternatively, one can consider the density function $u_t$ of the real and imaginary parts
of the matrix
elements as being evolved by the {\it generator} of the OU process:
\be
   \pt_t u_t = A u_t, \qquad A:= \frac{1}{4}\frac{\pt^2}{\pt x^2} -\frac{x}{2} 
  \frac{\pt}{\pt x},
\label{OU}
\ee
where the initial condition $u_0(x)$ is the density (with respect to the reversible
Gaussian measure) of the distribution 
of the real and imaginary parts of the matrix elements of $\sqrt{N}\wh H$.
For the diagonal elements, an OU process with a slightly different normalization is used.
The OU process \eqref{OU} keeps the expectation zero and
variance $\frac{1}{2}$ if the  initial $u_0$ has these properties.

\bigskip

The joint distribution of the eigenvalues
of Gaussian divisible hermitian random matrices of the form \eqref{HaV} 
still has a certain determinantal structure. The formula 
is somewhat simpler if we write
$$
  H = \sqrt{1-\e}\big( \wh H + a V\big)
$$
with $a= \sqrt{\e/(1-\e)}$, i.e., we use the standard Gaussian convolution
\be
   \wt H =  \wh H + a V
\label{hh}
\ee
and then rescale at the end. Note that \eqref{hh} can be generated
by evolving the matrix elements by standard Brownian motions $\beta$ up to time
$t=a^2$, i.e. by solving  
\be
  \rd \wt H_t = \frac{1}{\sqrt{N}}\, \rd \beta_t, \qquad \wt H_0 =\wh H.
\label{DBMoriginal}
\ee
Moreover, to be in line with the normalization convention of \cite{EPRSY} that
follows \cite{J}, we assume
that the matrix elements of the Wigner matrix
 $\wh H$ and the GUE matrix $V$ have variance $\frac{1}{4N}$ instead of $1/N$
as in Definition \ref{def:gen}. This means that the eigenvalues are
scaled by a factor $\frac{1}{2}$ compared with the convention in the previous sections,
 and  the semicircle law \eqref{sc} is modified to $2\pi^{-1}\sqrt{(1-x^2)_+}$.
This convention applies only up to the end of this section.

Let $\by = (y_1, \ldots , y_N)$ denote the eigenvalues of $\wh H$
and $\bx=(x_1, \ldots, x_N)$ denote the eigenvalues of $\wt H$.
Then we have the following representation formulae (a slight variant of these formulae
were given and used by Johansson in \cite{J} and they were motivated by similar formulae
by Br\'ezin and Hikami \cite{BH}):

\begin{lemma}  \cite[Proposition 3.2]{EPRSY}
Let $V$ be a GUE matrix. For any fixed hermitian matrix $\wh H$ with eigenvalues $\by$, the
density function of the eigenvalues $\bx$ of $\wt H =  \wh H + a V$
is given by
\be
    q_S(\bx ; \by) : = \frac{1}{(2\pi S)^{N/2}} \frac{\Delta_N(\bx)}{\Delta_N(\by)}
  \det\big( e^{-(x_j-y_k)^2/2S}\big)_{j,k=1}^N, 
\label{def:qs}
\ee
with $S=a^2/N$ and we recall that $\Delta_N$ denotes the Vandermonde determinant \eqref{vd}.
 The $m$-point correlation functions of the eigenvalues 
of $\wt H =  \wh H + a V$,
$$
  p_{N,\by}^{(m)}(x_1, \ldots , x_m): = \int_{\R^{N-m}} q_S(x_1, x_2, \ldots, x_N; \by)
 \rd x_{m+1} \ldots \rd x_N,
$$
are given by the following formula
\be
   p_{N,\by}^{(m)}(x_1, \ldots , x_m)= \frac{(N-m)!}{N!} \det
\big( \cK_N^{S}(x_i, x_j; \by)\big)_{i,j=1}^m, \qquad S= a^2/N.
\label{def:R}
\ee
Here we define
\be
\begin{split}\label{ck}
  \cK_N^S (u,v;\by):= & \frac{1}{(2\pi i)^2 (v-u)S}
  \int_\gamma \rd z\int_\Gamma \rd w (e^{-(v-u)(w-r)/S} -1) \prod_{j=1}^N 
  \frac{w-y_j}{z-y_j} \\
 &  \times \frac{1}{w-r}\Big( w-r+z-u - S\sum_j \frac{y_j-r}{(w-y_j)(z-y_j)}\Big)
  e^{(w^2-2uw -z^2+2uz)/2S},
\end{split}
\ee
where $r\in \bR$ is an arbitrary constant. The integration curves $\gamma$ and $\Gamma$ in the
complex plane are given by
 $\gamma= \gamma_+\cup\gamma_-$ as the union of two lines $\gamma_+:\tau\to -\tau + i\om$ and
 $\gamma_-:\tau \to \tau -i\om$ ($\tau\in \bR$)
for any fixed $\om>0$ and $\Gamma$ is $\tau\to i\tau$, $\tau\in \bR$.
\end{lemma}
We note that $\Gamma$ 
can be shifted to any vertical line since the
integrand is an entire function in $w$ and has a Gaussian decay
as $|\im \; w| \to \infty$. 
The constants $r \in \bR$ and $\om>0$  (appearing in the
definition of the contour $\gamma$)  can be arbitrary
and can be appropriately chosen in the contour integral estimates.

\bigskip

The key step behind the proof of  \eqref{def:qs} is the Harish-Chandra-Itzykson-Zuber
integral \cite{IZ}
$$
   \int_{U(N)} e^{\tr (UAU^*B)} \rd U = (\mbox{const.})
\frac{\det \big( e^{a_ib_j}\big)_{i,j=1}^N}{
  \Delta_N(\ba)\Delta_N({\bf b})},
$$
where $A, B$ are two  hermitian matrices with eigenvalues $\ba = (a_1, \ldots , a_N)$
and ${\bf b}=(b_1, \ldots, b_N)$ and the integration is over the unitary group
$U(N)$ with respect to the Haar measure. We note that this is the
step where the unitary invariance (or the hermitian character of $H$) 
is crucially used; analogous simple formula is not known for
other symmetry groups, see \cite{BH2}.

For any testfunction $f$, and for a fixed matrix $\wh H$, we have
$$
   \int f(\bx) q_S(\bx; \by)\rd \bx= (\mbox{const.})
 \int  f(\bx)  e^{-\frac{1}{2}N\tr (\wt H - \wh H)^2} \rd \wt H,
$$
where we used  that $V= a^{-1}(\wt H - \wh H)$ is a GUE matrix with distribution
$$
   \cP(V)\rd V=e^{-\frac{1}{2}N\tr V^2}\rd V. 
$$
We set $\wt H = U X U^*$ to be the diagonalization of $\wt H$
with $X=\mbox{diag}(\bx)$, then we have, using \eqref{VDM},
\begin{align}
 \int f(\bx) q_S(\bx; \by)\rd \bx
  = & (\mbox{const.}) \int_{\R^N} \int_{U(N)}
  f(\bx) e^{-\frac{1}{2a^2}N\tr (UXU^* - \wh H)^2} \rd U
   \Delta_N^2(\bx) \rd \bx 
  \non \\
 = & (\mbox{const.}) \int_{\R^N}  \Bigg[ \int_{U(N)} 
  e^{\frac{N}{a^2}\tr UXU^*\wh H} \rd U\Bigg] f(\bx) 
 e^{-\frac{N}{2S} \sum_i (x_i^2+y_i^2)}
 \Delta_N^2(\bx) \rd \bx  \non \\ 
 = &  (\mbox{const.}) \int_{\R^N} 
 f(\bx) \frac{\det \big( e^{x_iy_j/S}\big)_{i,j=1}^N}{
  \Delta_N(\bx)\Delta_N({\bf y})}   e^{-\frac{N}{2S} \sum_i (x_i^2+y_i^2)}
  \Delta_N^2(\bx) \rd \bx  \non \\ 
 = & (\mbox{const.}) \int_{\R^N} 
 f(\bx) \frac{\Delta_N(\bx)}{\Delta_N(\by)}
  \det \big( e^{-\frac{1}{2S} (x_i-y_j)^2}\big)_{i,j=1}^N   \rd \bx , \non
\end{align}
which proves \eqref{def:qs} (apart from the constant). This shows how the
Vandermonde determinant structure emerges for Gaussian convolutions.
The proof of the contour integral representation
 \eqref{ck} from \eqref{def:qs} is a bit more involved, see 
Proposition 3.2 of \cite{EPRSY} (or Proposition 2.3
\cite{J}) for the details.

\bigskip

Once \eqref{ck} is given, the key idea is to view it as a complex integral
suited for Laplace asymptotics or saddle point calculation. More precisely, 
after some straightforward algebraic steps, it can be brought in the
following form (see Section 3.1 of \cite{EPRSY}), where we already 
changed variables in the argument to detect the microscopic structure. For any fixed $|u|<1$
and $t=a^2$ we find
from (\ref{ck}) that 
\be
  \frac{1}{N\varrho(u)}\cK_N^{t/N}\Big(u,u+\frac{\tau}{N\varrho(u)}; \by\Big)
  = N \int_\gamma \frac{\rd z}{2\pi i}\int_\Gamma \frac{\rd w}{2\pi i}
  h_N(w) g_N(z,w) e^{N(f_N(w)-f_N(z))}
\label{repr}
\ee
with
\be
\begin{split}
   f_N(z)  &:= \frac{1}{2t}(z^2-2u z) +\frac{1}{N}\sum_j\log(z-y_j)
\label{def:fN} \\
  g_N(z,w) & := \frac{1}{t(w-r)}[w-r+z-u] -
\frac{1}{N(w-r)}\sum_j \frac{y_j-r}{(w-y_j)(z-y_j)}
\non\\
   h_N( w)  & : =
\frac{1}{\tau} \Big( e^{-\tau (w-r)/t\varrho(u)}
  - 1 \Big). \non
\end{split}
\ee
Notice the $N$ factor in front of the exponent in \eqref{repr}, indicating
that the main contribution to the integral comes from the saddle points,
i.e., from  $z$ and $w$ values where $f'_N(z)=f'_N(w)=0$. Note that
\be
   f_N'(z) = \frac{z-u}{t} +\frac{1}{N}\sum_j\frac{1}{z-y_j},
\label{fn}
\ee
i.e., it is essentially given by the empirical Stieltjes transform \eqref{St} of
the eigenvalues of $\wh H$. Suppose that the Wigner semicircle law or, equivalently, \eqref{limst}
 holds, then the saddle point $z_N$, $f'_N(z_N)=0$ can be well approximated by the
solution to 
\be
    \frac{z-u}{t} + 2\big( z-\sqrt{z^2-1}\big)=0
\label{fnlim}
\ee
(the formula for the Stieltjes transform slightly differs from \eqref{limst} because
of the different normalization of $\wh H$). 
It is easy to check that there are two solutions, $z^\pm$, with imaginary part
given by $\pm 2ti\sqrt{1-u^2} + O(t^2)$ for small $t$. 

Once the saddle points
are identified, the integration contours $\gamma$ and $\Gamma$ in \eqref{repr} can be shifted
to pass through the saddle points from a direction where $f_N$ is real and its
second derivative has the ``good sign'', so that usual saddle point approximation 
holds. There are altogether four pairs of saddles $(z^\pm_N, w^\pm_N)$, but only
two give contributions to leading order. Their explicit  evaluation gives the
sine kernel \eqref{sine} for $\cK_N$ in the $N\to\infty$ limit.

The key technical input is to justify the semicircle approximation used in 
passing from \eqref{fn} to \eqref{fnlim} near the saddle point (some estimate
is needed away from the saddle as well, but those are typically easier to obtain).
The standard argument for the semicircle law presented in Section \ref{sec:moment}
holds for any {\it fixed} $z$ with $\im z>0$, independently of $N$, especially important is
that $\im z >0$ uniformly in $N$.  Therefore this argument can be used
to justify \eqref{fnlim} only for a fixed $t>0$. 
Recall that $t=a^2$ is the variance of the Gaussian convolution.
 This was the path essentially followed by Johansson
who proved \cite{J} that sine-kernel universality \eqref{sineres} holds if
the Wigner matrix has a Gaussian component comparable with its total size.

 One implication of the local semicircle law
 explained in 
Section \ref{sec:mainlsc} is that the approximation
from \eqref{fn} to \eqref{fnlim} could be justified even for very short 
times $t$. Essentially any time of order $\gg 1/N$ (with some logarithmic
factor) are allowed, but for
technical reasons we carried out the estimates only for $t= N^{-1+\e}$
for any $\e>0$ and we showed that the contour integral \eqref{repr}
is given by the sine-kernel even for such short times \cite{EPRSY}. This proved
the sine-kernel universality in the form of \eqref{sineres}
for any fixed $E$ in the bulk spectrum for hermitian
matrices with a Gaussian component of variance $N^{-1+\e}$.

\subsubsection{Gaussian convolutions for arbitrary matrices: the local relaxation 
flow}\label{sec:lrf}

The method presented in Section \ref{sec:conv} heavily relies on the
Br\'ezin-Hikami type formula \eqref{ck} that is available only for
hermitian matrices. For ensembles with other symmetries,
in particular for symmetric Wigner matrices, a new method was necessary.
 In a series of papers \cite{ERSY, ESY4, ESYY, EYY, EYY2, EYY3}
we developed an approach based upon hydrodynamical ideas 
from interacting particle systems. We will present it in more
details in Section \ref{sec:sine}, here we summarize the key points.

The starting point is  a key observation of Dyson \cite{Dy} from 1962. 
Let $\wh H$ be an arbitrary fixed matrix and consider the solution $\wt H_t$
to \eqref{DBMoriginal}. Recall that for each 
fixed $t$, $\wt H_t$ has the same distribution as 
 $\wh H+ \sqrt{t} V$, where 
 $V$ is a standard GUE matrix (independent of $\wh H$).
Dyson noticed that 
 the evolution of the eigenvalues of the flow $\wt H_t$
is given by a coupled system of
stochastic differential equations, commonly called  the  Dyson Brownian motion (DBM in short). 
For convenience,  we will replace the Brownian motions by  
OU  processes to keep the variance constant, i.e., we
will use \eqref{process} instead of \eqref{DBMoriginal}  to generate the matrix flow.
 The Dyson Brownian motion we will use (and we will still call it DBM) is  given by
the following system of stochastic differential equations
 for the eigenvalues ${\bla}(t) = (\lambda_1(t), \ldots, \lambda_N(t))$,
see, e.g. Section 4.3.1 of \cite{AGZ},
\be\label{sde}
  \rd  \la_i  =    \frac{\rd B_i}{\sqrt{N}} +  \left [ -  \frac{\beta}{4}
 \la_i+  \frac{\beta}{2 N}\sum_{j\ne i} 
\frac{1}{\la_i - \la_j}  \right ]  \rd t, \qquad 1\leq i\leq N,
\ee
where $\{ B_i\; : \; 1\leq i\leq N\}$ is a collection of
independent Brownian motions. 
The initial condition $\bla(0)$ is given by the eigenvalues of $\wh H$.
The choice of parameter $\beta=2$ corresponds to the hermitian case, but
the process
 \eqref{sde} is well  defined for any $\beta\ge 1$ and
the eigenvalues do not cross due to the strong repulsion among them.
The threshold $\beta=1$ is critical for the non-crossing property.
As $t\to \infty$, the distribution of $\bla(t)$
converges to the  Gaussian $\beta$-ensemble distribution
\eqref{expli2} as the global invariant measure; for example,
 for $\beta=2$, it converges to the GUE.

Using Dyson Brownian Motion, the question of universality for
Gaussian divisible ensembles can be translated into the
question of the time needed for DBM to reach equilibrium. 
The time scale to approach the {\it global} equilibrium
is of order one but, as we eventually proved in \cite{EYY3},
 the decay to the {\it local} equilibrium is much faster, it occurs
already in time scale of order $t\sim N^{-1}$. 
Since the local statistics of 
eigenvalues  depend exclusively on  the  local equilibrium,
this means that
the local statistics of Gaussian divisible ensembles with a tiny Gaussian component
of size $N^{-1+\e}$ are already universal.

We remark that using the relation between Gaussian divisible ensembles 
and the relaxation time of DBM,
 the result of Johansson \cite{J} can be interpreted as stating that the 
local statistics of GUE are reached via DBM after  time at most of order one.
Our result from \cite{EPRSY}, explained in the previous section, indicates
that the decay to the  local equilibrium  occurs
already in time $t\sim N^{-1+\e}$. This is, however, only a
reinterpretation of the results since neither \cite{J} nor \cite{EPRSY}
used hydrodynamical ideas. In particular, these proofs are
valid only for hermitian matrices since they used some
version of the Br\'ezin-Hikami formula.

\medskip

To establish universality in full generality,
we have developed a purely hydrodynamical approach based
upon the relaxation speed of DBM.
The key point in this approach is that
 there is no restriction on the symmetry type; the argument works equally
for symmetric, hermitian or quaternion self-dual ensembles, moreover,
with some obvious modifications, it also works for random covariance matrices.

\medskip

Our first paper  that used hydrodynamical ideas is \cite{ERSY}.
 In this paper we extended Johansson's result \cite{J} 
  to hermitian ensembles with a Gaussian component of size $t\gg N^{-3/4}$ 
by capitalizing on the fact  that the local statistics of 
eigenvalues  depend exclusively on the approach to {\it local} equilibrium
which in general is faster than reaching global equilibrium.
Unfortunately, the identification of  local equilibria in \cite{ERSY} still used 
explicit representations of correlation functions by  orthogonal polynomials
(following e.g. \cite{PS}), and the extension to other ensembles 
in this way is not a simple task (see \cite{Shch} for extension of \cite{PS}
to symmetric matrices to prove edge universality).

To depart from using orthogonal polynomials, we introduced new
hydrodynamical tools in \cite{ESY4} which
entirely eliminated  explicit formulas and  it gave a unified proof for the  universality 
of {\it symmetric} and hermitian  Wigner matrices with a small Gaussian convolution.
The size of the Gaussian component, equivalently, the
time needed to reach the local equilibrium in a sufficiently strong sense
has increased from $N^{-3/4}$ to  $N^{-\xi}$  with a small positive $\xi$
but the method has become general. The result  was further  generalized  in \cite{ESYY}
 to  quaternion self-dual 
Wigner matrices  and  sample covariance matrices and even 
to {\it generalized Wigner matrices} in \cite{EYY, EYY2} (see Definition \ref{def:gen}).
Finally, in \cite{EYY3} we showed that the local equilibrium is already
reached after time $t\ge N^{-1+\e}$, which is essentially optimal.
More importantly,  the hydrodynamical method not only applies to all these specific 
ensembles,    but it also gives a {\it conceptual
 interpretation} that the occurrence of the universality 
is due to the relaxation to local equilibrium of the DBM. 

\bigskip

Our hydrodynamical approach consists of two parts. First, we have a general theorem
stating that under certain structural and convexity conditions on the Hamiltonian 
$\cH$ of the equilibrium
measure of the DBM (see \eqref{H} for a special case) and under  a fairly strong
control on the local density of eigenvalues, the local equilibrium is reached
within a short time $t\sim N^{-\xi}$, $\xi>0$, in the sense that
the local correlation functions rescaled in the form of \eqref{sineres}
coincide with the same correlation functions in  equilibrium.
By the general Bakry-Emery \cite{BE} criterion,
the speed of convergence to global equilibrium for DBM depends on the lower bound on
the Hessian of the Hamiltonian $\cH$, which in our case is of order one. The key
idea is to speed up this convergence by modifying the Hamiltonian.
We add to $\cH$ an auxiliary potential of the form 
$$
   W(\bla) = \frac{1}{2 R^2} \sum_j (\la_j-\gamma_j)^2,
$$
where $R\ll 1$ is a parameter, depending on $N$, and $\gamma_j$'s 
are the classical location of the eigenvalues, given
 by 
\be
  N \int_{-\infty}^{\gamma_j} \varrho(x) \rd x = j.
\label{def:gamma}
\ee
Here $\varrho(x)$ is the limiting density, e.g., $\varrho(x)= \varrho_{sc}(x)$
for Wigner matrices. The  Hamiltonian $\wh  \cH: = \cH  + W$ generates a new stochastic
flow of the eigenvalues, called the {\it local relaxation flow}.
The equilibrium Gibbs measure given by
 $\wh\cH$ will be called {\it pseudo equilibrium measure}.
The convergence to equilibrium for this flow is faster, it occurs in a 
time scale $R^2$. In fact, due to the strong repulsion between eigenvalues
(reflected by a singular repulsion potential \eqref{sde}), 
the convergence is even faster for observables that depend 
only on {\it eigenvalue differences}.
Then we show that the modified dynamics
is, in fact, not far from the original one, using that the
typical size of the auxiliary potential $W$ is small.
More precisely, we will 
 need to prove that the eigenvalues $\la_j$ lie near
 $\gamma_j$ with a precision  $N^{-1/2-\e}$, i.e., that
\be
  \E \frac{1}{N} \sum_{i=1}^N (\la_i- \gamma_i)^2 =   N^{-1-2\e}
\label{El}
\ee
holds with some $\e>0$. This is
the key input condition to our general theorem and it
will be proved by a strong control on the local density.
The exponent $\xi$ in the time scale $t\sim N^{-\xi}$  is essentially $2\e$ appearing
in the estimate \eqref{El}.

The second part of the hydrodynamical approach
 is to prove the necessary input conditions on the local density
for the general theorem, especially \eqref{El}. This is the step where specific properties of
the matrix ensemble come into the game. To obtain relaxation to local equilibrium on 
time scale $t\sim N^{-\xi}$, $\xi>0$, we need to locate the eigenvalues 
with a precision at least $N^{-1/2-\e}$, $\e=\xi/2$. To obtain the optimal relaxation time,
$t\gg N^{-1}$, the eigenvalues need to be located essentially with a $N^{-1}$ precision,
similarly to \cite{EPRSY}. Very crudely, the precision of locating eigenvalues
corresponds to the scale $\eta$, on which the local
semicircle law  holds, so this will be the key input to verify \eqref{El}.
  The technical difficulty is that
we need a fairly good control on the local density near the spectral edges as well,
since \eqref{El} involves {\it all} eigenvalues. Although we are interested
only in the bulk universality, i.e., local behavior
away from the edges, we still need the global speed of convergence for
the modified dynamics that is influenced by the eigenvalues at the edge as well.
Recall that the control on the density near the edges becomes weaker since
eigenvalues near the edge  tend to fluctuate more.

A good  control on the local density has been developed in our previous work on Wigner
matrices \cite{ESY1, ESY2, ESY3},  but the edge behavior   was not optimal.
Nevertheless, in \cite{ESYY}  we succeeded in proving \eqref{El} in a 
somewhat complicated way,
relying on some improvements of estimates from  \cite{ESY1, ESY2, ESY3}.
In \cite{EYY}, we found a more direct  way to control the
local density and prove \eqref{El} more efficiently and a  more streamlined
version is given in \cite{EYY2} which we will sketch in Section \ref{sec:sc}.
The strongest result \cite{EYY3}, to be explained in 
Section \ref{sec:best}, gives \eqref{El}
with essentially $2\e=1$.

\bigskip

We  mention that these proofs
 also apply to generalized Wigner matrices 
where the variances satisfy \eqref{VV}.
In this case, we prove the local semicircle law down to essentially the 
smallest possible energy scale $N^{-1}$ (modulo $\log N$ factors). 
This is sufficient to prove \eqref{El} and thus we can 
apply our general theorem and prove the bulk universality of local statistics
for these matrices.
A much more difficult case is the {\it Wigner band matrices}
 \eqref{BM} where, roughly speaking, 
$\sigma_{ij}^2 = 0$ if $|i-j|> W$ for some 
$W \ll N$. 
In this case, we obtain \cite{EYY, EYY2}
the local semicircle law to the energy scale $W^{-1}$
which is not strong enough to prove \eqref{El} if $W$ is much smaller than $N$
(the case $W\ge N^{1-\delta}$ with some small $\delta$ still works).

\subsubsection{Removing the Gaussian convolution I. The reverse heat flow}

In the previous two sections we discussed how to prove bulk universality
for matrices with a small Gaussian convolution. The method of \cite{EPRSY}
(Section \ref{sec:conv})
required only a very small Gaussian component (variance $N^{-1+\e}$)
but it was restricted to the hermitian case. The  hydrodynamical method \cite{EYY3}
(Section  \ref{sec:lrf}) works in general
(the earlier versions \cite{EYY, EYY2}
assumed  a larger Gaussian component with variance $\sim N^{-\xi}$).
Both methods, however, need to be complemented
by a perturbative step to remove this small Gaussian component.

\medskip

There have been two independent approaches developed to remove
the restriction on Gaussian divisibility.
The first method is the {\it reverse heat flow} argument that appeared 
first in \cite{ESY4} and was streamlined in Section 6 of \cite{ESYY}.
The  advantage of this method is that it can prove universality for
a fixed energy $E$ as formulated in \eqref{sineres}, moreover,
it is also very simple. The disadvantage
is that it requires some smoothness of the distribution $\nu$ of $\sqrt{N}h$,
the rescaled entries 
of the Wigner matrix. We always assume  that $\nu$ has the subexponential decay, i.e.,
 that
there are constants $C, \ttau>0$ such that for any $s$
\be
   \int {\bf 1}\big( |x|\ge s \big)\rd\nu(x) \le C \exp\big({-s^\ttau}\big).
\label{subexp}
\ee

The second method, appeared slightly after the first,
 is the {\it Green function comparison theorem}
via a perturbation argument  with the {\it four moment condition}. 
 The advantage of this 
approach is that it  holds for any distribution with the only condition
being the subexponential decay \eqref{subexp}.
The disadvantage is that it proves universality \eqref{sineres} only after some
averaging over $E$. 

The four moment condition  was originally introduced by Tao and Vu
 \cite{TV} and used later in \cite{TV2, TV3} in their  study of eigenvalue perturbation
which focused on the joint statistics of eigenvalues with fixed indices.
 In Section \ref{sec:4mom}
we will present our approach \cite{EYY} based on  resolvent perturbation.
Our result does not identify  fixed eigenvalues
but it is sufficiently strong to identify the local statistics and
its proof is much simpler than in \cite{TV} (see Section \ref{sec:remove4}
for more explanation).

In this section we sketch the method of the reverse heat flow
and in the next section we explain the four moment comparison principles.

\bigskip

For simplicity of the presentation, we consider  the hermitian case
but we emphasize that this method applies to other symmetry classes
as well unlike the method outlined in Section  \ref{sec:conv}.
Consider the OU process defined in \eqref{OU}
 that keeps the variance $\frac{1}{2}$ fixed and let
 $\gamma(\rd x) = \gamma(x)\rd x:=\pi^{-1/2}e^{- x^2} \rd x$
denote the reversible measure for this process. Let 
$\nu_0(\rd x)=u(x)\gamma(\rd x)$ be
the initial measure of the real and imaginary parts of the rescaled entries
of the Wigner matrix $\wh H$ (note that in most of this paper $\nu$
denotes the distribution of $\sqrt{N} h_{ij}$; for this discussion
we introduced the notation $\nu_0$ for the common distribution of
$\sqrt{N} \re h_{ij}$ and $\sqrt{N}\im h_{ij}$, $i\ne j$).
 We let the OU process \eqref{OU} act on the matrix elements, i.e.,
we consider $H_t$, the solution to \eqref{process}. For a fixed $t>0$, 
the distribution of $H_t$ is given by 
\be
    e^{-t/2}\wh H + (1-e^{-t})^{1/2} V,
\label{DBMOU}
\ee
where $V$ is a GUE matrix, independent of $\wh H$.
The distribution of the real and imaginary parts of
the matrix elements of $H_t$ is then given by $u_t(x)\gamma(\rd x)$,
 where $u_t$ is the solution to \eqref{OU}
with initia data $u_0=u$ (strictly speaking, these formulas hold for
the offdiagonal elements, the diagonal element has twice bigger variance
and it is subject to a slightly different OU flow).

The main observation is that the arguments in Section \ref{sec:conv} or Section \ref{sec:lrf}
guarantee
that sine kernel holds for {\it any} hermitian matrix that has a Gaussian
component of variance $N^{-1+\e}$ or $N^{-\xi}$, respectively
(the method of Section \ref{sec:conv} applies only to the hermitian case,
while Section \ref{sec:lrf} works in general). Given a Wigner matrix $\wh H$, we do not
necessary have to compare $\wh H$ with its Gaussian convolution \eqref{DBMOU}; 
it is sufficient to find {\it another} Wigner matrix $\wt H$ such that
\be
  \wh H \approx  e^{-t/2}\wt H + (1-e^{-t})^{1/2} V
\label{form1}
\ee
with a very high precision. In fact, $\wt H$ can even be chosen $t$-dependent.
The following lemma shows that {\it any} Wigner matrix $\wh H$ with 
a sufficiently smooth distribution can  be arbitrary well approximated 
by Gaussian divisible matrices of the form  \eqref{form1} if $t\sim N^{-\delta}$
for some $\delta>0$.

We assume that the initial density is positive, $u(x)>0$, and it can be written as
\be
   u(x) = e^{-V(x)}, \qquad \mbox{with} \qquad \sum_{j=1}^{2K} |V^{(j)}(x)|
 \le C_K(1+x^2)^{C_K}
\label{Vcond}
\ee
with any $K\in\N$ and with sufficiently large constants  $C_K$. Moreover,
we assume that the initial single entry distribution $\rd\nu_0 =u\rd\gamma$ has a 
subexponential decay \eqref{subexp}.
The key technical lemma is the following approximation statement.

\begin{lemma}\label{meascomp1} [Proposition 6.1 \cite{ESYY}]
Suppose that for some $K>0$, the  measure
$\rd\nu_0 = u \rd\gamma $ satisfies \eqref{subexp} and \eqref{Vcond}.
Then there is a small constant $\alpha_K$ depending on $K$ such
that for   any  $t  \le  \alpha_K $
there exists a probability density $g_t$ with mean zero and variance $\frac{1}{2}$ such that
\be\label{gtilde}
\int  \left  | e^{t A}g_t  - u \right |
\rd \gamma  \le C\; t^{K}
\ee
for  some $C>0$ depending on $K$.

Furthermore, let  $\cA=A^{\otimes n}$, $ F=  u^{\otimes n}$ with some $n\le CN^2$.
Denote by $G_t=  g_t^{\otimes n}$. Then we also have
\be\label{FFtilde}
\int  \left  | e^{t\cA}G_t  - F \right |
\rd \gamma^{\otimes n}   \le C\; N^2 t^{K}
\ee
for  some $C>0$ depending on $K$.
\end{lemma}

{\it Sketch of the proof.}  
Given $u$, we want to solve the equation
$$
  e^{At}g_t = u,
$$
i.e., formally $g_t = e^{-At}u$. However, the operator $e^{-At}$
is like running a heat flow (with an OU drift) in reverse time
which is typically undefined unless $u$ is analytic. But we can define
an approximate solution to the backward heat equation, i.e., we set
$$
    g_t:= \Big( I - At  + \frac{t^2A^2}{2!} - \ldots + \frac{(-tA)^{K-1}}{(K-1)!} \Big) u.
$$
Since $A$ is a second order differential operator and $u$ is sufficiently smooth, this
expression is well defined, moreover
$$
   e^{At}g_t = O\Big(t^K A^{K}u\Big)  = O(t^K).
$$
This proves \eqref{gtilde} and
 \eqref{FFtilde} directly follows from it. \qed

\bigskip

Armed with Lemma \ref{meascomp1},
we can prove the sine kernel universality
in the form of \eqref{sineres} for any hermitian Wigner matrix
satisfying  \eqref{subexp} and \eqref{Vcond} 
for any fixed $|E|<2$.
We  choose $n\sim N^2$ to be the number of independent
OU processes needed to generate the flow of the matrix elements.
By choosing $K$ large enough, we can compare  the two measures $e^{t\cA}G_t$ and $ F$
in the total
variational norm; for any observable $J:\bR^n\to \bR$ of the matrix
elements, we have
\[
\left| \int J ( e^{t\cA}G_t  - F )
\rd \gamma^{\otimes n} \right| \le   \|J\|_\infty C\; N^2 t^{K}.
\]
In order to prove \eqref{sineres}, appropriate observables
$J$ need to be chosen that depend on the matrix elements via the eigenvalues
and they express local correlation functions.
It is easy to see that for such $J$, its norm $\|J\|_\infty$ may grow at most 
polynomially in $N$.
But we can always choose $K$ large enough to compensate for it 
with the choice $t= N^{-\delta}$. Since the sine kernel  holds for the distribution
$e^{t\cA}G_t$ with $t=N^{-1+\e}$ (Section \ref{sec:conv})
or $t=N^{-\xi}$ (Section \ref{sec:lrf})
 it will also hold for the Wigner measure $F$.

For symmetric matrices
the reverse heat flow argument is exactly the same,
but then only Section \ref{sec:lrf} is available to obtain
universality for short time; in particular, $E$ in \eqref{sineres} needs to be averaged.

\subsubsection{Removing the Gaussian convolution II. The Green function comparison 
theorem}\label{sec:remove4}

Let $H$ and $H'$ be two Wigner ensembles such that the first four moments
of the single entry distribution, $\nu$ and $\nu'$, coincide:
\be
        m_j = m_j', \qquad j=1,2,3,4
\label{mj}
\ee
where 
$$
   m_j :=\int_\R x^j \rd \nu (x) \quad \mbox{and}  \quad m_j': =\int_\R x^j \rd \nu' (x).
$$
For complex entries one has to take
the collection of all $j$-moments, i.e., $m_j$ represents the collection
of all $\int_\C x^a\bar x^b \rd \nu(x)$
with $a+b=j$. Recall that $\nu$ is the distribution of $\sqrt{N}h_{ij}$.
By our normalization of Wigner matrices, the first moment is
always zero and the second moment is one, so \eqref{mj} is really a 
condition on the third and fourth moments.

Our main result is the following comparison theorem
for the  joint distribution of Green functions.
Here we only state the result in a somewhat simplified form,
a more detailed presentation will be given in Section~\ref{sec:4mom}.

\begin{theorem}[Green function comparison theorem] \cite[Theorem 2.3]{EYY}
\label{thm:4mom} Consider two Wigner matrices,  
  $H$ and  $H'$,  with single entry
distributions $\nu$ and $\nu'$. Assume that \eqref{subexp} 
and \eqref{mj} hold for $\nu$ and $\nu'$. Let $G(z) = (H-z)^{-1}$
and $G'(z) = (H'-z)^{-1}$ denote the resolvents. Fix $k$ and suppose
that the  function $F:\R^k\to\R$ satisfies
\be
  \sup_{x\in\R^k} |\nabla^j F(x)|\le N^{\e'}, \qquad 0\le j\le 5.
\label{fj}
\ee
 Fix small parameters $\kappa$ and $\e$.
Then, for sufficiently small $\e'$ there is $c_0>0$ such that for any 
integers $\ell_1, \ldots, \ell_k$ and spectral parameters $z_j^m= E_j^m
\pm i\eta$, $1\le j\le \ell_m$, $m=1,2,\ldots, k$ with $E_j^m \in [-2+\kappa, 2-\kappa]$
and $\eta \ge N^{-1-\e}$, we have
\be
  \Bigg| \E F\Bigg( 
\frac{1}{N}\tr   \Bigg[\prod_{j=1}^{\ell_1} G(z_j^1)\Bigg], \ldots , 
\frac{1}{N}\tr   \Bigg[\prod_{j=1}^{\ell_k} G(z_j^k)\Bigg] \Bigg) - 
  \E' F( G\to G')\Bigg|\le N^{-c_0}.
\label{4momres}
\ee
\end{theorem}
Here the shorthand notation $F\left ( G \to  G' \right ) $
means that we consider the same argument of $F$ as in 
the first term in \eqref{4momres}, but all $G$
terms are replaced with $G'$.

In fact, the condition \eqref{mj} can be weakened
 to require that the third and fourth moment be only close;
\be
        m_j = m_j', \qquad j=1,2, \quad\mbox{and}
\quad  |m_3-m_3'|\le N^{-\frac{1}{2}-\delta}, \quad  |m_4-m_4'|\le N^{-\delta}
\label{mjuj}
\ee
with some $\delta>0$. Then \eqref{4momres} still holds, but $\e$ and
$\e'$ have to be sufficiently small, depending on $\delta$ and
 $c_0$ will also depend on $\delta$. The precise 
estimate will be stated in Theorem~\ref{comparison}.

In other words, under the four moment matching condition
for  two Wigner ensembles,
 the expectations of traces of any combination of resolvent products 
coincide if the spectral parameters in the resolvents are not
closer than $\eta=N^{-1-\e}$ to the real axis. Such a small distance
corresponds to spectral resolution on scale $N^{-1-\e}$, i.e.,
it can identify local correlation functions of individual eigenvalues.
It is an easy algebraic
identity to express correlation functions from traces of resolvents, 
for example the one point correlation function (density) on scale $\eta$
 is approximated by 
$$
   p^{(1)}_N(E)\sim \frac{1}{\pi N}\im \tr G(E+i\eta) = \frac{1}{2\pi}\Big[ 
  \frac{1}{N}\tr G(E+i\eta) -
  \frac{1}{N}\tr G(E-i\eta)\Big]
$$
and higher point correlation functions involve higher order polynomials
of resolvents. Thus Theorem \ref{thm:4mom} directly compares
correlation functions (for the precise statement, see Theorem \ref{com}).
We remark that taking traces is not essential in \eqref{4momres}, a similar
comparison principle works for matrix elements of the resolvents as well
(see \cite{EYY} for the precise formulation). 
In fact, the proof of Theorem \ref{thm:4mom} is a perturbation
argument directly involving matrix elements of the resolvent.
The key ingredient is a stronger form of the local semicircle
law that directly estimates $G_{ii}$ and $G_{ij}$, $i\ne j$, and
not only the normalized trace, $m(z) =\frac{1}{N}\sum_i G_{ii}$
(see \eqref{Gii}--\eqref{Gij} and \eqref{Lambdaodfinal} for the strongest result).

\bigskip

A related theorem for eigenvalues was  proven earlier by Tao and Vu \cite{TV}.
Let $\lambda_1<\lambda_2 <\ldots < \la_N$ and
$\lambda_1'<\lambda_2' <\ldots < \la_N'$ denote the
eigenvalues of $H$ and $H'$, respectively. The following theorem
states that the joint distribution of any $k$-tuple of
eigenvalues on scale $1/N$ is very close to each other.

\begin{theorem}[Four moment  theorem for eigenvalues]\cite[Theorem 15]{TV}\label{thm:TV}
Let $H$ and  $H'$ be two Wig\-ner matrices and 
 assume that  \eqref{subexp} and \eqref{mj}  hold for their single entry distributions
$\nu$ and $\nu'$.
 For any sufficiently small positive $\e$ and $\e'$
and for any function $F:\R^k\to\R$ satisfying \eqref{fj},
and for any selection of $k$-tuple of indices $i_1, i_2, \ldots, i_k\in[\e N,
(1-\e)N]$ away from the edge, we have
\be
   \Bigg| \E F\Big( N\la_{i_1}, N\la_{i_2}, \ldots, N\la_{i_k}\Big) 
  -  \E' F\Big( N\la'_{i_1}, N\la'_{i_2}, \ldots, N\la'_{i_k}\Big) \Bigg|\le N^{-c_0}
\label{FF}
\ee
with some $c_0>0$.
The condition \eqref{mj} can be relaxed to \eqref{mjuj},
but $c_0$ will depend on $\delta$.
\end{theorem}

\bigskip

Note that the arguments in \eqref{FF} are magnified by a factor $N$
and $F$ is allowed to be concentrated on a scale $N^{-\e'/5}$, so
the result is sufficiently precise to detect eigenvalue correlations on scale 
$N^{-1-\e'/5}$, i.e., even somewhat smaller than the eigenvalue spacing.
Therefore Theorem \ref{thm:4mom} or 
\ref{thm:TV} can prove bulk universality for a Wigner matrix $H$
if another $H'$ is found, with matching four moments, for which universality is
already proved. In the {\it hermitian} case, the GUE matrices, or more generally the
 Gaussian divisible matrices
\eqref{HaV} provide a good reference ensemble. Matching with a GUE matrix requires
that the third and the
 fourth moments match, $m_3=0$, $m_4=3$. Since
the {\it location} of the eigenvalues for GUE is known very precisely \cite{Gu, OR},
 \eqref{FF} can be translated into the limit
of correlation functions as \eqref{sineres} even at a fixed energy $E$.
If one aims only at the limiting gap distribution \eqref{maineq2}
instead of \eqref{sineres}, then one can directly use the Gaussian divisible
matrix \eqref{HaV} for matching.
 It is easy to check \cite{CF}
that for any probability distribution $\nu$ with $m_1=0$ and $m_2=1$  that is
supported on at least  three points, there is a distribution
with an order one Gaussian component so that the first four moments match.
Therefore $H$ can be matched with a Gaussian divisible matrix 
for which Johansson \cite{J} has proved universality.
Using the result of \cite{EPRSY} on the universality of hermitian Wigner matrices
with a tiny Gaussian convolution (conclusion of Section \ref{sec:conv}),
and using that exact moment matching \eqref{mj} can be relaxed to \eqref{mjuj}, 
one can compare any Wigner matrix $H$ with its very short  time $t\sim N^{-1+\e}$
 Ornstein-Uhlenbeck  convolution \eqref{HaV}. This removes the requirements $m_3=0$  and 
that the support has at least three points and proves universality  of correlation functions
for any hermitian Wigner matrix  in the sense of \eqref{sineres} after
a little averaging in $E$ \cite{ERSTVY}. The only technical condition
is the subexponential decay of $\nu$ \eqref{subexp}.

In the symmetric case, the analogue of Johansson's result is not available
(unless one uses \cite{ESY4}), and the
only reference ensemble is GOE. Theorem \ref{thm:TV} thus implies \cite{TV} universality
for symmetric Wigner matrices whose single entry distribution has
first four moments matching with GOE in the sense of \eqref{mjuj}.

\bigskip

The careful reader may notice a subtle difference between the observable
in \eqref{FF} and the local correlation functions \eqref{sineres}.
While both detect the structure on scale $1/N$, in \eqref{FF} 
the {\it indices} of the eigenvalues are fixed, while in \eqref{sineres}
their {\it location}. Roughly speaking, \eqref{FF} can answer the 
question, say, ``where are the $N/2$-th and the $(N/2-1)$-th eigenvalue''. The local
correlation function asks ``what is the probability of the simultaneous
event that there is an eigenvalue at $E$ and another one at $E'= E+\al/N$''.
These questions can be related only if some a-priori information is
known about the location of the eigenvalues.

Prior to \cite{TV}, apart from the GUE case
\cite{Gu,OR}, for no other ensembles could the eigenvalues 
be located with a precision $N^{-1+c_0}$ for small $c_0$,
and such precision is needed to translate \eqref{FF} into
\eqref{sineres} for a fixed $E$. Using a three-moment matching
version of Theorem \ref{thm:TV}, one can locate the eigenvalues
of any Wigner ensembles with such precision, provided that the third moment vanishes
(i.e. matches with GUE). Given this information, one
can proceed to match the fourth moment by choosing
an appropriate Gaussian divisible matrix. This is possible
if the original distribution is supported on at least three points.
This is why eventually \eqref{sineres} was proven  in \cite{TV} under the condition
 that the third moment vanishes and the support contains at least three points.
If one accepts that \eqref{sineres} will be proven after some averaging in $E$,
then the necessary information on the location of the eigenvalues is
much weaker and it can typically be obtained from the local semicircle law.

\medskip

In fact, tracking individual eigenvalues can be a difficult task; note that
 Theorem  \ref{thm:TV}
in itself does not directly imply convergence of correlation functions,
one still needs some information about the location of the $i$-th eigenvalue.
On the other hand,  Theorem  \ref{thm:TV} contains information about
eigenvalues with {\it fixed indices} which was not contained
in Theorem \ref{thm:4mom}. We remark that the local semicircle
law is an essential input for both theorems.

 The main reason why the proof of Theorem \ref{thm:4mom} is shorter
is due to that fact
 that correlation
functions can be identified from observables involving traces
of resolvents $(H-z)^{-1}$ with $\im z \sim N^{-1-\e}$ and these resolvents
have an a-priori bound of order $|\im z|^{-1}\le N^{1+\e}$, so perturbation
formulas involving resolvents do not blow up. On the other hand, 
the individual eigenvalues tracked by 
Theorem  \ref{thm:TV}  may produce resonances which could render
some terms even potentially infinite (we will sketch the proof of
Theorem \ref{thm:TV} in  Section~\ref{sec:4mom}).
 While level repulsion is a general
feature of Wigner ensembles and it strongly suppresses resonances,
the direct proof of the level repulsion is not an easy task.
In fact, the most complicated technical estimate in \cite{TV} is
the lower tail estimate on the gap distribution (Theorem 17 of \cite{TV}).
It states that for any $c_0>0$ there is a $c_1$ such that
\be
   \P \big(\lambda_{i+1}-\lambda_i\le N^{-1-c_0}\big) \le N^{-c_1}
\label{lowtail}
\ee
if the index $i$ lies in the bulk ($\e N\le i \le (1-\e)N$).

\subsubsection{Summary of the new  results on bulk universality}

Even the expert reader may find the recent developments slightly confusing since
there have  been many papers on bulk universality of Wigner matrices
under various conditions and with different methods.
Their interrelation was not always optimally presented in the
research publications since it was, and still is, a fast developing
story. In this section we try to give some orientation
to the reader for the recent literature.

As mentioned in the introduction, the main guiding principle behind these proofs
of universality of the local eigenvalue statistics  is to compare the local statistics of a
Wigner matrix with
another matrix with some Gaussian component. More precisely,
our approach consists of three main steps: 

\begin{itemize}
\item[(1)]   the local semicircle law;
\item[(2)]   universality for
Gaussian divisible ensembles, i.e., if the probability law of matrix elements contains a small
Gaussian component.  
\item[(3)]  universality for general ensembles; approximation by a Gaussian divisible ensemble
to remove  the small Gaussian component.
\end{itemize}

It was clear to us from the very beginning that a good local semicircle
law must be the first step in any proof of universality. In fact, all proofs of universality
rely heavily on the details of the estimates one can obtain for the local semicircle law.
We now summarize the existing results according to these three components.

\medskip

Although the proof of the  local semicircle law down to the shortest scale $\eta\sim 1/N$
is rather simple now, it was only gradually achieved. In our first paper, \cite{ESY1},
we gave an {\it upper bound} on the local density essentially
down to the optimal energy scale $\eta\sim (\log N)/N$, but
the local semicircle law itself was proven only on scale $\eta\gg N^{-2/3}$.
In the second paper, \cite{ESY2}, we proved the local semicircle
law down to the scale $\eta\ge (\log N)^8/N$, almost optimal but still
off by a logarithmic factor. The tail probability to violate the local
semicircle law was also far from optimal.  
Both defects were remedied in \cite{ESY3} (see Theorem \ref{wegnerlsc} below)
 where, additionally,
 an optimal delocalization result for eigenvectors was also proven
(Theorem~\ref{deloc:trad}).
In the first paper \cite{ESY1}
 we  assumed  a strong (Gaussian) decay condition and some convexity property
of the single entry distribution that implies concentration (either via
Brascamp-Lieb or logarithmic Sobolev inequalities).
These technical conditions were subsequently removed and the Gaussian decay condition
was replaced by a subexponential decay.
Finally, in \cite{EYY} and in its improved and streamlined version
in \cite{EYY2}, we obtained a much stronger error estimate to the
local semicircle law, see
\eqref{mmm3}--\eqref{GIJ}, but these estimates still deteriorate at the edge.
 The optimal result \cite{EYY3},
which we call the strong local semicircle law (Theorem \ref{45-1}),
holds uniformly in the energy parameter.

\medskip

As for Step (2), the main point is that the Gaussian component
enables one to exhibit the universal behavior. There are two ways to implement
this idea:

\begin{itemize}
\item[(i)] the contour integral representation following
Johansson \cite{J} and Ben-Arous, P\'ech\'e \cite{BP}
but this option is available only for the hermitian case (or for
the complex sample covariance case, \cite{BP});
\item[(ii)] the hydrodynamical approach, where
a small Gaussian component (equivalently, a small time evolution of the OU process)
already drives the system to local equilibrium.  This approach  applies to all
ensembles, including symmetric, hermitian,  sympletic and sample covariance ensembles,
and it also gives the conceptual interpretation that the universality arises from the
Dyson Brownian motion.
\end{itemize}

Both approaches require a good local semicircle law. Additionally,
the earlier  papers on the hydrodynamical methods, \cite{ESY4} and \cite{ESYY},
also assumed the logarithmic Sobolev inequality (LSI) for the
single entry distribution $\nu$, essentially in order to verify \eqref{El}
from the local semicircle law.
In \cite{EYY2} we removed this last  condition
by using a strengthening of the  local semicircle law which
gave a simpler and more powerful proof of \eqref{El} (Theorem 
\ref{prop:lambdagamma}). Finally, 
the strong local semicircle  law in \cite{EYY3} (Theorem \ref{45-1}) 
provided the  optimal exponent $2\e=1$  in \eqref{El}.

Summarizing the first two steps,  we  thus obtained bulk
 universality for generalized  Wigner
matrices  \eqref{VV}
with a small Gaussian convolution
under the sole condition of subexponential decay.
This condition  can
be relaxed to a high order polynomial decay, but we have not worked out
the details.
 Furthermore, although the extension of the strong local semicircle
law to  sample covariance matrices
is straightforward, these details have not been carried out
either
(the earlier detailed proof \cite{ESYY}  required LSI).

The first two steps provide a large class of matrix ensembles with universal
local statistics. In Step (3) it remains to approximate arbitrary
matrix ensembles by these matrices so that the local statistics 
are preserved. The approximation step  can be done in two ways
\begin{itemize}
\item[(i)]  via the reverse heat flow;
\item[(ii)] via the Green function comparison theorem.
\end{itemize}
The reverse heat flow argument is very simple, but it requires smoothness on the 
single entry distribution.
This approach was used   in \cite{EPRSY}, \cite{ESY4} and  \cite{ESYY} and this leads to
universality for all ensembles mentioned under the smoothness condition.
This smoothness condition was then removed in 
\cite{EYY} where   the Green function comparison theorem was
first proved. Unfortunately, we still needed the LSI and
universality was established for matrices 
whose distribution  $\nu$ is supported on at least three points.

A stronger version of the local semicircle law was proved in  \cite{EYY2}
and  all smoothness and support conditions on the distributions were removed.
In summary, in \cite{EYY2}  we obtained
bulk universality of correlation functions \eqref{sineres} and gap distribution
\eqref{maineq2} for all classical Wigner ensembles
(including the generalized Wigner matrices, \eqref{VV}).
The universality in \eqref{sineres}
is understood after a small averaging in the energy parameter $E$.
The only condition on the single entry distribution $\nu$ has
the subexponential decay \eqref{subexp}.

\bigskip

The  approach of Tao and Vu \cite{TV} uses a similar strategy of the three Steps (1)--(3)
mentioned at the beginning of this section. For Step (2),
the  universality for {\it hermitian} Wigner matrices and
{\it complex} sample covariance matrices were previously proved by
Johansson \cite{J}  and  Ben-Arous and P\'ech\'e \cite{BP}.
Step (3) follows from  Tao-Vu's  four moment theorem  (Theorem \ref{thm:TV})
whose proof
uses is the local semicircle law, Step (1).
This leads to the universality for the  {\it hermitian} Wigner matrices  \cite{TV} 
and {\it complex} sample covariance matrices  \cite{TV3}
 satisfying the condition that the support
of the distribution contains at least three points (and, if
one aims at a fixed energy result in \eqref{sineres}, then
 the third moment has also to vanish).
For the symmetric case, the matching of the first four
moments was required. The Tao-Vu's approach can also be applied to prove
edge universality \cite{TV2}.
In     \cite{TV3}   the subexponential decay was replaced with
a condition with a sufficiently strong
 polynomial decay. The support and the third moment condition condition can be removed
by combining \cite{TV}
with the result from our approach \cite{EPRSY}  and this has led to the universality
of hermitian matrices \cite{ERSTVY} for any distribution  including the Bernoulli  measure. 
On the other hand, even for hermitian matrices, the variances
of the matrix elements are required to be constant in this approach.

Historically, Tao-Vu's first paper on the universality \cite{TV} appeared shortly after the
paper \cite{EPRSY} on the universality of hermitian matrices. 
A common ingredient for both \cite{EPRSY} and \cite{TV}
is the local semicircle law and the eigenfunction delocalization
estimates that were essentially available from \cite{ESY2, ESY3},
but due to certain technical conditions, they were reproved in \cite{TV}.
The local semicircle law for sample covariance matrices was first proved in \cite{ESYY} and 
a slightly different version was given in \cite{TV3} with some change in the 
technical assumptions tailored to the application.

The four moment condition
 first appeared in the four moment theorem  by Tao-Vu \cite{TV} 
(Theorem \ref{thm:TV}) and  it was used in the Green function comparison theorem
\cite{EYY} (Theorem \ref{thm:4mom}).
 The four moment theorem  concerns individual eigenvalues and thus it  contains
information about the eigenvalue gap distribution directly.
In order to translate this information
into correlation functions, the locations of individual
eigenvalues  of the comparison ensemble are required. The
Green function comparison theorem, on the other hand, 
can be used to compare the correlation functions directly, but the
information on the individual eigenvalues is  weaker.
Nevertheless, a standard exclusion-inclusion
principle argument (like the one presented in \eqref{ex})
concludes the universality of the gap distribution as well.
Since individual eigenvalues tend to fluctuate and Green functions are
more stable, this explains why the  proof of the four moment theorem for eigenvalues
is quite involved but the Green function comparison theorem is very simple.
Furthermore, the Green function comparison theorem yields not only 
spectral  information, but information 
on matrix elements as well.

\subsubsection{New results on edge universality}\label{sec:newedge}

Recall that $\lambda_N$ is the largest eigenvalue of the random matrix.  The 
probability distribution  functions of $\lambda_N$ for the classical Gaussian ensembles are
identified by Tracy and Widom  \cite{TW, TW2} to be  
\be\label{Fb}
\lim_{N \to \infty} \P( N^{2/3} ( \lambda_N -2) \le s ) =  F_\beta (s), 
\ee
where the function $F_\beta(s)$ can be computed in terms 
of Painlev\'e equations and $\beta=1, 2, 4$
 corresponds to the standard classical  ensembles. The distribution of $\lambda_N$ 
is believed to be universal and independent of the Gaussian structure. 
 The strong local semicircle law, Theorem \ref{45-1},
combined with a modification of the Green function comparison theorem,
Theorem \ref{thm:4mom}, taylored to spectral edge
implies  the following 
version of universality of the extreme eigenvalues:

\bigskip

\begin{theorem}[Universality of extreme eigenvalues]\cite[Theorem 2.4]{EYY3} \label{twthm}  
Suppose that we have 
two  $N\times N$ generalized Wigner matrices, $H^{(v)}$ and $H^{(w)}$, with matrix elements $h_{ij}$
given by the random variables $N^{-1/2} v_{ij}$ and 
$N^{-1/2} w_{ij}$, respectively, with $v_{ij}$ and $w_{ij}$ satisfying
the subexponential decay condition \eqref{subexp} uniformly 
for any $i,j$. Let $\P^\bv$ and
$\P^\bw$ denote the probability and $\E^\bv$ and $\E^\bw$ 
the expectation with respect to these collections of random variables.
 If 
the first two moments of
 $v_{ij}$ and $w_{ij}$ are the same, i.e.
\be\label{2m}
    \E^\bv \bar v_{ij}^l v_{ij}^{u} =  \E^\bw \bar w_{ij}^l w_{ij}^{u},
  \qquad 0\le l+u\le 2,
\ee
then there is an $\e>0$ and $\delta>0$
depending on $\ttau$ in \eqref{subexp}
 such that 
for any $s \in \R $ 
we have
 \be\label{tw}
 \P^\bv ( N^{2/3} ( \lambda_N -2) \le s- N^{-\e} )- N^{-\delta}  
  \le   \P^\bw ( N^{2/3} ( \lambda_N -2) \le s )   \le 
 \P^\bv ( N^{2/3} ( \lambda_N -2) \le s+ N^{-\e} )+ N^{-\delta}  
\ee 
for $N$ sufficiently  large independently of $s$. 
Analogous result holds for the smallest eigenvalue $\lambda_1$.

\end{theorem}

Theorem \ref{twthm} can be extended to finite correlation functions of  extreme eigenvalues.
 For example, 
we have the following extension to \eqref{tw}:
 \begin{align}\label{twa}
&  \P^\bv \Big ( N^{2/3}  ( \lambda_N -2) \le s_1- N^{-\e}, \ldots, N^{2/3} ( \lambda_{N-k} -2) 
\le s_{k+1}- N^{-\e} \Big )- N^{-\delta}   \nonumber \\
& 
 \le   \P^\bw \Big ( N^{2/3} (  \lambda_N -2) \le s_1,  \ldots, N^{2/3} ( \lambda_{N-k} -2) 
\le s_{k+1}  \Big )  \\
 &  \le 
 \P^\bv \Big ( N^{2/3} ( \lambda_N -2) \le s_1+ N^{-\e}, \ldots, N^{2/3} ( \lambda_{N-k} -2)
 \le s_{k+1}+ N^{-\e}   \Big )+ N^{-\delta}  \nonumber 
\end{align}  
for all $k$ fixed and $N$ sufficiently  large.

The edge universality for Wigner matrices was first  proved 
via the moment method by Soshnikov \cite{Sosh} (see also the earlier work \cite{SS})
for hermitian and symmetric  ensembles with  symmetric single entry distributions $\nu$
to ensure that all odd moments vanish.  
By combining the moment 
method and Chebyshev polynomials \cite{FSo}, Sodin 
proved edge universality of band matrices and some special class
of sparse matrices \cite{So1, So2}.

The removal of the symmetry assumption was not straightforward. 
The approach of  \cite{So1, So2} is  restricted to ensembles with symmetric distributions. 
The symmetry assumption  was partially removed in 
\cite{P1, P2} and significant progress was made in  \cite{TV2} which assumes only  
that the first three moments of two Wigner ensembles  are identical. In other words, 
the symmetry assumption was replaced 
by the vanishing third moment condition for Wigner matrices.  For a special class of
 ensembles, the Gaussian divisible  hermitian ensembles, edge universality was proved \cite {J1} 
under the sole condition that the second moment is finite. By a
 combination of methods from \cite{J1} 
and \cite{TV2}, the same result can be proven for  all  hermitian Wigner ensembles
with finite second moment \cite{J1}.

In comparison with these results,  Theorem 
\ref{twthm} does not imply  the edge universality of band matrices or sparse matrices 
\cite{So1, So2},  but it implies in particular that, for the purpose  to identify 
the distribution of the top 
eigenvalue for a generalized Wigner matrix, 
it suffices to consider generalized Wigner ensembles with Gaussian distribution. 
Since the distributions 
of the top eigenvalues of the  Gaussian Wigner ensembles 
are given by $F_\beta$ \eqref{Fb},  Theorem 
\ref{twthm} shows the edge universality of the standard 
Wigner matrices under the subexponential decay assumption alone. 
We remark that one can use Theorem \ref{7.1} as an input
in the approach of \cite{J1} to prove that the distributions of the top eigenvalues of
  generalized hermitian Wigner ensembles with Gaussian distributions  are given by $F_2$. But for
 ensembles in a different symmetry class, 
there is no corresponding  result to identify  the distribution 
of the top eigenvalue with $F_\beta$.

Finally, we comment that  the subexponential  decay 
assumption in our approach, though can be weakened,   is far from optimal 
for edge universality \cite{ABP, BBP, P2}.

\subsection{Level repulsion and Wegner estimate on very short scales}\label{sec:wegner}

One of our earlier local semicircle laws for Wigner matrices,
\be\label{mmm2}
  |m(z)-m_{sc}(z)|\lesssim \frac{C}{\sqrt{N\eta\kappa}}, \qquad \kappa=\big| |E|-2\big|,
\ee
proven in Theorem 4.1 of \cite{ERSY},
 can be turned into a direct
estimate on the empirical density in the form
$$
   |\varrho_\eta(E) - \varrho_{sc}(E)|\lesssim \frac{C}{\sqrt{N\eta\kappa}}, \quad 
 \kappa = \big| |E|-2\big|.
$$
Here $\varrho_\eta$ denotes the empirical density $\varrho(x) =\frac{1}{N}\sum_i
\delta (\lambda_i- x)$ smoothed out on a scale $\eta$.
This result asserts that the empirical density
on scales  $\eta\gg O(1/N)$  is close to the semicircle density.
On even smaller scales $\eta \le O(1/N)$, the empirical density fluctuates, but
its average, $\E \, \varrho_\eta (E)$, remains bounded
uniformly in $\eta$.  This is  
a type of Wegner estimate that plays a central role
in the localization theory of random Schr\"odinger operators.
In particular, it says that the probability of finding
at least one eigenvalue in an interval $I$ of size $\eta =\e/N$
is bounded by $C\e$ uniformly in $N$ and $\e\le 1$, i.e., no eigenvalue
can stick to any energy value $E$.
Furthermore, if the eigenvalues were independent (forming a Poisson process), then the probability
of finding $n=1, 2,3,\ldots $ eigenvalues in $I$ were proportional with
$\e^n$. For random matrices in the bulk of the
spectrum this probability is much smaller. 
This phenomenon is known
as level repulsion and the precise statement is the following:

\begin{theorem}\cite[Theorem 3.4 and 3.5]{ESY3}\label{thm:repul} Consider symmetric
or hermitian Wigner matrices with a single entry distribution $\nu$ 
that has a Gaussian decay.
Suppose  $\nu$ is absolutely continuous
with a  strictly positive and smooth density. Let $|E| < 2$ and
$I= [E-\eta/2, E+\eta/2]$ with $\eta = \e/N$ and let $\cN_I$ denote
the number of eigenvalues in $I$.
 Then  for any fixed $n$,
\be
   \P (\cN_I \ge n) \le
 \left\{  \begin{array}{cl} C_n\e^{n^2} &  \quad \mbox{[hermitian case]}\\
  C_n \e^{n(n+1)/2} & \quad\mbox{[symmetric case]}
 \end{array} \right.
\label{rep}
\ee
uniformly in $\e\le 1$ and for all sufficiently large $N$.
\end{theorem}

The exponents are optimal as one can easily see
from the Vandermonde determinant
in the joint probability density \eqref{expli} for invariant
ensembles. The sine kernel behavior \eqref{sinesimple}
indicates level repulsion and even a lower bound on $\P (\cN_I \ge n)$,
 but usually not on arbitrarily small
scales since sine kernel is typically proven 
only as a weak limit (see \eqref{sineres}).

We also mention that \eqref{lowtail}  (Theorem 17 from \cite{TV})
is also a certain type of level repulsion bound, but  the exponents
are not optimal  and it does not hold on arbitrary
small scales. However, the virtue of \eqref{lowtail} is
that it assumes no smoothness of the distribution, in 
particular it holds for discrete distributions as well.
Clearly, say, for Bernoulli distribution,  even the Wegner estimate, \eqref{rep} for
$n=1$, cannot hold on
superexponentially small scales $\e\sim 2^{-N^2}$.

\bigskip

{\it Sketch of the proof.} 
The first step of the proof is to provide an
upper bound on $\cN_I$. Let $H^{(k)}$ denote the $(N-1)\times (N-1)$
minor of $H$ after removing the $k$-th row and $k$-th column. Let
 $\lambda_\al^{(k)}$, $\al=1,2, \ldots N-1$ denote the eigenvalues of $H^{(k)}$
and $\bu_\al^{(k)}$ denote its eigenvectors. Computing the $(k,k)$ 
diagonal element of the resolvent $(H-z)^{-1}$ we easily obtain the following expression
for $m(z)=m_N(z)$
\be
   m(z) = \frac{1}{N}\sum_{k=1}^N \frac{1}{H-z}(k,k)
  = \frac{1}{N}\sum_{k=1}^N \Bigg[ h_{kk} -z-\frac{1}{N}
  \sum_{\al=1}^{N-1} \frac{\xi_\al^{(k)}}{\lambda_\al^{(k)}-z}\Bigg]^{-1},
\label{id}
\ee
where 
\be
\xi_\al^{(k)}: = N |\ba^{(k)}\cdot \bu_\al^{(k)}|^2,
\label{xidef}
\ee
 and
$\ba^{(k)}$ is the $k$-th column of $H$ without the diagonal element $h_{kk}$. 
Taking the imaginary part, and using 
\be
\cN_I\le CN\eta\, \im \; m(z), \qquad z= E+i\eta,
\label{cNN}
\ee 
we have
\be
  \cN_I \le CN\eta^2 \sum_{k=1}^N \Big| \sum_{\al\;: \; \lambda_\al^{(k)}\in I} 
  \xi_\al^{(k)}\Big|^{-1}.
\label{key}
\ee
It is an elementary fact that, for
each fixed $k$, the eigenvalues $\la_1\le \la_2\le \ldots \le\la_N$
 of $H$ and the eigenvalues $\mu_1\le \mu_2\le \ldots \le \mu_{N-1}$ of $H^{(k)}$
are {\it interlaced}, meaning that
\be
  \la_1\le \mu_1\le \la_2\le \mu_2 \le \ldots \le \mu_{N-1}\le \la_N.
\label{interlace}
\ee
This can be seen by analyzing the equation for the eigenvalues $\la$ in terms
of the eigenvalues $\mu$'s
$$
   \lambda - h_{ii} = \sum_{\al=1}^{N-1} \frac{ |\ba^i\cdot \bu_\al|^2}{\la - \mu_\al}
$$
where $\bu_\al$ is the normalized
eigenvector of the minor $H^{(i)}$ belonging to $\mu_\al$. 

The interlacing property clearly implies that
 the number of $\lambda_\al^{(k)}$ in $I$ is at least $\cN_I-1$.
For each fixed $k$ the random variables $\{ \xi_\al^{(k)} \; : \; \al=1, 2, \ldots N-1\}$
are almost independent and have expectation value one, thus
the probability of the event
$$
   \Omega_k : =   \Big\{ \sum_{\al\;: \; \lambda_\al^{(k)}\in I} 
  \xi_\al^{(k)} \le \delta (\cN_I-1)\Big\}
$$
is negligible for small $\delta$ \cite[Lemma 4.7]{ESY3}. On the complement
of all $\Omega_k$ we thus have from \eqref{key} that
$$
  \cN_I \le \frac{CN^2\eta^2}{\delta (\cN_I-1)},
$$
from which it follows that $\cN_I \le C N\eta$ with very high probability.
One of the precise results of this type is:

\begin{lemma}\cite[Theorem 4.6]{ESY3}\label{lm:upper}
Assuming the single entry distribution $\nu$ has a Gaussian decay, then for any interval $I$ with 
$|I|\ge (\log N)/N$ we have
$$
   \P \big( \cN_I\ge KN|I|\big) \le Ce^{-c\sqrt{KN|I|}}.
$$
\end{lemma}
We remark that the Gaussian decay condition can be weakened and a somewhat weaker result
holds also for even shorter intervals $|I|\ge 1/N$ (see Theorem 5.1 \cite{ESY3}).

\bigskip

The proof of Theorem \ref{thm:repul} also starts with 
\eqref{id} and \eqref{cNN}.
They imply
\be
 \cN_I
\leq C \eta\sum_{k=1}^N \frac{1}{(a_k^2 + b_k^2)^{1/2}}
\label{out}
\ee
with
$$
 a_k := \eta + \frac{1}{N}\sum_{\al=1}^{N-1}
\frac{\eta \xi_\al^{(k)}}{(\lambda_\al^{(k)}-E)^2 + \eta^2}, \qquad
b_k:= h_{kk} - E - \frac{1}{N}\sum_{\al=1}^{N-1}
\frac{ (\lambda_\al^{(k)}-E)
\xi_\al^{(k)}}{(\lambda_\al^{(k)}-E)^2 + \eta^2}\; ,
$$
i.e., $a_k$ and $b_k$ are the imaginary and real
part, respectively, of the reciprocal of the summands in \eqref{id}.
The proof of  Lemma \ref{lm:upper}  relied only on
the imaginary part, i.e.,  $b_k$
in \eqref{out} was  neglected in the estimate \eqref{key}.
In the proof of Theorem \ref{thm:repul}, however, we
make an essential use of $b_k$ as well. Since typically
$1/N \lesssim |\lambda_\al^{(k)}-E|$, we
note that $a_k^2$ is much smaller
than $b_k^2$ if $\eta\ll 1/N$
and this is the relevant regime for the Wegner estimate and
for the level repulsion.

Assuming a certain smoothness  on the single entry distribution $\rd \nu$,
 the distribution of
the variables $\xi_\al^{(k)}$ will also be smooth even 
if we fix an index $k$ and we condition on the minor $H^{(k)}$, i.e., if we fix
the eigenvalues $\lambda_\al^{(k)}$ and the eigenvectors $\bu_\al^{(k)}$.
 Although the random variables $\xi_\al^{(k)}= N|\ba^{(k)}\cdot \bu_\al^{(k)}|^2$
are not independent for different $\al$'s, they are sufficiently
uncorrelated so that the distribution of $b_k$ inherits some 
smoothness from $\ba^{(k)}$. Sufficient smoothness on the distribution of $b_k$
makes the expectation value $(a_k^2 + b_k^2)^{-p/2}$
finite for any $p>0$. This will give a bound on the $p$-th moment
on $\cN_I$ which will imply \eqref{rep}. 

We present this idea for hermitian matrices and for the simplest
case $n=1$. {F}rom \eqref{out} we have
$$
   \P (\cN_I\ge 1) \le \E \, \cN_I^2 \le C(N\eta)^2 \E \frac{1}{a^2_1 + b^2_1} .
$$
Dropping the superscript $k=1$ and introducing the notation
$$
    d_\al = \frac{N(\lambda_\al -E)}{N^2 (\lambda_\al-E)^2 +\e^2},
  \qquad c_\al =\frac{\e}{N^2 (\lambda_\al-E)^2 +\e^2},
$$
we have
\be
   \P (\cN_I\ge 1) \le C\e^2 \, \E \Bigg[ \Big( \sum_{\al=1}^{N-1}
   c_\al\xi_\al\Big)^2 + \Big( h - E - \sum_{\al=1}^{N-1}
   d_\al\xi_\al\Big)^2 \Bigg]^{-1}.
\label{PN}
\ee
{F}rom one version of the local semicircle law (see Theorem \ref{wegnerlsc} below),
 we know that with a very high probability,
there are several eigenvalues 
$\lambda_\al$ within a distance of $O(1/N)$ of $E$. Choosing four such
eigenvalues, we can guarantee that  for some index $\gamma$ 
\be
  c_\gamma,\, c_{\gamma+1} \ge C\e, \quad  d_{\gamma+2},\, d_{\gamma+3} \ge C
\label{cbound}\ee
for some positive constant $C$. If $\xi_\al$'s were indeed independent 
and distributed according to the square of a complex random variable $z_\al$
with a smooth and decaying density $\rd\mu(z)$ on the complex plane, then
the expectation in \eqref{PN} would be bounded by
\be
   \sup_{E}\int \frac{1}{ \big(c_\gamma |z_\gamma|^2 + c_{\gamma+1} |z_{\gamma+1}|^2\big)^2
  +  \big( E - d_{\gamma+2} |z_{\gamma+2}|^2 -d_{\gamma+3} |z_{\gamma+3}|^2 \big)^2  }
  \prod_{j=0}^3 \rd \mu(z_{\gamma+j}).
\label{exint}\ee
Simple calculation shows that this integral is bounded by $C\e^{-1}$ 
assuming the lower bounds \eqref{cbound}. 
Combining this bound with \eqref{PN}, we obtain \eqref{rep} 
for $n=1$. The proof for the general $n$ goes by induction.
The difference between the hermitian and the symmetric cases
manifests itself in the fact that $\xi_\al$'s are squares
of complex or real variables, respectively. This gives different estimates
for integrals of the type \eqref{exint}, resulting in different
exponents in \eqref{rep}. \qed 

\bigskip

In this proof we used the following version of the local semicircle law:

\begin{theorem}\label{wegnerlsc}[Theorem 3.1 \cite{ESY3}]
Let $H$ be an $N\times N$ hermitian or symmetric Wigner matrix 
with a single entry distribution having a Gaussian decay.
Let $\kappa>0$ and
fix an energy $E\in [-2+\kappa, 2-\kappa]$.
Then there exist positive constants $C$, $c$, depending
only on $\kappa$, and a universal constant $c_1>0$ such that the following hold:
\begin{itemize}
\item[(i)] For any $\delta \le c_1 \kappa$  and $N\ge 2$ we have
\be
 \P (  |m(E+i\eta)-m_{sc}(E+i\eta)|\ge \delta)
\leq C\, e^{-c\delta\sqrt{N\eta}}
\label{sc:new}
\ee
for any $K/(N\sqrt{E}) \leq \eta \le 1$, where $K$ is a large universal constant.

\item[(ii)] Let
$\cN_{\eta^*}(E)= \cN_{I^*}$
denote the number of eigenvalues in
the interval $I^*:=[E-\eta^*/2, E+\eta^*/2]$. Then
for any $\delta \leq c_1 \kappa$ there is a constant $K_\delta$, depending only on
$\delta$, such that
\be
\P \Big\{  \Big| \frac{\cN_{\eta^*}(E)}{N\eta^*}
- \varrho_{sc}(E)\Big|\ge \delta\Big\}\leq
C\, e^{-c\delta^2 \sqrt{N\eta^*}}
\label{ncont}
\ee
holds  for all $\eta^*$ satisfying $K_\delta/N \leq \eta^* \leq 1$
and for all $N\ge 2$.
\end{itemize}
\end{theorem}

\section{Local semicircle law and delocalization}\label{sec:sc}

Each approach that proves bulk universality for generalized Wigner matrices
requires first to analyze the local density of eigenvalues.
The Wigner semicircle law \cite{W} (and its analogue for
Wishart matrices, the Marchenko-Pastur law \cite{MP}) has traditionally been
among the first results established on random matrices.
Typically, however, the empirical density  is shown to
converge weakly on macroscopic scales, i.e.,
on intervals that contain $O(N)$ eigenvalues. 
Based upon our results \cite{ESY1, ESY2, ESY3, ERSY, EYY, EYY2, EYY3}, here we show that
the semicircle law holds on much smaller scales as well. 
In Section \ref{sec:crude} we follow the formalism of \cite{ESY3}, 
while in Section \ref{sec:refined} we use  \cite{EYY, EYY2}.
The former formalism directly aims at the Stieltjes transform,
or the trace of the resolvent;  the
latter formalism is designed to establish the semicircle
law for individual diagonal elements of the resolvent and it
also gives an estimate on the off-diagonal elements.
The strongest result \cite{EYY3} that holds uniformly in the  energy
parameter is presented in Section \ref{sec:best}.
Finally, in Section \ref{sec:deloc}, we indicate how 
to prove delocalization of eigenvectors from
local semicircle law.

\subsection{Resolvent formulas}\label{sec:res}

For definiteness, we present the proof for the hermitian case,
but all formulas below carry over to the other symmetry classes
with obvious modifications.
We first collect a few useful formulas about resolvents. Their proofs 
are elementary results from linear algebra.

\begin{lemma}\label{basicf}
Let $A$, $B$, $C$ be $n\times n$, $m\times n$ and $m\times m$
 matrices. We define $(m+n)\times (m+n)$
 matrix $D$ as 
\be 
D:=\begin{pmatrix}
    A & B^*  \\
  B& C 
\end{pmatrix}
\ee
and  $n\times n$ matrix $\widehat D$ as
\be\label{defhatD}
\widehat D:=A -B^*C^{-1}B.
\ee
Then for any $1\leq i,j\leq n$, we have 
\be
(D^{-1})_{ij}= {(\widehat D^{-1})}_{ij}
\label{Dinv}
\ee
for the 
corresponding matrix elements. \qed
\end{lemma}
\bigskip

Recall that $G_{ij}=G_{ij}(z)$ 
denotes the  matrix element of the resolvent
$$
G_{ij}=\left(\frac1{H-z}\right)_{ij}.
$$
Let $G^{(i)}$ denote the resolvent of $H^{(i)}$, which is the $(N-1)\times (N-1)$ minor
of $H$ obtained by removing the $i$-th row and column. Let $\ba^i =(h_{1i}, h_{2i},
\ldots h_{Ni})^t$ be the
$i$-th column of $H$, sometimes after removing one or more elements.
We always keep the original labelling of the rows and columns, so
there will be no confusion: if $\ba^i$ is multiplied by a matrix whose
$j$-th column and row are removed, then we remove the $j$-th entry from $\ba^i$ as
well.  With similar conventions, we can define
$G^{(ij)}$ etc.  The superscript in parenthesis for resolvents always means ``after removing
the corresponding row and column'', in particular, by independece of matrix
elements, this means that the matrix $G^{(ij)}$, say, is
independent of the $i$-th and $j$-th row and column of $H$. This
helps to decouple dependencies in formulae.

Using Lemma \ref{basicf} for $n=1$, $m=N-1$, we have
\be
     G_{ii} = \frac{1}{ h_{ii} - z- \ba^i\cdot \frac{1}{H^{(i)}-z}\ba^i}=
\frac{1}{ h_{ii} - z- \ba^i\cdot G^{(i)}\ba^i}.
\label{1row}
\ee
where $\ba^i$ is $i$-th column with the $i$-th entry $h_{ii}$ removed.

For the offdiagonal elements, one has to do a two-row expansion. 
In this case, let $\ba^1$ and $\ba^2$ denote the first and the second column
of $H$ after removing the first and second elements, i.e.,
$h_{11}, h_{21}$ from the first column and $h_{12}, h_{22}$ from
the second.
With the notation $D=H-z$, $B=[\ba^1, \ba^2]$ and $C=H^{12}-z$
 in  Lemma \ref{basicf} for $n=2$, $m=N-2$,
we can compute the $\wh D$ matrix which in this case we will call $K^{(12)}$:
\be
\label{Km}
     \wh D =  \begin{pmatrix} h_{11}-z - \ba^1 \cdot G^{(12)}\ba^1 &  
  h_{12} - \ba^1 \cdot G^{(12)}\ba^2 \cr &&\cr
  h_{21} - \ba^2 \cdot G^{(12)}\ba^1 &  h_{22}-z - \ba^2 \cdot G^{(12)}\ba^2
\end{pmatrix} =: \begin{pmatrix}
   K_{11}^{(12)} & K_{12}^{(12)} \cr &&\cr K_{21}^{(12)} & K_{22}^{(12)}
\end{pmatrix}
\ee
where we conveniently introduced
\be
   K^{(12)}_{ij}: = h_{ij} - z\delta_{ij} - \ba^i \cdot G^{(12)}\ba ^j, \qquad i,j=1,2.
\label{Kdef}
\ee
Thus, from Lemma \ref{basicf}, we have, e.g.
\be\label{G11}
   G_{11}= \frac{K_{22}^{(12)}}{K_{22}^{(12)}K_{11}^{(12)} - K_{12}^{(12)}K_{21}^{(12)}}, \qquad 
\ee
and
\be\label{G12}
G_{12}= -\frac {\wH^{(12)}_{12} } { \wH^{(12)}_{22}\wH^{(12)}_{11}
-\wH^{(12)}_{12}\wH^{(12)}_{21} }
 = -G_{22}\frac{\wH^{(12)}_{12}}{\wH^{(12)}_{11}}=-G_{22}G_{11}^{(2)}\wH^{(12)}_{12}. 
\ee
In the last step we used
\be
   G_{11}^{(2)} = \frac{1}{K^{(12)}_{11}}
\label{Gjji}
\ee
which is exactly the one-row expansion \eqref{1row} applied to the $H^{(2)}$ minor of $H$
after removing the second row/column.

\bigskip

There is another set of formulas, that express how to compare resolvents of $H$ and $H^{(1)}$,
for example, for any $i\ne j$.
\be
   G_{ii} = G_{ii}^{(j)} + \frac{G_{ij}G_{ji}}{G_{jj}}.
\label{gii}
\ee
This can be easily checked on a two by two matrix and its inverse:
$$
    M = \begin{pmatrix} a& b\cr c&d\end{pmatrix}, \qquad M^{-1} =\frac{1}{\Delta}
  \begin{pmatrix} d& -c\cr -b &a\end{pmatrix}, \qquad \mbox{with $\quad\Delta = ad-bc$,}
$$
so checking \eqref{gii}, e.g. for $i=1$, $j=2$ boils down to the identity
$$
   \frac{d}{\Delta} = \frac{1}{a} + \frac{\frac{c}{\Delta}\frac{b}{\Delta}}{\frac{a}{\Delta}}.
$$
For larger matrices, one just uses \eqref{Dinv}. Note that
in formulas \eqref{G11}, \eqref{G12} and \eqref{Gjji} we already expressed
all resolvents appearing in \eqref{gii} in terms of the matrix
elements of the two by two $K^{(12)}$ matrix \eqref{Km} which can play the role of $M$ above.
Similarly one has for any three different indices $i,j,k$ that
\be
    G_{ij} = G_{ij}^{(k)} + \frac{G_{ik}G_{kj}}{G_{kk}}.
\label{gij}
\ee
This identity can be checked on 3 by 3 matrices and then proved by induction
in the general case.

\subsection{Semicircle law via resolvents:
 Sketch of a crude method}\label{sec:crude}

In this section we sketch the proof of
\begin{theorem}\label{lm:crude} Let $z=E+i\eta$, $1/N\ll \eta \ll 1$ and
$\kappa:= \big| |E|-2\big|$. Let $H$ be a Wigner matrix and
$G(z) = (H-z)^{-1}$ its resolvent and set  $m(z):=\frac{1}{N}\tr G(z)$. We assume
that the single entry
distribution $\nu$ has Gaussian decay ($\ttau=2$ in \eqref{subexp}).
  Then we have
the following approximation 
\be
    |m(z)-m_{sc}(z)|\le \min\Big\{ \frac{(\log N)^C}{\sqrt{N\eta \kappa}},
 \; \frac{(\log N)^C}{(N\eta)^{1/4}}\Big\}
\label{eq:crude}
\ee
with a very high probability.
\end{theorem}
We proved the local semicircle law in this form 
in Proposition 8.1 of \cite{ESYY} for sample covariance matrices
(replacing semicircle with the Marchenko-Pastur distribution),
but the same (or even easier) proof applies to Wigner matrices.
The original proof was presented with a Gaussian decay condition,
but it can easily be relaxed to subexponential decay, this affects only the estimate of 
the probability that the event \eqref{eq:crude} is violated.
[For technical experts: in our previous papers, up to \cite{ESYY},
we typically used the Hanson-Wright theorem \cite{HW} 
to estimate large deviation probabilities of quadratic forms.
This gives a very good control for the tail, but requires Gaussian
decay. In our more recent papers we use Lemma \ref{generalHWT}
based upon martingale inequalities, which requires only subexponential
decay, and in fact can be relaxed to polynomial decay as well,
but the tail probability estimate is weaker.]

\medskip

For the proof, we
start with the identity \eqref{1row} and express $G_{ii}$ 
\be
  G_{ii} =\frac{1}{h_{ii}- z - \E_i \ba^i \cdot G^{(i)} \ba^i - Z_i},
\label{idg}
\ee
where we split
$$
  \ba^i \cdot G^{(i)} \ba^i = \E_i \ba^i \cdot G^{(i)} \ba^i + Z_i, \qquad 
  Z_i: =  \ba^i \cdot G^{(i)} \ba^i - \E_i \ba^i \cdot G^{(i)} \ba^i
$$
into its expectation and fluctuation, where $\E_i$ denotes
the expectation with respect to the variables in the $i$-th column/row.
In particular, $G^{(i)}$ is independent of $\ba^i$, so we need to compute
expectations and fluctuations of quadratic functions.

The expectation is easy
$$
   \E_i  \ba^i \cdot G^{(i)} \ba^i = \E_i \sum_{k,l\ne i} \ov{\ba^i_k} G^{(i)}_{kl} \ba^i_l
   = \sum_{k,l\ne i} \E_i \bar h_{ik} G^{(i)}_{kl} h_{il} = \frac{1}{N}\sum_{k\ne i} G^{(i)}_{kk},
$$
where in the last step we used that different matrix elements are independent, i.e.
$\E_i \bar h_{ik} h_{il} = \frac{1}{N} \delta_{kl}$. The summations always
run over all indices from 1 to $N$, apart from those that are explicitly excluded.

Similarly to 
$$
m(z)= \frac{1}{N}\tr G(z) = \frac1N\sum_{k=1}^NG_{kk}(z).
$$
we define
$$
m^{(i)}(z):= \frac{1}{N-1}\tr G^{(i)}(z) = \frac{1}{N-1}\sum_{k\ne i} G^{(i)}_{kk}(z),
$$
and we have the following lemma to compare the trace of $G$ and $G^{(i)}$:

\begin{lemma} For any $1\le i\le N$,
\be
   |m(z)-m^{(i)}(z)|\le \frac{C}{N\eta}, \qquad \eta = \im z>0.
\label{mm}\ee
\end{lemma}

{\it Proof.}  Let
$$
    F(x): = \frac{1}{N}\# \{ \lambda_j\le x\}, \qquad 
    F^{(i)}(x): = \frac{1}{N-1}\# \{ \mu_j\le x\}
$$
denote the normalized counting functions of the eigenvalues. The interlacing
property of the eigenvalues of $H$ and $H^{(i)}$ 
(see \eqref{interlace}) in terms of these functions means that
$$
  \sup_x |NF(x) - (N-1)F^{(i)}(x)|\le 1.
$$
Then, after integrating by parts,
\begin{align}
    \Big| m(z) - \Big(1-\frac{1}{N}\Big) m^{(i)}(z)\Big| = &
  \Big| \int \frac{\rd F(x)}{x-z} - \Big(1-\frac{1}{N}\Big) \int \frac{\rd F^{(i)}(x)}{x-z}
  \Big| \non\\
   = & \frac{1}{N}\Big| \int \frac{NF(x) - (N-1)F^{(i)}(x)}{(x-z)^2}\rd x\Big| \non\\
  \le & \frac{1}{N} \int \frac{\rd x}{|x-z|^2} \le \frac{C}{N\eta},
\end{align}
and this, together with the trivial bound $|m^{(i)}|\le \eta^{-1}$,
proves \eqref{mm}. \qed

\bigskip

Returning to \eqref{idg}, we have thus
\be
  G_{ii} =\frac{1}{- z - m(z) + \Omega_i},
\label{id2}
\ee
where 
\be
   \Omega_i: = h_{ii}-Z_i + O \Big(\frac{1}{N\eta}\Big).
\label{Omdef}
\ee
Summing up \eqref{id2}, we get
\be
  m = \frac{1}{N}\sum_{i=1}^N \frac{1}{- z - m(z) + \Omega_i}.
\label{sumsum}
\ee
Suppose that $\Omega: =\max_i |\Omega_i|$ is small and $|z+m|\ge C>0$, thus an expansion is
possible, i.e.,
$$
   \frac{1}{- z - m(z) + \Omega_i} =  \frac{1}{- z - m(z)} + O( \Omega).
$$
Then we have the following {\it self-consistent equation} for $m$
\be
   m + \frac{1}{z+m} = O(\Omega).
\label{selfm}
\ee
We recall that the Stieltjes transform of the semicircle law
$$
   m_{sc}(z) =\int_\R \frac{\varrho_{sc}(x)\rd x}{x-z}
$$
can  be characterized as the only solution to the quadratic equation
\be
   m_{sc}(z) + \frac{1}{z+ m_{sc}(z) }= 0
\label{selfcons11}
\ee
with $\im m_{sc}(z)>0$ for $\im z>0$. We can thus use the
stability of the equation \eqref{selfcons11} to identify
$m$, the solution to \eqref{selfm}. The stability deterioriates
near the spectral edges $z\sim \pm 2$ and we have 
the following precise result that can be proved by elementary calculus:

\begin{lemma}\cite[Lemma 8.4]{ESYY} \label{lm:scsc}
Fix $z=E+i\eta$, $\eta>0$, and set $\kappa:= \big| |E|-2\big|$.
Suppose that $m=m(z)$ has a positive imaginary part and
$$
   \Big| m + \frac{1}{z+m} \Big|\le \delta.
$$
Then
$$
   |m-m_{sc}(z)|\le \frac{C\delta}{\sqrt{\kappa+\delta}}.
$$
\qed
\end{lemma}
Applying this result, we obtain  
\be
   |m(z)-m_{sc}(z)|\le \frac{C\Omega}{\sqrt{\kappa+\Omega}}.
\label{mmmom}
\ee

We now give a rough bound on the size of $\Omega$.
Clearly $|h_{ii}|\lesssim N^{-1/2}$. If the single entry distribution
has subexponential decay, then we can guarantee that {\it all} diagonal
elements simultaneously satisfy essentially this bound with a very high probability.
Recall that \eqref{subexp}
implies
$$
  \P \Big( |h_{ii}|\ge M^\al N^{-1/2} \Big)\le C e^{-M}
$$
for each $i$. Choosing $M=(\log N)^{1+\e}$, we have
\be
  \P \Big( \exists i\; : \; |h_{ii}|\ge (\log N)^{(1+\e)\al}
   N^{-1/2} \Big)\le C N e^{-(\log N)^{1+\e}} \le Ce^{- (\log N)^{1+\e}},
\label{hest}
\ee
which is faster than any polynomial decay.

 To estimate $Z_i$, we 
compute its second moment
\be
 \E |Z_i|^2 = \sum_{k,l\ne i}\sum_{k',l'\ne i} \E_i \Bigg( \Big[ \ov h_{ik} G^{(i)}_{kl} h_{il} - 
  \E_i \ov h_{ik} G^{(i)}_{kl} h_{il}\Big] \Big[  h_{ik'} \ov G^{(i)}_{k'l'} \ov h_{il'} - 
  \E_i  h_{ik'} \ov G^{(i)}_{k'l'} \ov h_{il'}\Big]\Bigg).
\label{Zib}
\ee
Since $\E h=0$, the non-zero contributions to this sum come from
index combinations when all $h$ and $\ov h$ are paired.
For pedagogical simplicity, assume that $\E h^2=0$, this can be achieved, for example, if
the distribution of the real and imaginary parts are the same. Then $h$ factors in the above
expression have to be paired in such a way that $h_{ik}=h_{ik'}$ and $h_{il}=h_{il'}$,
i.e., $k=k'$, $l=l'$. Note that pairing $h_{ik}=h_{il}$  would give zero because
the expectation is subtracted. The result is
\be
   \E |Z_i|^2 = \frac{1}{N^2}\sum_{k,l\ne i} |G^{(i)}_{kl}|^2  + \frac{m_4-1}{N^2} \sum_{k\ne i}
  |G_{kk}^{(i)}|^2,
\label{ZZ}
\ee
where $m_4 =\E |\sqrt{N}h|^4$ is the fourth moment of the single entry
distribution. 
The first term can be computed
\be
    \frac{1}{N^2}\sum_{k,l\ne i} |G^{(i)}_{kl}|^2 =  \frac{1}{N^2}\sum_{k\ne i} 
 (|G^{(i)}|^2)_{kk} 
= \frac{1}{N\eta} \frac{1}{N}
  \sum_k \im G^{(i)}_{kk}  =   \frac{1}{N\eta} \im m^{(i)},
\label{egy}
\ee
where $|G|^2 = GG^*$ and we used the identity
$$
   |G|^2 = \frac{1}{|H-E|^2 + \eta^2} = \frac{1}{\eta} \im G.
$$
To estimate this quantity, we need
an upper bound on the local density on scale $\eta\gg 1/N$.
For any interval $I\subset \R$, let
$\cN_I: = \#\{ \lambda_j \in I\}$
denote the number of eigenvalues in $I$.
Lemma \ref{lm:upper}
from Section \ref{sec:wegner} shows that $\cN_I \lesssim N|I|$ with
a very high probability.
Using this lemma for the matrix $H^{(i)}$ with eigenvalues $\mu_1, \mu_2, \ldots, \mu_{N-1}$,
 we have
\be
    \im m^{(i)} = \frac{1}{N-1}\sum_\al \frac{\eta}{(E-\mu_\al)^2+\eta^2}
  \lesssim \int \frac{\eta}{(x-E)^2+\eta^2}\varrho_{sc}(x) \rd x  \lesssim O(1).
\label{loccalc}
\ee
Recall that the sign $\lesssim $ here means ``up to $\log N$ factors''.

The second term in \eqref{ZZ}  we can estimate by using the trivial bound
$|G_{kk}^{(i)}|\le \eta^{-1}$ and thus
\be
   \frac{1}{N^2}\sum_{k\ne i} |G_{kk}^{(i)}|^2 \le  \frac{1}{N^2\eta} \sum_{k\ne i} |G_{kk}^{(i)}| \le 
  \frac{1}{N^2\eta}\sum_{k\ne i}\sum_{\al=1}^{N-1} \frac{|\bu_\al(k)|^2}{|\lambda_\al-z|}
  \le \frac{1}{N\eta} \frac{1}{N} \sum_{\al=1}^{N-1} \frac{1}{|\lambda_\al-z|} 
\lesssim \frac{1}{N\eta},
\label{ketto}
\ee
where $\bu_\al$ is the (normalized) eigenvector to $\mu_\al$ and in the last step we used 
an estimate similar to \eqref{loccalc}.

\medskip

The estimates \eqref{egy} and \eqref{ketto}  confirm that the size of $Z_i$ is roughly
\be
  |Z_i|\lesssim \frac{1}{\sqrt{N\eta}}
\label{Zest}
\ee
at least in second moment sense. One can compute higher moments or
use even stronger concentration results (if, for example, logarithmic
Sobolev inequality is available), to strengthen \eqref{Zest} so that it holds
in the sense of probability.

Altogether \eqref{Omdef}, \eqref{hest} and \eqref{Zest} give
$$
  \Omega \lesssim \frac{C}{\sqrt{N}} + \frac{C}{\sqrt{N\eta}} + \frac{C}{N\eta}.
$$
Since we are interested in $1/N\ll \eta \le 1$, we get that
$\Omega\lesssim (N\eta)^{-1/2}$. Combining this with the stability bound, \eqref{mmmom}, we
have
$$
    |m(z)-m_{sc}(z)|\lesssim \min\Big\{ \frac{1}{\sqrt{N\eta \kappa}}, \; 
\frac{1}{(N\eta)^{1/4}}\Big\},
$$
which proves Lemma \ref{lm:crude}. \qed

\bigskip

The remarkable feature is that the method works down to scales $\eta\sim 1/N$.
The factor $\kappa$ expresses the fact that the estimate deterioriates
near the edge. 
The exponents are not optimal, $m-m_{sc}$ can be compared
with a precision of order $(N\eta)^{-1}$. This will be presented
in the next section.
 The gain will come from the fact
that the  main error term $Z_i$ in \eqref{Omdef} is fluctuating
and it is possible to show \cite{EYY} that its contributions to $m-m_{sc}$ cancel
to leading order and it is smaller than the 
size of $Z_i$ predicted by the variance calculation
(this effect was first exploited in Theorem 4.1 of \cite{ERSY}
and substantially improved in \cite{EYY} and \cite{EYY2}).

We emphasize that the presentation  was very sketchy, many
technical issues were neglected.

\subsection{Semicircle via resolvents: refined method}\label{sec:refined}

\subsubsection{Statement of the theorem and consequences}

In this section we present a more refined method that can 
estimate matrix elements of the resolvent
and it also applies to universal Wigner matrices (Definition~\ref{def:gen}).
 This is also
a key ingredient for the improved precision
on the semicircle law both in terms of $(N\eta)$-power
and edge behavior. The main ingredient is 
to analyze a self-consistent equation for 
the vector of the diagonal elements
of the resolvent, $(G_{11}, G_{22}, \ldots , G_{NN})$,
instead of their sum which has led to \eqref{sumsum}.
Again, for definiteness, we formulate the result for
generalized hermitian Wigner matrices 
only; the extension to symmetric matrices is straightforward.

A key quantity will be the matrix of variances, $\Sigma$, introduced
in \eqref{def:Sigmamatrix}. Recall that 
 $\Sigma$ is symmetric,  doubly stochastic by \eqref{sum}, and in particular it  satisfies
$-1\leq \Sigma \leq 1$. 
Let the spectrum of $\Sigma$ be supported in 
\be\label{de-de+}
\mbox{Spec}(\Sigma)\subset [-1+\delta_-, 1-\delta_+]\cup\{1\}
\ee  
with some nonnegative constants $\delta_\pm$. 
We will always have the following spectral assumption
\be\label{speccond}
\mbox{\it  1 is a simple eigenvalue of $\Sigma$ and
$\delta_-$ is a positive constant,  independent of $N$.}
\ee
For Wigner matrices, all entries of $\Sigma$ are identical and
 $\delta_\pm=1$.  It is easy to prove (see Lemma A.1 of \cite{EYY})
that \eqref{speccond}
holds for {\it random band matrices}, see \eqref{BM} for the definition,
with $\delta_->0$, depending only on $f$.
For {\it generalized Wigner matrices}, i.e., Wigner
matrices with comparable variances, see  \eqref{VV} for the
definition, it is easy to check that
$$
  \delta_\pm\ge C_{inf}>0.
$$
The fact that $\delta_+>0$ for generalized Wigner matrices allows
better control in terms of the edge behavior of the estimates.
This is the main reason why the statement below is
different for universal Wigner matrices (see \eqref{def:gen})
and for generalized Wigner matrices, \eqref{VV}.

The precision of the local semicircle law depends on three factors.
The first factor is the resolution (scale on which the semicircle holds),
 this is given by
the imaginary part $\eta=\im z$ of the spectral parameter $z$ in
the Stieltjes transform. The second factor is the distance to the
edge, measured by $\kappa = \big| |E|-2\big|$.
The last factor is the size of the typical matrix elements, measured
by the quantity
\be
   M: = \frac{1}{\max_{ij}\sigma_{ij}^2}
\label{defM}
\ee
called the {\it spread} of the matrix.
For example, for Wigner matrices $M=N$, for generalized Wigner matrices,
\eqref{def:gen}, $M\sim N$ and for random band matrices \eqref{BM}
we have $M\sim W$.

\begin{theorem}[Local semicircle law for universal Wigner matrices]\cite[Theorem 2.1]{EYY2} \label{lsc}{}
Let $H$ be a hermitian  $N\times N$ random matrix
with $\E\, h_{ij}=0$, $1\leq i,j\leq N$,  and assume that the variances $\sigma_{ij}^2$ 
satisfy \eqref{sum} and \eqref{speccond}.
 Suppose that the distributions of the matrix elements have a uniformly 
  subexponential decay
in the sense that  there exist constants $C$, $\ttau>0$, independent 
of $N$, such that for  any $x> 0$  and  for each $(i,j)$  we have 
\be\label{subexpuj}
\P(|h_{ij}|\geq x |\sigma_{ij}|)\leq C \exp\big( - x^\ttau\big).
\ee
We consider universal Wigner matrices and its  special class, the generalized Wigner matrices 
in parallel. The parameter $A$ will distinguish between the two cases;
we set $A=2$ for universal Wigner matrices, and $A=1$ for generalized Wigner
matrices, where the results will be stronger.

Define the following domain in $\C$ 
\be\label{fakerelkaeta}
    D : = \Big\{ z=E+i\eta\in \C\; : \;  |E|\le 5 , \;  0<\eta<10, \; \; 
  \sqrt{M\eta}\ge (\log N)^{C_1} (\kappa +\eta)^{\frac{1}{4}-A}
\Big\}
\ee
where  $\kappa : = \big| \, |E|-2 \big|$.
Then there exist constants $C_1$, $C_2$, and $c>0$, 
depending only on $\ttau$ and $\delta_-$ in \eqref{speccond}, such that 
for any $\e>0$ and $K>0$  the Stieltjes transform of the empirical 
eigenvalue distribution of  $H $  satisfies 
\be\label{mainlsresult}
\P\left(\bigcup_{z\in D}\Bigg\{ |m(z)-m_{sc}(z)|\geq 
 \frac{N^\e}{{M\eta}\,(\kappa+\eta)^A}\Bigg\}\right)\leq \frac{C(\e, K)}{N^K}
\ee
for sufficiently large $N$.
The diagonal matrix elements of
the Green function $G_{ii}(z) = (H-z)^{-1}(i,i)$ satisfy 
\be\label{Gii}
\P\left(\bigcup_{z\in D}\Bigg\{ \max_i | G_{ii}(z)-m_{sc}(z)|\geq 
 \frac{ (\log N)^{C_2}} {\sqrt{M\eta}} \,
  (\kappa+\eta)^{\frac{1}{4}-\frac{A}{2}}\Bigg\}\right)\leq CN^{-c(\log \log  N)},
\ee
and for the off-diagonal elements we have
\be\label{Gij}
\P\left( \bigcup_{z\in D} \Bigg\{ \max_{i\ne j} | G_{ij}(z)|\geq
 \frac{ (\log N)^{C_2}} {\sqrt{M\eta}} \,  (\kappa+\eta)^{\frac{1}{4}} \Bigg\}
 \right)\leq CN^{-c(\log \log  N)}
\ee
for any sufficiently large $N$.
\end{theorem}

{\it Remark 1.} These estimates are optimal in the power of $M\eta$,
but they are not optimal as far as the edge behavior (power of $\kappa$)
and the control on the probability is
concerned. Under stronger decay assumptions on the single entry
distributions it is possible to get subexponential bounds
on the probability, e.g. Theorem \ref{wegnerlsc} (see  Theorem 3.1 \cite{ESY3}).
On the other hand,  the subexponential decay condition \eqref{subexpuj} can be
 easily weakened  if we are not aiming 
at error estimates faster than any power law of $N$.

{\it Remark 2.} Concerning the edge behavior, 
we remark that our first two papers \cite{ESY1, ESY2} we simply assumed $\kappa\ge \kappa_0$
for some positive constant $\kappa_0$.
The edge behavior was effectively treated first in \cite{ESY3}
and substantially improved in Theorem 4.1  of \cite{ERSY}, 
but the bounds were not optimal. The best result 
for universal Wigner matrices is Theorem \ref{lsc}.
For generalized Wigner matrices, Theorem \ref{45-1} (proved in \cite{EYY3})
gives an optimal  estimate uniform in $\kappa$.

\medskip

The local semicircle estimates imply that the empirical counting function of
the eigenvalues is close to the semicircle counting function (Theorem~\ref{prop:count})
and that the location of the eigenvalues are close to their classical 
location in mean square deviation sense (Theorem \ref{prop:lambdagamma}).
This latter result will be used to verify \eqref{El}, or more
precisely Assumption III \eqref{assum2} later, that will be 
the key input to our hydrodynamical approach for universality.

\medskip

To formulate these statements precisely,
let $\la_1\le \la_2 \le \ldots \le \la_N$ be the ordered
eigenvalues of a universal Wigner matrix.
We define the {\it normalized empirical counting function}  by
\be
 {\frak n}(E):= \frac{1}{N}\# \{ \lambda_j\le E\}
\label{deffn}
\ee
and the {\it  averaged counting function} by
\be\label{defnlambda}
n(E)=\frac1N\E \#[{\lambda_j\leq E}].
\ee
Finally, let
\be
 n_{sc}(E) :  = \int_{-\infty}^E \varrho_{sc}(x)\rd x
\label{nsc}
\ee
be the distribution function
of the semicircle law which is very close to
the counting function of the $\gamma$'s,  $\frac{1}{N}\#[\ga_j\le E] \approx n_{sc}(E)$.
Recall that $\gamma_j$'s are the classical location 
of the eigenvalues, determined by the semicircle law, see \eqref{def:gamma}.

With these notations, we have the following theorems:

\begin{theorem}\label{prop:count}\cite[Theorem 6.3]{EYY2}
Let $A=2$  for universal Wigner matrices, satisfying \eqref{sum}, \eqref{defM} and 
 \eqref{subexpuj} with  
$M\geq (\log N)^{24+6\al}$.
 For generalized Wigner matrices, satisfying \eqref{sum}, \eqref{defM}, \eqref{VV} 
and \eqref{subexpuj},
we set $A=2$ and recall $M=N$ in this case.
Then for any $\e>0$ and $K\ge 1$
there exists a constant $C(\e,K)$ such that
$$
   \P\Big\{ \sup_{|E|\le 3} \big| {\frak n} (E)-n_{sc}(E)\big| [\kappa_E]^A \le
 \frac{CN^{\e}}{M} \Big\}\ge 1- \frac{C(\e, K)}{N^K},
$$
where the $\fn(E)$ and $n_{sc}(E)$ were defined in \eqref{deffn} and \eqref{nsc}
and $\kappa_E=\big| |E|-2\big|$.
\end{theorem}

\begin{theorem}\label{prop:lambdagamma} \cite[Theorem 7.1]{EYY2}
Let $H$ be a generalized Wigner matrix  with 
subexponential decay, i.e., assume that \eqref{sum}, \eqref{defM}, \eqref{VV} and 
\eqref{subexpuj} hold.
 Let $\lambda_j$ denote the eigenvalues of $H$ and 
$\gamma_j$ be their semiclassical location, defined by \eqref{def:gamma}.
Then for any  $\e<1/7$ and for any $K>1$ there exists a constant $C_K$ such that
\be
    \P \Big\{ \sum_{j=1}^N |\lambda_j-\gamma_j|^2 \le N^{-\e}\Big\}
  \ge 1-\frac{C_K}{N^K}.
\label{lambdaminusgamma}
\ee
and 
\be
    \sum_{j=1}^N \E |\lambda_j-\gamma_j|^2 \le CN^{-\e}.
\label{lambdaminusgammaE}
\ee
\end{theorem}

These theorems are consequences
of the local semicircle law, Theorem \ref{lsc}. We will not give the detailed
proof, but we mention a  useful formula that allows one to translate Stieltjes
transforms to densities. In this context this formula first appeared in \cite{ERSY}.

\begin{lemma} \cite[Helffer-Sj\"ostrand formula]{Dav}
 Let $f$ be a real valued $C^1$ function
on $\R$.
Let $\chi(y)$ be a smooth cutoff function with 
 support in $[-1,1]$, with $\chi(y)=1$ for  $|y|\leq 1/2$
and with bounded
derivatives.
Let
$$ 
  \wt f(x+iy):= (f(x) + iyf'(x))\chi(y),
$$
then
\be
   f(\lambda) =\frac{1}{2\pi}\int_{\bR^2}
\frac{\partial_{\bar z} \wt f(x+iy)}{\lambda-x-iy} \rd x  \rd y 
=\frac{1}{2\pi}\int_{\bR^2}
\frac{iy f''(x)\chi(y) +i(f(x) + iyf'(x))\chi'(y) }{\lambda-x-iy} \rd x 
\rd y.
\label{1*}
\ee
\end{lemma}

We will apply this lemma in the following form.
Let $\varrho \in L^1(\R)$ be a real function and let $m(z)$ be its Stieltjes transform
$$
   m(z): = \int_\R \frac{\varrho(\la)\rd \la}{\la-z}.
$$
Since $f$ is real, we have
\begin{equation}\label{eq:1+2}
\begin{split} 
\Big| \int_\R f(\la) \varrho(\rd \la) \Big| = & \Big|
\text{Re } \int_\R f(\la) \varrho(\rd \la) \Big| \\
 = & \Big|\frac{1}{2\pi} \text{Re } \int_{\R^2}
 \partial_{\bar z} \wt f(x+iy) m(x+iy) \rd x \rd y \Big| \\
 \leq &
\left| \frac{1}{2\pi}\int_{\bR^2} \, yf'' (x) \,\chi (y) \, 
\text{Im} \, m(x+iy) \rd x  \rd y \right|  \\ 
&+ C \int_{\R^2}  \left(|f(x)| + |y| |f'(x)| \right) |\chi'(y)| 
\left|  m (x+iy) \right| \rd x  \rd y \, .
\end{split}\end{equation}

In order to get counting function, we will choose
$$
   f(\la)=f_{E,\eta}(\la),
$$
where $f_{E,\eta}$ is the characteristic function of the semi-axis
$(-\infty, E)$, smoothed out on a scale $\eta$ (i.e.,
$f_{E,\eta}(\la) \equiv 1$ for $\la\le E-\eta$, $f_{E,\eta}(\la) \equiv 0$
for $\la\ge E+\eta$, and $|f'|\le C\eta^{-1}$, $|f''|\le C\eta^{-2}$
in the interval $[E-\eta, E+\eta]$).

The second term in \eqref{eq:1+2} is typically harmless, since $\chi'$ is supported
at $|y|\ge 1/2$, i.e. this term requires information on the
Stieltjes transform far away from the real axis.
In the first term,  we have only the imaginary part of
the Stieltjes transform and it is easy to see that
$$
  y \im m(x+iy) \le \int |\varrho(x)|\rd x.
$$
One can perform and integration by parts bringing the second derivative
of $f$ down to the first derivative, which, after integration is 
under control even if $\eta$ is very small.

Using these ideas for the  measure $\varrho$ being
the difference of the empirial density and $\varrho_{sc}$,
one can control $\int f(\la)\varrho(\rd\la)$,
i.e. essentially the difference of the counting functions,
in terms of the size of $m-m_{sc}$.
 For the details, see
\cite{ERSY} and \cite{EYY2}.

\subsubsection{Sketch of the proof of the semicircle law
for matrix elements}\label{sec:sk:elem}

For pedagogical reasons we will neglect the edge problem
for the presentation, i.e., we will assume that $E=\re z$
always satisfies $\kappa=\big||E|-2\big|\ge \kappa_0$
for some fixed $\kappa_0>0$ and we do not follow
the dependence of the constants on $\kappa_0$.
We will thus prove the following partial version of Theorem \ref{lsc}:

\begin{theorem}\label{partial} Assume the conditions of Theorem \ref{lsc}.
For some $\kappa_0>0$,
define the following domain in $\C$ 
\be\label{fakerelkaetapart}
    D := D_{\kappa_0} = \Big\{ z=E+i\eta\in \C\; : \;  |z|\le Q, \;  \eta>0, \; \; \big| |E|-2\big|\ge
  \kappa_0, \;\; 
  \sqrt{M\eta}\ge (\log N)^{C_1},
\Big\}
\ee
where 
$C_1$ and $Q$ are sufficiently large.
Then there exist constants $C_1$, $C_2$, $C$ and $c>0$, 
depending only on $\ttau$, $\kappa_0$ and $\delta_-$ in \eqref{speccond}, such that 
 the diagonal matrix elements of
the Green function $G_{ii}(z)$ satisfy  
\be\label{Giipart}
\P\left(\bigcup_{z\in D} \Bigg\{ \max_i | G_{ii}(z)-m_{sc}(z)|\geq 
 \frac{ (\log N)^{C_2}} {\sqrt{M\eta}} \,\Bigg\}
  \right)\leq CN^{-c(\log \log  N)},
\ee
and for the off-diagonal elements we have
\be\label{Gijpart}
\P\left( \bigcup_{z\in D}\Bigg\{ \max_{i\ne j} | G_{ij}(z)|\geq
 \frac{ (\log N)^{C_2}} {\sqrt{M\eta}} \Bigg\} 
 \right)\leq CN^{-c(\log \log  N)}
\ee
for any sufficiently large $N$.
\end{theorem}

We start with  a system of self-consistent  equations for the
diagonal matrix elements of the resolvent.  The following lemma
is a simple combination of the resolvent identities from Section \ref{sec:res}.

\begin{lemma} The diagonal resolvent matrix elements $G_{ii}= (H-z)^{-1}(i,i)$ satisfy
the following system of self-consistent equations
\be\label{mainseeq}
G_{ii} = \frac{1}{-z- \sum_{j}\sigma^2_{ij}G_{jj}+\Upsilon_i},
\ee
where
\be\label{seeqerror}
\Upsilon_j:=
 A_i +h_{ii}-Z_i.
\ee
with
\be
   A_i: =\sigma^2_{ii}G_{ii}+\sum_{j\neq i}\sigma^2_{ij}\frac{G_{ij}G_{ji}}{G_{ii}}
\label{defA}
\ee
\be
  Z_i:=Z_{ii}^{(i)}-\E_i Z_{ii}^{(i)}, \qquad Z_{ii}^{(i)}:= \ba^i \cdot G^{(i)}\ba^i
=  \sum_{k,l\ne i} 
 \ov{\ba^i_k} G^{(i)}_{kl}\ba^i_l.
\label{Zdef}
\ee
\end{lemma}

{\it Proof.}
Introduce
$$
  Z^{(ij)}_{ij}:= \ba^i \cdot G^{(ij)}\ba ^j, \qquad 
 Z^{(i)}_{ii} : = \ba^i \cdot G^{(i)}\ba^i,
$$
and recall from \eqref{Kdef}
$$
  K^{(ij)}_{ij}  =  h_{ij} - z\delta_{ij}
  - Z^{(ij)}_{ij}, \quad
  K^{(i)}_{ii} = h_{ii} -z -  Z^{(i)}_{ii}.
$$
We can write  $G_{ii}$  as follows \eqref{Gjji}
\be
   G_{ii} = (\wH^{(i)}_{ii})^{-1}=\frac{1}{\E_{\ba^i}\wH^{(i)}_{ii}+\wH^{(i)}_{ii}-
\E_{\ba^i}\wH^{(i)}_{ii}},
\label{Gid}
\ee
where $\E_{\ba^i} = \E_i$ denotes the expectation with respect to 
the elements in the $i$-th column of the matrix $H$.

Using the fact that $G^{(i)}=(H^{(i)}-z)^{-1}$ is  independent of $\ba^i$ and $\E_{\ba^i}
 \overline {\ba^i_k}\ba^i_l=\delta_{kl}\sigma^2_{ik}$, we obtain
$$
\E_{\ba^i}\wH^{(i)}_{ii}=-z- \sum_{j\neq i}\sigma^2_{ij}G^{(i)}_{jj}
$$
and
\be
  \wH^{(i)}_{ii}-\E_{\ba^i}\wH^{(i)}_{ii}
  = h_{ii} - Z_i.  \qquad
\label{hZ}
\ee
Use
$$
   G_{kl} = G_{kl}^{(i)} + \frac{G_{ki}G_{il}}{G_{ii}}
$$
from \eqref{gii} and the notation from \eqref{defA}   to express
$$
\E_{\ba^i}\wH^{(i)}_{ii}=-z- \sum_{j\neq i}\sigma^2_{ij}G^{(i)}_{jj}
  = -z- \sum_{j\neq i}\sigma^2_{ij}G_{jj} +
 \sum_{j\neq i}\sigma^2_{ij}\frac{G_{ji}G_{ij}}{G_{ii}}= -z- \sum_{j}\sigma^2_{ij}G_{jj} +A_i.
$$
Combining this with \eqref{hZ},
from \eqref{Gid} we eventually obtain
\eqref{mainseeq}. \qed

\bigskip
Introduce the notations
$$
  v_i := G_{ii}-m_{sc}, \qquad m:=\frac{1}{N}\sum_i G_{ii}, \qquad \barv:=\frac{1}{N}
  \sum_i v_i = \frac{1}{N}\sum_i (G_{ii}-m_{sc}).
$$
We will estimate the following key quantities
\be\label{defLambda}
  \Lambda_d:=\max_k |v_k| = \max_k |G_{kk}-m_{sc}|, \qquad
 \Lambda_o:=\max_{k\ne \ell} |G_{k\ell}|,
\ee
where the subscripts refer to ``diagonal'' and ``offdiagonal'' matrix elements.
All the quantities defined so far depend on the spectral parameter $z=E+i\eta$, but
we will mostly omit this fact from the notation. The real part $E$ will always be kept
fixed. For the imaginary part we will use a continuity argument at the end of the proof
and then the dependence of $\Lambda_{d,o}$ on $z$ will be indicated.

\bigskip

Both quantities $\Lambda_d$ and $\Lambda_o$ will be typically small for $z\in D$; 
eventually we will prove that their size is less than $(M\eta)^{-1/2}$,
modulo logarithmic corrections. We thus define the exceptional event
$$
   \Omega_\Lambda =\Omega_\Lambda(z):= \Big\{ \Lambda_d(z) + \Lambda_o(z)
\ge (\log N)^{-C} \Big\}
$$
with some $C$ (in this presentation, we will not care about matching all exponents).
We will always work in $\Omega_\Lambda^c$, and, in particular,  we will have
$$
   \Lambda_d(z) + \Lambda_o(z)
\ll 1.
$$
 It is easy to check from the explicit formula on $m_{sc}$
that
\be
   c\le |m_{sc}(z)|\le C, \qquad z\in D.
\label{msclowup}
\ee
with some positive constants, so from $G_{ii} = m_{sc}(z) + O(\Lambda_d)$ we have
\be
 c\le |G_{ii}(z)|\le C, \qquad z\in D.
\label{lowerbound}
\ee

Recalling \eqref{gij}
$$
   G_{kl}^{(i)}= G_{kl} - \frac{G_{ki}G_{il}}{G_{ii}}, \qquad  i\ne l,k,
$$
together with \eqref{lowerbound}, it
implies that for any $i$ and $z\in D$
$$
    \max_{k\ne l} |G_{kl}^{(i)}|\le \Lambda_o + C \Lambda_o^2 \le C\Lambda_o
 \qquad \mbox{in $\Omega_\Lambda^c$},
$$
\be
   c\le |G_{kk}^{(i)}|\le C,  \qquad \mbox{for all 
$k\ne i$ and in $\Omega_\Lambda^c$}
\label{Gkk}
\ee
\be
 |G_{kk}^{(i)}-m_{sc}|\le \Lambda_d +  C\Lambda_o^2
 \qquad \mbox{for all $k\ne i$ and in $\Omega_\Lambda^c$}
\label{Gkkm}
\ee
and (see \eqref{defA})
\be
   |A_i|\le \frac{C}{M} + C\Lambda^2_o \qquad \mbox{in $\Omega_\Lambda^c$}.
\label{Aest}
\ee
Similarly, with one more expansion step and still for $z\in D$, we get 
$$
  \max_{ij}\max_{k\ne l} |G_{kl}^{(ij)}|\le C\Lambda_o,  \qquad
   \max_{ij}\max_{k} |G_{kk}^{(ij)}|\le C \qquad \mbox{in $\Omega_\Lambda^c$}
$$
and
\be
 |G_{kk}^{(ij)}-m_{sc}|\le \Lambda_d +  C\Lambda_o^2
 \qquad \mbox{for all $k\ne i,j$ and in $\Omega_\Lambda^c$}.
\label{Gkkim}
\ee
\medskip

Using these estimates, 
the following lemma shows that $Z_i$ and  $Z_{ij}^{(ij)}$ are small
assuming  $\Lambda_d+\Lambda_o$ is small and the $h_{ij}$'s are not
too large. 
These bounds hold uniformly in $D$.

\begin{lemma}\label{selfeq1} 
Define the exceptional events 
\begin{align} 
\Omega_1 &:= \left \{  \max_{1\leq i,j\leq N}|h_{ij}|\geq
 (\log N)^C|\sigma_{ij}| \right \} \non \\
\Omega_d(z) &:= \left \{   \max_{i}|Z_i(z)|\ge \frac{(\log N)^C}{\sqrt{M\eta}}\right \} 
 \non \\
\Omega_o(z) &:= \left \{    \max_{i\ne j}|Z_{ij}^{(ij)}(z)| \ge \frac{(\log N)^C}{\sqrt{M\eta}}
\right \} \non  \\
\end{align}
and we let 
\be
  \Omega:= 
\Omega_1 \cup \bigcup_{z\in D}
\Big[\big( \Omega_d(z) \cup \Omega_o(z) \big)\cap \Omega_\Lambda^c(z) \Big]
\label{defOmega}
\ee 
to be the set of all exceptional events. 
Then we have 
\be\label{B12}
\P (\Omega)    \le CN^{-c(\log \log  N)}.
\ee
\end{lemma}

\bigskip

{\it Proof}: Under the assumption of \eqref{subexpuj}, analogously to \eqref{hest}, we have 
\be\label{resboundhij}
\P\left(\Omega_1 \right)\leq CN^{-c\log\log N},
\ee
so we can work in the complement set $\Omega_1^c$
and assume that
\be
 \max_{ij} |h_{ij}|\le \frac{(\log N)^C}{\sqrt M}.
\label{hupper}
\ee
We now prove that for any fixed $z\in D$, we have
\be\label{POdz}
\P\Big(\Omega_\Lambda^c(z)\cap  \left \{   \max_{i}|Z_i(z)|\ge
  \frac{(\log N)^C}{\sqrt{M\eta}}\right \}  \Big)\leq CN^{-c\log\log N}
\ee
and
\be\label{Poz}
\P\Big(\Omega_\Lambda^c(z)\cap 
 \left \{    \max_{i\ne j}|Z_{ij}^{(ij)}(z)| \ge \frac{(\log N)^C}{\sqrt{M\eta}}
 \right \}\Big)\leq CN^{-c\log\log N}.
\ee
Recall that $Z^{(ij)}$ is a quadratic form in the components of 
the random vectors $\ba^i$ and $\ba^j$.
For such functions, we have the following general large deviation result.
The proof relies on the Burkholder martingale inequality and it will be 
given in Appendix \ref{sec:lde}.
\begin{lemma}\label{generalHWT}\cite[Lemma B.1, B.2]{EYY}
Let $a_i$ ($1\leq i\leq N$) be $N$ independent random complex  variables with mean zero, 
variance $\sigma^2$  and having the uniform  subexponential decay
\be\label{assumdelta2}
\P(|a_i|\geq x^\al)\leq C e^{-x}
\ee
for some positive  $\alpha$ and for all $x$.
Let $A_i$, $B_{ij}\in \C$ ($1\leq i,j\leq N$). 
 Then we have that 
 \begin{align}
\P\left\{\left|\sum_{i=1}^N a_iA_i\right|\geq 
(\log N)^{\frac32+\al} \sigma \Big(\sum_{i}|A_i|^2\Big)^{1/2}\right\}\leq & CN^{-\log\log N},
\label{resgenHWTD} \\
\P\left\{\left|\sum_{i=1}^N\overline a_iB_{ii}a_i-\sum_{i=1}^N\sigma^2 B_{ii}\right|\geq 
(\log N)^{\frac32+2\al} \sigma^2 \Big( \sum_{i=1}^N|B_{ii}|^2\Big)^{1/2}\right\}\leq &
 CN^{-\log\log N},\label{diaglde}\\
\P\left\{\left|\sum_{i\neq j}\overline a_iB_{ij}a_j\right|\geq (\log N)^{3+2\al} \sigma^2 
\Big(\sum_{i\ne j} |B_{ij}|^2 \Big)^{1/2}\right\}\leq & CN^{-\log\log N}. \label{resgenHWTO}
\end{align}
\end{lemma}

To see \eqref{POdz}, we
apply the estimate \eqref{diaglde}
 and we obtain that 
 \be\label{BI23}
|Z_{i}|
 \leq(\log N)^C\Big(\sum_{k,l\not= i }\left|\sigma_{ik}
G^{(i)}_{kl}\sigma_{li}\right|^2\Big)^{1/2}
\ee
holds with a probability larger than $1-CN^{-c(\log \log  N)}$ for sufficiently large $N$.

Denote by $u^{({i})}_\alpha$ and $\lambda_\al^{({i})}$ ($\al=1,2,\ldots ,N-1$)
 the eigenvectors and eigenvalues of $H^{({i})}$. Let $u^{({i})}_\alpha(k)$ denote the
 $k$-th coordinate of $u^{({i})}_\alpha$. 
 Then, using $\sigma_{il}^2 \le 1/ M$, \eqref{Gkkm} and \eqref{Gkk}, we have
\begin{align} \label{B14}
\sum_{k, l\not= i }\left|\sigma_{ik}
G^{(i)}_{kl}\sigma_{li}\right|^2 &\le \frac{1}{M}\sum_{k\ne i} \sigma_{ik}^2
\sum_{l\ne i} |G^{(i)}_{kl}|^2 \non\\
&\le \frac1M \sum_{k\not= i } \sigma_{ik}^2
 \left(|G^{(i\,)}|^2\right)_{kk}  \nonumber \\
&=
\frac1M \sum_{k\not= i } \sigma_{ik}^2 \sum_{\alpha}
\frac{|u^{(i\,)}_\alpha(k)|^2}{|\lambda_\alpha^{(i\,)}-z|^2}
\le 
\frac1M \sum_{k\not= i } \sigma_{ik}^2 \frac{\mbox{\rm Im } G^{(i\,)}_{kk}(z)}{\eta}\non\\
&\le \frac{C}{M\eta} \qquad \mbox{in $\Omega_\Lambda^c$}.
\end{align}
Here we defined $|A|^2:= A^* A$ for any matrix $A$.
Together with \eqref{BI23} we have proved \eqref{POdz} for a fixed $z$.
The offdiagonal estimate \eqref{Poz}, for  $i\ne j$, is proven similarly,
using \eqref{resgenHWTO} instead of \eqref{diaglde}.

We thus proved that for each fixed $z\in D$, the sets 
$\Omega_d(z)$, $\Omega_o(z)$ have a very small probability.
In order to prove \eqref{B12}, in principle, we have to
take the union of uncountable many events. But it is
easy to see that the quantities $Z_i(z)$ and $Z_{ij}^{(ij)}(z)$
defining these events are Lipschitz continuous
functions in $z$ with a Lipschitz constant $\eta^{-1}\le N$.
Thus controlling them on a sufficiently dense
but finite net of points will control them for every $z\in D$.
The number of necessary points is only a power of $N$ while
the probability bounds are smaller than any inverse power  of $N$.
This proves   Lemma \ref{selfeq1}.   
\qed

\bigskip 

The key step in the proof of Theorem~\ref{partial} is the following lemma, which says
that if $\Lambda$ is somewhat small, like $(\log N)^{-C}$ then 
the estimate can be boosted to show that it is much smaller,
like $(M\eta)^{-1/2}$.

\begin{lemma}\label{thm: stepone}
Recall $\Lambda_d$, $\Lambda_o$ and $\Omega$ defined in  \eqref{defLambda} 
and \eqref{defOmega} and recall the set $D$ from \eqref{fakerelkaetapart}.
 Then  for any $z\in D$ and in the event $\Omega^c$, we have the following implication: if 
\be\label{alamb}
\Lambda_o(z) +\Lambda_d(z)\leq (\log N)^{-C}
\ee
then 
\be 
\label{alamb2}
\Lambda_o(z)+\Lambda_d(z)\leq \frac{(\log N)^C}{\sqrt{M\eta}\,\,}.
\ee
\end{lemma}
\bigskip

\textit{Proof of Lemma \ref{thm: stepone}}.
Choosing $C_1$ in \eqref{fakerelkaetapart} sufficiently large, we can ensure
from Lemma \ref{selfeq1} that
$$
  Z_{ij}^{(ij)}\ll 1, \qquad Z_i\ll 1 \qquad \mbox{in $\Omega^c$}.
$$
We first estimate  the offdiagonal term $G_{ij}$. Under the condition
\eqref{alamb}, from \eqref{G12}, \eqref{lowerbound}
and \eqref{Gkk} we have
$$
  |G_{ij}|  = |G_{ii}| |G_{jj}^{(i)}| |K_{ij}^{(ij)}|
  \le C\left(|h_{ij}| + |Z_{ij}^{(ij)}|\right), \qquad i\ne j.
$$
By Lemma \ref{selfeq1} and under  the condition
\eqref{alamb} we have
\be
\label{gijest}
|G_{ij}|\le \frac{(\log N)^C}{\sqrt{M}}+\frac{(\log N)^C}{\sqrt{M\eta}} \le
 \frac{(\log N)^C}{\sqrt{M\eta}} \qquad \mbox{in $\Omega^c$}.
\ee
 This
proves the estimate \eqref{alamb2} for the summand $\Lambda_o$.

\bigskip

Now we estimate the diagonal terms. 
 Recalling $\Upsilon_i= A_i +h_{ii}-Z_i$ from \eqref{seeqerror}, with \eqref{Aest}, 
\eqref{gijest}
and  Lemma \ref{selfeq1} we have, 
\be
\Upsilon:=\max_i   |\Upsilon_i|\le  
 C\frac{(\log N)^C}{\sqrt{M}}+\frac{(\log N)^C}{\sqrt{M\eta}} \le
 \frac{(\log N)^C}{\sqrt{M\eta}} \ll 1
\qquad \mbox{in  $\Omega^c$} . 
\label{defUpsilon}
\ee
(in the last step we used $z\in D$ and that $C_1$ is large).
{F}rom \eqref{mainseeq} we have the identity 
\be
  v_i = G_{ii} - m_{sc} 
 = \frac{1}{-z- m_{sc}- \Big(\sum_{j}\sigma^2_{ij}v_j-\Upsilon_i\Big)}
  - m_{sc} .
\label{idd}
\ee
Using that $(m_{sc}+z)=-m_{sc}^{-1}$, the fact that $|m_{sc}+z|\ge 1$ and
that  $\Lambda_d +\Upsilon\ll 1 $,
we can expand \eqref{idd} as
\be
   v_i = m_{sc}^2\Big(\sum_{j}\sigma^2_{ij}v_j-\Upsilon_i\Big)
  + O\Big(\sum_{j}\sigma^2_{ij}v_j-\Upsilon_i\Big)^2
  = m_{sc}^2\Big(\sum_{j}\sigma^2_{ij}v_j-\Upsilon_i\Big) + 
 O\Big((\Lambda_d+\Upsilon)^2 \Big).
\label{expand}
\ee
Summing up this formula for all $i$ and
recalling the definition
 $\barv:=\frac1N\sum_{i}v_i=m-m_{sc}$ 
yield
$$
   \barv = m_{sc}^2(z) \barv - \frac{m_{sc}^2(z)}{N}\sum_i \Upsilon_i 
  +  O\Big((\Lambda_d+\Upsilon)^2 \Big).
$$
Introducing the notations $\zeta: = m_{sc}^2(z)$, $[\Upsilon]:= \frac{1}{N}\sum_i\Upsilon_i$
 for simplicity, we have
(using $\Lambda_d\le 1$ and $|\zeta|\le 1$)
\be\label{vi09}
(1-\zeta)\barv=-\zeta[\Upsilon]
+O\left( (\Lambda_d+\Upsilon)^2 \right)
=O\left(\Lambda^2_d+\Upsilon\right).
\ee
Recall that $\Sigma$ denotes the matrix of covariances,
 $\Sigma_{ij}=\sigma^2_{ij}$, and we know that $1$ is 
a simple eigenvalue of $\Sigma$ with the constant vector  ${\bf e}=N^{-1/2}(1,1,\ldots,1)$ as the eigenvector.
Let $Q:=I-|{\bf e}\rangle\langle {\bf e}|$
be the projection onto the orthogonal complement of ${\bf e}$, note that $\Sigma$ and $Q$ commute.
  Let  $\| \cdot \|_{\infty\to \infty}$ denote the $\ell^\infty\to \ell^\infty$ matrix norm.
With these notations we can combine \eqref{vi09} and \eqref{expand} to get
$$
   v_i-\barv = \zeta \sum_{j}\Sigma_{ij} (v_j-\barv) -
\zeta\Big( \Upsilon_i -[\Upsilon]\Big)
+O\left(\Lambda^2_d+\Upsilon\right).
$$
When summing up for all $i$,  all three explicit terms sum up to zero, hence
so does the error term. Therefore $Q$ acts as identity on the vector of error terms, and we have
\begin{align}
    v_i-\barv = &  -\sum_j\Big(\frac{\zeta}{1-\zeta\Sigma}\Big)_{ij} 
\left(\Upsilon_j- [\Upsilon]\right) + 
   O\Bigg( \Big\| \frac{\zeta Q}{1-\zeta \Sigma}\Big\|_{\infty\to \infty}
 (\Lambda_d^2+\Upsilon)\Bigg) \label{vi10} \\
= & \Big\| \frac{\zeta Q}{1-\zeta \Sigma}\Big\|_{\infty\to \infty}
O\big( \Lambda^2_d+\Upsilon\big). \non
\end{align}

Combining \eqref{vi09} with \eqref{vi10}, we have 
\be\label{temp3.49}
\Lambda_d=\max_{i}|v_i|\leq C \left(\Big\| \frac{\zeta Q}{1-\zeta \Sigma}
\Big\|_{\infty\to \infty}+\frac{1}{|1-\zeta|}\right)(\Lambda_d^2+\Upsilon).
\ee 

From this calculation it is clear how the estimates deteriorate at the edge.
It is easy to check that
$$ 
  1-\zeta = 1- m_{sc}^2(z) \sim \sqrt{\kappa+\eta}, \qquad z= E+i\eta, \quad \kappa
  = \big| |E|-2\big|.
$$
However, as we remarked at the beginning of the proof, we assume $\kappa\ge \kappa_0>0$
for simplicity,
so we will not follow this deterioration.

\medskip

To estimate the norm of the resolvent, we recall  
the following elementary lemma:

\begin{lemma} \label{lemma:msc}\cite[Lemma 5.3]{EYY}
Let $\delta_->0$ be a given constant. Then there exist small real numbers 
$\tau\geq 0$ and $c_1>0$, depending only on $\delta_-$, such that 
 for any positive number 
$\delta_+$, we have 
\be
    \max_{x\in [-1+\delta_-,1-\delta_+]}\left\{\Big| \tau + x\, m_{sc}^2(z)\Big|^2 
\right\}\le \left(1-c_1\, q(z) \right)(1+\tau)^2
\label{al}
\ee
with 
\be\label{defq}
q(z):=\max\{\delta_+,|1- \re  m_{sc}^{2}(z)|\}. 
\ee
\qed
\end{lemma}

\begin{lemma}\label{lm:Qnorm} Suppose that $\Sigma$ satisfies \eqref{de-de+}, i.e.,
$\mbox{Spec}(Q\Sigma)\subset [-1+\delta_-, 1-\delta_+]$. Then we have
\be
\Big\| \frac{Q}{1- m_{sc}^2(z) \Sigma}\Big\|_{\infty\to\infty} 
 \le \frac{C(\delta_-) \log N}{ q(z)} 
\label{inftynorm}
\ee
with some constant $C(\delta_-)$ depending on $\delta_-$ and 
with $q$ defined in \eqref{defq} 
\end{lemma}

{F}rom this lemma one can see that the deterioration of the
estimates at the edge caused by the first term in the right hand side of
\eqref{temp3.49} can be offset by assuming $\delta_+>0$.

\medskip

{\it Proof of Lemma \ref{lm:Qnorm}}: 
Let $\| \cdot \|$ denote the usual $\ell^2\to\ell^2$ matrix norm
and recall $\zeta=m_{sc}^2(z)$. Rewrite 
$$
  \Big\| \frac{Q}{1-\zeta \Sigma}\Big\| 
  = \frac{1}{1+\tau} \Big\| \frac{Q}{1- \frac{\zeta \Sigma +\tau}{1+\tau}}\Big\|
$$
with $\tau$ given in \eqref{al}.  
By \eqref{al}, we have 
$$
   \Big\| \frac{\zeta \Sigma +\tau}{1+\tau}Q\Big\|  \le     \sup_{x\in[-1+\delta_-,
1-\delta_+]}
  \Big| \frac{\zeta x+\tau}{1+\tau} \Big|
\le  (1- c_1q(z))^{1/2} .
$$
To estimate the  $\ell^\infty\to\ell^\infty$ norm of this matrix,  recall that 
$|\zeta| \le 1$  and 
$\sum_j | \Sigma_{ij}| =\sum_{j}\sigma_{ij}^2=1$.
Thus we have 
$$
   \Big\|  \frac{\zeta \Sigma +\tau}{1+\tau} Q\Big\|_{\infty\to\infty}
    = \max_i \sum_j 
   \Big| \Big( \frac{\zeta \Sigma +\tau}{1+\tau} \Big)_{ij} \Big| 
   \le \frac{1}{1+\tau} \max_i \sum_j |\zeta \Sigma_{ij} + \tau \delta_{ij}|
  \le \frac{ |\zeta |+\tau}{1+\tau}\le 1.
$$

To see \eqref{inftynorm}, we can expand, up to an arbitrary threshold $n_0$,
\begin{align*}
\Big\| \frac{1}{1- \frac{\zeta \Sigma +\tau}{1+\tau}}Q\Big\|_{\infty\to
  \infty} 
  & \le \sum_{n< n_0} 
 \Big\| \frac{\zeta \Sigma +\tau}{1+\tau}Q\Big\|_{\infty\to\infty}^{n}
 + \sum_{n\ge n_0}  \Big\| \Big( \frac{\zeta \Sigma +\tau}{1+\tau}\Big)^nQ
 \Big\|_{\infty\to\infty} \\
 &  \le n_0 + \sqrt{N} \sum_{n\ge n_0}  \Big\| \Big( \frac{\zeta \Sigma +\tau}{1+\tau}\Big)^n
 Q\Big\|
  = n_0 + \sqrt{N} \sum_{n\ge n_0} (1-c_1 q(z))^{n/2}\\
&  = n_0 + C\sqrt{N} \frac{ (1-c_1q(z))^{n_0/2}}{q(z)} \le \frac{C\log N}{q(z)}.
\end{align*}
Choosing $n_0= C\log N/q(z)$, 
we have  proved  Lemma~\ref{lm:Qnorm}. 
\qed

\bigskip

We now return to the proof of Lemma \ref{thm: stepone}. Recall that we are in the set $\Omega^c$
and $\kappa\ge \kappa_0$, i.e., $1-\re m_{sc}^2(z)$ and thus $q(z)$ are separated away from zero.
  First, inserting 
\eqref{inftynorm} into \eqref{temp3.49}, we obtain
$$
  \Lambda_d
   \le  C (\Lambda^2_d+\Upsilon)\log N .
$$
By the  assumption  \eqref{alamb}, we have $C\Lambda_d \log N\le 1/2$, 
 so the quadratic term can be absorbed in the linear term on the right hand size,
and we get
$$
\Lambda_d   \le   C\Upsilon\log N.
$$
Using the bound \eqref{defUpsilon}
 on  $\Upsilon$, we obtain 
\be
 \Lambda_d \leq \frac{(\log N)^C}{\sqrt{M\eta}},
\label{Lambdad}\ee
which, together with \eqref{gijest},  completes the proof of \eqref{alamb2}
and thus  Lemma \ref{thm: stepone} is proven.
\qed

\bigskip

\textit{Proof of Theorem \ref{partial}}. Introducing the functions
\[
R(z): =  (\log N)^{-C}, \qquad S(z) :=  
\frac{ (\log N)^C}{\sqrt{M\eta} },
\]
 Lemma \ref{thm: stepone} states that, in the event $\Omega^c$,
if $\Lambda_d(z)+\Lambda_o(z) \le R(z)$ holds for some $z\in D$,
then  $\Lambda_d(z)+\Lambda_o(z) \le S(z)$.
By   assumption \eqref{fakerelkaetapart} of Theorem \ref{partial}, 
 we have $S(z)< R(z)$ for any $z\in D$.
Clearly, $\Lambda_d(z)+\Lambda_o(z)\le 3/\eta \le 3/10$ for $\im z =10$.
Using this information, one can mimic the proof of Lemma \ref{selfeq1}
and Lemma \ref{thm: stepone} to get an apriori bound
$\Lambda_d(z)+\Lambda_o(z) \le R(z)$ for $\eta =10$.
Using that $R(z)$, $S(z)$, $\Lambda_d(z)$  and $\Lambda_o(z)$
are continuous, moving $z$ towards the real axis,
by a continuity argument, we get that
 $\Lambda_d(z)+ \Lambda_o(z)\le S(z)$ in  $\Omega^c$, as long as the condition
 \eqref{fakerelkaetapart} for $z$ is satisfied. 
This proves  Theorem \ref{partial}.  \qed

\subsubsection{Sketch of the proof of the  semicircle law
for Stieltjes transform}\label{sec:sk:strong}

In this section we strenghten the estimate of Theorem \ref{partial} for
the Stieltjes transform $m(z) = \frac{1}{N}\sum_i G_{ii}$. The key improvement is
that $|m-m_{sc}|$ will be estimated with a precision $(M\eta)^{-1}$
while the $|G_{ii}-m_{sc}|$ was controlled by a precision $(M\eta)^{-1/2}$ only
(modulo logarithmic terms and terms expressing the deterioriation 
of the estimate near the edge). In the following theorem  we  prove a partial version
of \eqref{mainlsresult} of Theorem \ref{lsc}:

\begin{theorem}\label{prop:mmsc} Assume the conditions of Theorem \ref{lsc},
let $\kappa_0>0$ be fixed 
and recall the definition of the domain $D= D_{\kappa_0}$ from \eqref{fakerelkaetapart}.
 Then for any $\e>0$  and $K>0$
there exists a constant $C=C(\e,K,\kappa_0)$
 such that
\be
   \P \Big( \bigcup_{z\in D}\Big\{  |m(z)-m_{sc}(z)|\ge 
 \frac{N^\e }{M\eta } \Big\}
\Big) \le \frac{C(\e, K, \kappa_0)}{N^K}.
\label{eq:mmsc}
\ee
\end{theorem}

{\it Proof of Theorem \ref{prop:mmsc}.} We mostly follow the
proof of Theorem~\ref{partial}.
Fix $z\in D$ and we can assume
that we work in the complement of the small probability events
estimated in \eqref{Giipart} and \eqref{Gijpart}. In particular, the
estimate \eqref{alamb2} is available.
As in \eqref{vi09} we have that 
$$
\barv = m-m_{sc}=-\frac{\zeta}{1-\zeta}\frac1N \sum_{i}\Upsilon_i
+O\Big((\Lambda_d+\Upsilon)^2 \Big)
$$
holds with a very high probability, but now we keep the first
term and estimate it better than the most trivial bound
used in   \eqref{vi09} to extract some cancellation from the fluctuating sum.
Then with \eqref{defUpsilon} and   \eqref{Lambdad}
 we have 
$$
 m-m_{sc}=O\Big(\frac1N \sum_{j}\Upsilon_j\Big)
+O\Big(\frac{N^\e}{M\eta} \Big)
$$
holds with a very high probability for any small $\e>0$.
Recall that $\Upsilon_i = A_i + h_{ii} - Z_i$. 
We have, from \eqref{defA}, \eqref{lowerbound} and $\sigma_{ij}^2\le M^{-1}$,
$$
    A_j \le \frac{C}{M} + C\Lambda_o^2 \le \frac{CN^\e}{M\eta },
$$
where we used \eqref{alamb2} to bound $\Lambda_o$.

We thus obtain that
\be\label{tempd28}
m-m_{sc}=O\left(
\frac1N\sum_{i}Z_i-
\frac1N\sum_{i}h_{ii}\right) +O\Big(\frac{N^\e}{M\eta} \Big)
\ee
holds with a very high probability. 
Since $h_{ii}$'s are independent,  applying the 
first estimate in the large deviation Lemma \ref{generalHWT}, we have 
\be\label{sumhii}
\P\left(\Big|\frac1N\sum_{i}h_{ii}\Big|\geq (\log N)^{C_\al} 
\frac{1}{\sqrt{MN}} \right)\leq CN^{-c\log\log N}. 
\ee
On the complement event, the estimate $ (\log N)^{C_\al}(MN)^{-1/2}$ can be included in
the last error term in \eqref{tempd28}.
It only remains to bound 
$$
\barZ:=\frac1N\sum_{i}Z_i,
$$
whose moments are bounded in the next lemma, and we will comment on its proof below.

\begin{lemma}\label{motN}\cite[Lemma 5.2]{EYY2}, \cite[Lemma 4.1]{EYY3}
 Recalling the definition of $Z_i$ from \eqref{Zdef},
for any fixed $z$ in domain $D$  and any natural number $s$, we have
$$
\E\left|\frac1N\sum_{i=1}^NZ_i\right|^{2s}\le C_s \Big(\frac{(\log N)^C}{M\eta}\Big)^{2s}.
$$
\end{lemma}

Using this lemma, we have that for any $\e>0$ and $K>0$, 
\be
\P\left(\frac1N\left|\sum_{i=1}^NZ_i\right|\geq \frac{N^\e}{M\eta }\right)\leq N^{-K}
\label{OPU}
\ee
for sufficiently large $N$. Combining this with \eqref{sumhii} and \eqref{tempd28},
 we obtain \eqref{eq:mmsc} and complete the proof of Theorem~\ref{prop:mmsc}. 
\qed

\medskip
{\it Sketch of the proof of Lemma \ref{motN}.}
We have two different proofs for this lemma, both are quite
involved. Here we present the ideas of
the first proof from \cite{EYY2}.
The argument is a long
and carefully organized 
high moment calculation, similar to the second moment bound in \eqref{Zib},
but now we extract an additional factor from the sum.
Note that boosting the second moment calculation \eqref{Zib}
to higher moments (and recalling that for universal Wigner matrices
 $M$ replaces $N$) we can prove
\be
   |Z_i|\lesssim \frac{1}{\sqrt{M\eta}}.
\label{crudez}
\ee
(we will indicate the proof in Lemma \ref{y4} below).
If $Z_i$'s were independent, then the central limit theorem would imply
that
$$
  \Big| \frac1N\sum_{i}Z_i \Big|\lesssim \frac{1}{\sqrt{N}}\frac{1}{\sqrt{M\eta}},
$$
which would be more than enough.
They are not quite independent, but almost. The dependence among different $Z_i$'s 
can be extracted from using the resolvent formulas in Section \ref{sec:res}.
We will sketch the second moment calculation, i.e., the case $s=1$.
More details and higher moments are given in Sections 8--9 of \cite{EYY2}.

\medskip

Since $\E\, Z_i =0$,
the variance  of $\barZ$ is given by  
 \be\label{y2}
\frac{1}{N^2} \E\left|\sum_{i=1}^N Z_i \right|^2
= \frac{1}{N^2} \E\sum_{\al\neq \beta} 
\overline{Z_\al}Z_\beta +\frac{1}{N^2} \E\sum_{\al} 
 \left | Z_\al\right | ^2.
 \ee 
We start by estimating  the first term of \eqref{y2} for 
 $\al=1$ and $\beta=2$. The basic idea is to rewrite  $G^{(1)}_{kl}$ as
\be\label{expandG1P}
G^{(1)}_{kl}= P^{(12)}_{kl}+ P^{(1)}_{kl}
\ee  
with $P^{(12)}_{kl}$  independent of $\ba^1$, $\ba^2$ 
and $P^{(1)}_{kl}$  independent of $\ba^1$
(recall the notational convention: superscript indicate
independence of the corresponding column of $H$).
To construct this decomposition for $ k, l \notin \{1,2  \}$, 
 by \eqref{gii} or \eqref{gij} we  rewrite  $G^{(1)}_{kl}$  as 
\be\label{y13}
G^{(1)}_{kl}=G^{(12)}_{kl}+\frac{G^{(1)}_{k2}G^{(1)}_{2l}}{G^{(1)}_{22}},
 \qquad k, l \notin \{1,2  \}.
\ee
The first term on r.h.s is independent of $\ba ^2$.
Applying  Theorem \ref{partial} for the minors, we get that 
$G^{(1)}_{kl}\lesssim (M\eta)^{-1/2}$ for $k\ne l\ne 1$
 and $G^{(2)}_{kk}\ge c>0$, thus
\be\label{77x2}
\left|\frac{G^{(1)}_{k2}G^{(1)}_{2l}}{G^{(1)}_{22}}\right|\lesssim \frac{1}{M\eta}
\ee
holds with a very high probability.  Note that this bound
 is the square of the  bound $G^{(1)}_{kl}\lesssim
(M\eta)^{-1/2}$ from  Theorem \ref{partial}.

Now we define $P^{(1)}$  and $P^{(12)}$  (for $k,l \neq 1$) as
\begin{enumerate}
	\item If $ k, l \neq 2$, 
	\be\label{defP1}
	P^{(12)}_{kl}:=G^{(12)}_{kl},\,\,\,P^{(1)}_{kl}
 :=\frac{G^{(1)}_{k2}G^{(1)}_{2l}}{G^{(1)}_{22}}=G^{(1)}_{kl}- G^{(12)}_{kl}
	\ee
	\item if $k=2$ or $l=2$,
	\be\label{defP2}
	P^{(12)}_{kl}:=0,\,\,\,P^{(1)}_{kl}:=G^{(1)}_{kl}.
	\ee
\end{enumerate}
Hence \eqref{expandG1P} holds and $P^{(12)}_{kl}$ is independent of $\ba^2$. 

The size of the quadratic forms $ {\ba^{1}}  \cdot {P^{(1)}} {\ba^{1}}$
and $ {\ba^{1}}  \cdot {P^{(12)}} {\ba^{1}}$ is estimated in the following
lemma whose proof is postponed.
\begin{lemma}\label{y4}
For $N^{-1} \le \eta \le 1$ and fixed $p\in \N$, we have 
\be\label{y7}
\E\left |  {\ba^{1}}  \cdot {P^{(1)}} {\ba^{1}}  \right | ^p
\lesssim \frac{C_p}{(M\eta)^p}, \qquad \E\left | {\ba^{1}} \cdot {P^{(12)}}  {\ba^{1}}  \right |^p 
\lesssim \frac{C_p}{(M\eta)^{p/2}}.
\ee
\end{lemma}
 Note that the first quadratic form is smaller,
but the second one is independent of column (2) of $H$.

\medskip

Define an operator 
${\mathbb {IE}_i}:= \mathbb I-\E_{\ba^i}$,
where $\mathbb I$ is identity operator. 
With this convention, we have the following expansion of $Z_1$ 
\be\label{y14}
Z_1= \mathbb{IE}_1 {\ba^{1}}\cdot {P^{(12)}} {\ba^{1}}
  + \mathbb{IE}_1 {\ba^{1}}  \cdot {P^{(1)}}  {\ba^{1}}.
\ee

Exchange the index $1$ and $2$, we can define $P^{(21)}$ and $P^{(2)}$ and expand $Z_2$ as 
\be\label{y142}
Z_2= \mathbb{IE}_2 {\ba^{2}}\cdot {P^{(21)}} {\ba^{2}}  
+ \mathbb{IE}_2{\ba^{2}} \cdot  {P^{(2)}}  {\ba^{2}}
\ee
Here $P^{(21)}_{kl}$ is independent of $\ba^2$ and $\ba^1$;  $P^{(2)}_{kl}$
 is independent of $\ba^2$.
Combining \eqref{y142} with  \eqref{y14}, we have  for the $\al=1$, $\beta=2$
cross term in \eqref{y2} that
\be\label{y77}
 \E
 \overline Z_1
 Z_2
=
 \E \left [ 
  \left(  \mathbb{IE}_1  \left \{ \ov{ {\ba^{1}} \cdot {P^{(12)}} {\ba^{1}}} 
 +\ov { {\ba^{1}}  \cdot {P^{(1)}} {\ba^{1}} }\right \} \right) 
\left(   \mathbb{IE}_2  \left \{  {\ba^{2}}\cdot P^{(21)} \ba^{2}
  +{\ba^{2}}\cdot P^{(2)} \ba^{2} \right \}   \right) \right ].
\ee
Note that if $X^{(i)}$ is a random variable, independent of $\ba^i$, then
 for any random variable $Y$, we have
$$ \mathbb{IE}_i\big[ YX^{(i)}\big]=
 X^{(i)}\mathbb{IE}_i Y
$$
in particular $ \mathbb{IE}_i X^{(i)} =0$ (with $Y=1$).
Thus
$$
  \E \Big[ \big( \mathbb{IE}_1  X^{(2)}\big) \big( \mathbb{IE}_2 X^{(1)} \big)\Big]
  = \E \Big[ \mathbb{IE}_1 \Big[ \big( \mathbb{IE}_2 X^{(1)} \big)
  X^{(2)} \Big]\Big] = 0
$$
since $\E \, \mathbb{IE}_i=0$. Using this idea, one can easily see that
the only non-vanishing term on the right hand side  of \eqref{y77} is 
\be\label{tempd77}
 \E   \left ( \mathbb{IE}_1 \ov{ {\ba^{1}}   {P^{(1)}}   {\ba^{1}}}\right) 
  \left( \mathbb{IE}_2 {\ba^{2}} 
P^{(2)}  \ba^{2} \right) .
\ee

 By the Cauchy-Schwarz inequality and Lemma \ref{y4}
  we  obtain
\be\label{EZ1Z2}
|\E\overline Z_1Z_2|\lesssim \frac{1}{(M\eta)^2}.
\ee
Using \eqref{expandG1P}, 
Lemma~\ref{y4} also implies that 
\be
\label{EZK}
\E\left |Z_i \right |^p \lesssim \frac{C_p}{(M\eta)^{p/2}},
 \qquad 1\leq i\leq N
\ee
i.e. it also proves \eqref{crudez}
and it estimates the second term in \eqref{y2} by $N^{-1} (M\eta)^{-1}\le (M\eta)^{-2}$.
Since the indices $1$ and $2$ in \eqref{EZ1Z2} can be replaced by $\alpha \not = \beta$,
  together with \eqref{y2}  we have thus proved Lemma \ref{motN} 
for $s=1$.  \qed

\bigskip

{\it Proof of Lemma \ref{y4}. }
 First we rewrite  ${\ba^{1}} \cdot {P^{(1)}} {\ba^{1}} $ as follows
\be\label{y100}
 {\ba^{1}}  \cdot {P^{(1)}} {\ba^{1}}
=\sum_{k,l\neq 2}\overline {\ba^{1}_k}\cdot \left(\frac{G^{(1)}_{k2}G^{(1)}_{2l}}{G^{(1)}_{2,2}}
\right)\ba^{1}_l
+\sum_{k\neq 2}\overline {\ba^{1}_k}\cdot G^{(1)}_{k2}\ba^{1}_2
+\sum_{l\neq 2}\overline {\ba^{1}_2}\cdot G^{(1)}_{2l}\ba^{1}_l
+\overline {\ba^{1}_2}G^{(1)}_{22}\ba^{1}_2.
\ee
By the  large deviation estimate  \eqref{resgenHWTO} and \eqref{77x2}, we have 
\be\label{y72}
\P \left ( \left |\sum_{k,l\neq 2}\overline {\ba^{1}_k}
  \left(\frac{G^{(1)}_{k2}G^{(1)}_{2l}}{G^{(1)}_{22}}\right)\ba^{1}_l   \right | 
\ge \frac{(\log N)^C}{M\eta} \right ) \le N^{-c \log \log N}. 
\ee
Similarly,  from  \eqref{resgenHWTD} and the fact that
 $\|\ba^{1}\|_\infty \lesssim M^{-1/2}$ we get
that the second and third terms in \eqref{y100} are bounded by $\frac{1}{\sqrt{M}}
\frac{1}{\sqrt{M\eta}}\lesssim (M\eta)^{-1}$.
The last term is even smaller, this is of order $1/M$, with a very high probability.
We have thus proved that 
\be\label{y722}
\P \left ( \left | {\ba^{1}} \cdot {P^{(1)}} {\ba^{1}}  \right | 
\le \frac{(\log N)^C}{M\eta} \right ) \ge 1- N^{-c \log \log N}. 
\ee
This  inequality implies  the first  desired  inequality in \eqref{y7} except on the  exceptional set
with  probability less than any power of $1/N$. Since all Green functions are bounded by $\eta^{-1} \le N$, 
the contribution from the exceptional set is negligible and this proves the first estimate in  \eqref{y7}.
The second bound is proved similarly. 
\qed

\subsection{ Strong local semicircle law }\label{sec:best}

In this section we present our latest results from \cite{EYY3} which
remove the $\kappa$ dependence from Theorem \ref{lsc}, Theorem \ref{prop:count}
and Theorem \ref{prop:lambdagamma} for ensembles spread $M=N$, 
in particular for generalized Wigner matrices.
Recall the notations:
$$
  \Lambda_d: = \max_k |G_{kk}-m_{sc}|, \qquad \Lambda_o:= \max_{i\ne j} |G_{ij}|, \qquad
 \Lambda: = |m-m_{sc}|
$$
and recall that all these quantities depend on the spectral parameter $z$ and on
$N$.

\begin{theorem}[Strong local semicircle law] \label{45-1} \cite[Theorem 2.1]{EYY3}
Let $H=(h_{ij})$ be a hermitian or symmetric $N\times N$ random matrix, $N\ge 3$,
with $\E\, h_{ij}=0$, $1\leq i,j\leq N$,  and assume that the variances $\sigma_{ij}^2$ 
satisfy  \eqref{sum}, \eqref{de-de+} with some positive constants $\delta_\pm>0$ and 
the upper bound
\be
    \sigma_{ij}^2 \le \frac{C_0}{N}.
\label{C0bound}
\ee
 Suppose that the distributions of the matrix elements have a uniformly 
  subexponential decay
in the sense that  there exist  constants  $C, \ttau>0$, independent 
of $N$, such that for any $x\ge 1$ and $1\le i,j \le N$ we have
\be\label{subexpujabb}
\P(|h_{ij}|\geq x \sigma_{ij})\leq C\exp\big( -x^\ttau\big).
\ee
Then for any constant $\xi>1$ there exist positive constants   $L$,  $ C$ and $c$, 
depending only on $\xi$,   $\ttau$, on $\delta_\pm$ from \eqref{subexpujabb},
 \eqref{de-de+} and
on $C_0$ from \eqref{C0bound}, such that 
the Stieltjes transform of the empirical 
eigenvalue distribution of  $H $  satisfies 
\be\label{Lambdafinal} 
\P \Big ( \bigcup_{z\in \bS_L} \Big\{ \Lambda(z) 
 \ge \frac{(\log N)^{4L}}{N\eta} \Big\}   \Big )\le  C\exp{\big[-c(\log N)^{\xi} \big]}
\ee
with
\be
{\bf  S}:={\bf  S}_L=\Big\{ z=E+i\eta\; : \;
 |E|\leq 5,  \quad  N^{-1}(\log N)^{10L} < \eta \le  10  \Big\}.
\label{defS}
\ee
The individual  matrix elements of
the Green function  satisfy that 
\be\label{Lambdaodfinal}
\P \left  ( \bigcup_{z\in \bS_L} \left\{ \Lambda_d(z)  + \Lambda_o (z) \geq 
(\log N)^{4L} \sqrt{\frac{\im m_{sc}(z)  }{N\eta}} + \frac{(\log N)^{4L}}{N\eta}
   \right\}    \right)
\leq  C\exp{\big[-c(\log N)^{\xi} \big]}.
\ee
Furthermore, in the following set outside of  the limiting spectrum, 
\be
 \bO_L: =   \Big\{z= E+i\eta \; : \;  N\eta\sqrt{\kappa}\ge (\log N)^{4L} , 
\; \kappa \ge \eta,  \; |E| > 2\Big\}, \qquad \mbox{with}\quad \kappa: = \big| \, |E|-2\big|,
\label{defO}
\ee
 we have the stronger estimate  
\be  \label{8.1} 
\P \bigg( \bigcup_{z\in \bO_L} \Big\{ \Lambda (z) 
\ge \frac{(\log N)^{4L}}{N\kappa} \Big\} \bigg ) 
   \leq  C\exp{\big[-c(\log N)^{\xi} \big]}.
\ee
\end{theorem}

The subexponential decay condition \eqref{subexpujabb} can be
 weakened  if we are not aiming 
at error estimates faster than any power law of $N$. 
This can be easily carried out and we will not pursue  it in this paper. 

\medskip

Prior  to our results in \cite{EYY} and \cite{EYY2}, a central limit theorem  for
the semicircle law on macroscopic scale for band matrices was established
by Guionnet \cite{gui} and Anderson and Zeitouni \cite{AZ}; a
 semicircle law  for Gaussian band matrices  was proved 
by Disertori, Pinson and Spencer \cite{DPS}.
 For a review on band matrices, see  the recent  article \cite{Spe} by Spencer.

\bigskip

As before, the local semicircle estimates imply that the empirical counting function of
the eigenvalues is close to the semicircle counting function 
and that the locations of the eigenvalues are close to their classical 
location. We have the following improved results (cf. Theorems~\ref{prop:count}--\ref{prop:lambdagamma}):

\begin{theorem}\label{7.1}\cite[Theorem 2.2]{EYY3}
Assume the conditions of Theorem \ref{45-1}, i.e. 
 \eqref{sum}, \eqref{de-de+} with some positive constants $\delta_\pm>0$, \eqref{C0bound}
and \eqref{subexpujabb}.
Then for any constant $\xi>1$ there exist constants  $L_1$,  $C$ and $c>0$, 
depending only on $\xi$,  $\ttau$, $\delta_\pm$  and $C_0$ such that
\be\label{rigidity}
\P \Bigg\{  \exists j\; : \; |\lambda_j-\gamma_j| 
\ge (\log N)^{L_1}  \Big [ \min \big ( \, j ,  N-j+1 \,  \big) \Big  ]^{-1/3}   N^{-2/3} \Bigg\}
 \le  C\exp{\big[-c(\log N)^{\xi} \big]}
\ee
and
\be
   \P\Bigg\{ \sup_{|E|\le 5} \big| {\mathfrak n} (E)-n_{sc}(E)\big| \,  \ge
 \frac{(\log N)^{L_1}}{N} \Bigg\}\le  C\exp{\big[-c(\log N)^{\xi} \big]}.
\label{nn}
\ee
\end{theorem}

For Wigner matrices, \eqref{nn} with the factor $N^{-1}$ replaced by  $N^{-2/5}$
 (in a weaker sense with some modifications in the statement)  was established in \cite{BMT}
and  a  stronger $N^{-1/2}$ control was proven for the difference $\E \fn (E)-n_{sc}(E)$.
In Theorem 1.3 of a recent preprint \cite{TV4}, the following estimate (in our scaling)
\be\label{tv}
\Big (\E \big[  |\lambda_j-\gamma_j|^2 \big] \Big )^{1/2} \le 
\Big [ \min \big ( \, j ,  N-j+1 \,  \big) \Big  ]^{-1/3}   N^{-1/6 - \e_0},
\ee
with some small positive $\e_0$, 
was  proved for standard Wigner matrices
under the assumption that the third moment of the matrix element vanishes.
In the same paper, it was conjectured that the factor $N^{-1/6 - \e_0}$ on 
the right hand side of \eqref{tv} should be replaced by $N^{-2/3 + \e}$. 
Prior to the work \cite{TV4}, the estimate \eqref{rigidity} 
away from the edges with a slightly weaker probability estimate 
and with the $(\log N)^{L_1}$ factor replaced by $N^\delta$ for arbitrary $\delta>0$
was proved in \cite{EYY2} (see the equation before (7.8) in \cite{EYY2}).

We remark that all results are stated for both the hermitian or symmetric case, but
the statements and the proofs hold for quaternion self-dual random matrices as well
(see, e.g., Section 3.1 of \cite{ESYY}).

\bigskip

There are several improvements of  the argument presented  in Sections
\ref{sec:sk:elem} and
\ref{sec:sk:strong} that have led to the proof of the optimal Theorem \ref{45-1}
for the $M=N$ case.
Here we mention the most important ones.

First, notice that the end of the estimate \eqref{B14} can
be done more effectively for $M=N$ and $\sigma_{ik}^2\le C/N$, using that
$$
\frac{1}{N}\sum_k \im G_{kk}^{(i)} \le   \frac{1}{N}\sum_k \im G_{kk} + C\Lambda_o^2
=\im m + C\Lambda_o^2\le \Lambda+ \Lambda_o^2+\im m_{sc}
$$
The gain here is that $\im m_{sc}(z)\sim \sqrt{\kappa+\eta}$ which is
a better estimate near the edges than just $O(1)$. We therefore introduce
$$
  \Psi=\Psi(z): = \sqrt{\frac{\Lambda(z) + \im m_{sc}(z)}{N\eta}}
$$
as our main control parameter, and note that this is random, but
it depends only on $\Lambda$. Similarly to \eqref{gijest} and \eqref{defUpsilon}, one can then prove that
\be
  \Lambda_o +\max_i \Upsilon_i \le \Psi
\label{lpsi}
\ee
with very high probability.

Second, a  more careful algebra of the self-consistent equation
\eqref{expand} yields the following identity:
\be\label{661}
(1- m^2_{sc}) \barv =  m_{sc}^3 \barv^2+  m^2_{sc}\barZ   
 +  O\Big( \frac{\Lambda^2}{\log N}\Big)+   O\Big( (\log N)\Psi^2 \Big),
\ee
where $\barZ: = N^{-1} \sum_{i=1}^N Z_i$.
The advantage of this formula is that it allows not only to express $\barv$
from the left hand side (after dividing through with $1-m_{sc}^2$),
but in case of
$$
  \Big| (1- m^2_{sc}) \barv \Big| \ll \Big| m_{sc}^3 \barv^2 \Big|
$$
(which typically happens exactly near the edge, where $1-m_{sc}^2$ is small),
it is possible to express $\barv$ from the right hand side.
This makes a {\it dichotomy estimate} possible by noticing that
if
$$
  (1- m^2_{sc}) \barv =  m_{sc}^3 \barv^2 + \mbox{Small},
$$
or, in other words,
$$
    \al(z) \Lambda = \Lambda^2 + \beta(z), \qquad \al(z): =\Big|\frac{1-m_{sc}^2}{m_{sc}^3}\Big|
 \sim \sqrt{\kappa +\eta}, \qquad \beta = \mbox{Small}
$$
holds, then for some sufficiently large constant $U$ and another constant $C_1(U)$,
we have
\begin{align}\label{81}
\Lambda(z) & \le   U \beta(z)   \quad \text{ or }  \quad  \Lambda(z)  \ge  
 \frac {\alpha(z)}{  U }     & \text{ if } \;\; \al\ge U^2\beta \\
\Lambda(z) & \le C_1(U) \beta(z)     & \text{ if } \;\;  \al\le U^2\beta
 \label{82}
\end{align}
The bad case, $\Lambda \ge \al/U$, is excluded by
a continuity argument: we can easily show that for $\eta =10$
this does not happen, and then we reduce $\eta$ and 
see that we are always in the first case, as long as $\al\ge U^2\beta$,
or, if $\al\le U^2\beta$, then we automatically obtain $\Lambda \lesssim \beta$.
The actual proof is more complicated, since the ``Small'' term itself
depends on $\Lambda$, but this dependence can be absorbed into
the other terms. Moreover all these estimates hold only with
a very high probability, so exceptional events have to be tracked.

Finally, the estimate in Lemma \ref{motN} has to incorporate
the improved control on the edge. Inspecting the proof, one
sees that the gain $(M\eta)^{-2s}$ comes from the offdiagonal
resolvent elements, in fact Lemma  \ref{motN} is
better written in the following form
$$
\E\left|\frac1N\sum_{i=1}^NZ_i\right|^{2s}\le C_s \E\Big[ \Lambda_o^{2s} + N^{-s}\Big].
$$
As before, this can be turned into a probability estimate by taking a 
large power, $s\sim (\log N)^\xi$. 
Using  \eqref{lpsi} to estimate $\Lambda_o$
by $\Psi$ and the fact that $\Psi^2 \le o(\Lambda) +N^{-1}$ (since $N\eta\gg 1$),
one can show that the $m_{sc}^2 \barZ$ term  on the r.h.s of \eqref{661}
is also ``Small''. The details are given in Sections 3 and 4 of \cite{EYY3}.

\subsection{Delocalization of eigenvectors}\label{sec:deloc}

Let $H$ be a universal Wigner matrix with a subexponential decay \eqref{subexpuj}.
Let $\bv$ be an $\ell^2$-normalized eigenvector of $H$, then
the size of the $\ell^p$-norm of $\bv$, for $p>2$, gives
information about delocalization of $\bv$. 
We say that complete delocalization occurs
when $\|\bv \|_p \lesssim N^{-1/2+ 1/p}$ (note that  $\| \bv\|_p \ge 
CN^{-1/2+ 1/p}\|\bv\|_2$).
The following result
shows that for generalized Wigner matrices \eqref{VV},
the eigenvectors are fully delocalized with
a very high probability. For universal Wigner matrices with 
spread $M$ \eqref{defM}, the eigenvectors are delocalized 
on scale at least $M$.

\begin{theorem}\label{thm:deloc} Under the conditions
of Theorem \ref{lsc}, for any $E$ with $\kappa=|E-2|\ge \kappa_0$,
we have
\be
   \P\Bigg\{ \exists \bv\; : \; H\bv=\lambda\bv, \; |\lambda-E|\le \frac{1}{M},
   \; \|\bv\|_2=1, \; \|\bv\|_\infty \ge \frac{(\log N)^C}{\sqrt{M}}\Bigg\} 
  \le CN^{-c\log\log N}.
\label{linftydeloc}
\ee
\end{theorem}

We remark that $\|\bv\|_\infty \lesssim M^{-1/2}$ indicates that the
eigenvector has to be supported in a region of size at least $M$,
i.e. the localization length of $\bv$ is at least $M$.
We note that the delocalization conjecture predicts, that
the localization length is in fact $M^2$, i.e. the optimal
bound should be
$$
   \|\bv \|_\infty \lesssim \frac{1}{M}
$$
with a high probability. As it was explained in Section \ref{sec:mot},
this is an open question. Only some partial results are available, i.e.
we proved in \cite{EK, EK2} that for random band matrices \eqref{BM}
with band width $W\sim M$,
the localization length is at least $M^{1+ \frac{1}{6}}$.

\medskip

{\it Proof.} 
We again neglect the dependence of the estimate on $\kappa_0$
(this can be traced back from the proof). The estimate \eqref{Giipart}
guarantees that
$$
   |G_{ii}(z)| \le C
$$
for any $z=E+i\eta$ with $M\eta \ge (\log N)^C$ with a very high
probability. Choose $\eta = (\log N)^C/M$ .
 Let $\bv_\al$ be the eigenvectors of $H$
and let $\bv$ be an eigenvector with eigenvalue $\la$, where $|\la - E|\le 1/M$. Thus
$$
    |\bv(i)|^2\le \frac{2\eta^2 |\bv(i)|^2}{M^{-2}+\eta^2} \le 
 2 \sum_\al \frac{ \eta^2 |\bv_\al(i)|^2}{(\la_\al-E)^2+\eta^2} =2\eta  |\im G_{ii}|\le C\eta, 
$$
i.e.
$$
   \| \bv\|_\infty \le \frac{(\log N)^C}{\sqrt{M}}.
$$
\qed
Note that the proof was very easy since pointwise bounds on the
diagonal elements of the resolvent were available.
It is possible to prove this theorem
relying only on the local semicircle law, which is a conceptually
simpler input, in fact this was
our tradition path in \cite{ESY1, ESY2, ESY3}. 
For example, in \cite{ESY3} we proved

\begin{theorem}\cite[Corollary 3.2]{ESY3}\label{deloc:trad} Let $H$ be a Wigner
matrix with single entry distribution with a Gaussian decay.
 Then for any $|E|<2$, fixed $K$ and $2<p<\infty$
we have
$$
   \P\Bigg\{ \exists \bv\; : \; H\bv=\lambda\bv, \; |\lambda-E|\le \frac{K}{N},
   \; \|\bv\|_2=1, \; \|\bv\|_p\ge QN^{-\frac{1}{2}+\frac{1}{p}}\Bigg\} 
  \le Ce^{-c\sqrt{Q}}
$$
for $Q$ and $N$ large enough.
\end{theorem}

{\it Sketch of the proof.} We will give the proof of a weaker result,
where logarithmic factors are allowed.
Suppose that  $H\bv = \la \bv$ and $\la \in [-2+\kappa_0, 2-\kappa_0]$, with $\kappa_0>0$. 
 Consider the decomposition
\be
\label{Hd-old} H = \begin{pmatrix} h & \ba^* \\
\ba & H^{(1)}
\end{pmatrix}
\ee
introduced in Section \ref{sec:res}, i.e,
here  $\ba= (h_{1,2}, \dots h_{1,N})^*$  and  $H^{(1)}$ is the $(N-1)
\times (N-1)$ matrix obtained by removing the first row and first column
from $H$. Let  $\mu_\al$ and $\bu_\al$ (for $\al=1,2,\ldots , N-1$)
denote the eigenvalues and the normalized eigenvectors  of $H^{(1)}$.
{F}rom the eigenvalue equation $H \bv = \la \bv$
we find
$$
   (h-\la)v_1 + \ba\cdot v' = 0
$$
$$
   v_1 \ba + (H^{(1)}-\la)v' = 0
$$
where we decomposed the eigenvector $\bv =(v_1, v')$, $v_1\in \R$,
$v'\in \R^{N-1}$. Solving the second equation for $v'$ 
we get $v'=v_1 (\la-H^{(1)})^{-1}\ba$. From the normalization
condition, $\|\bv\|^2= v_1^2 + \|v'\|^2=1$ we thus obtain
for the first component of $\bv$ that
\be\label{v1} |v_1|^2 =
\frac{1}{1+ \ba\cdot(\la -H^{(1)})^{-2} \ba} = \frac{1}{1 + \frac{1}{N}
\sum_{\alpha} \frac{\xi_{\alpha}}{(\la - \mu_{\alpha})^2}}
\leq \frac{4 N \eta^2}{\sum_{\mu_\alpha \in I} \xi_{\alpha}} \, ,
\ee
where in the second equality we set
$\xi_{\alpha} = |\sqrt{N} \ba \cdot \bu_{\alpha}|^2$
and used
the spectral representation of $H^{(1)}$. We also chose an interval $I$
of length $\eta=|I|=Q/N$.
Is it easy to check that $\E \xi_\al=1$ and that different $\xi_\al$'s
are essentially independent and they satisfy
the following large deviation estimate:
$$
   \P \Big( \sum_{\al\in \cI} \xi_\al \le \frac{ m}{2}\Big) \le e^{-c\sqrt{m}}
$$
where $m=|\cI|$ is the cardinality of the index set. 
There are several proofs of this fact, depending on the condition on the
single site distribution.  For example,  under the Gaussian decay condition, it was
proved in Lemma 4.7 of \cite{ESY3} that relies on the Hanson-Wright theorem
\cite{HW}.

Let now $\cN_I$ denote the
number of eigenvalues of $H$ in $I$. From the local semicircle
(e.g. Theorem \ref{wegnerlsc})
we know that $\cN_I$ is of order $N|I|$ for any interval $I$ 
away from the edge and $|I|\gg 1/N$.
 We recall that the eigenvalues
of $H$, $\la_1\leq \la_2 \leq \ldots\leq \la_N$, and the eigenvalues
of $H^{(1)}$ are interlaced.
This means that
there exist at least $\cN_{I}-1$ eigenvalues  of $H^{(1)}$ in $I$.
Therefore, using that the components of any eigenvector are identically
distributed, we have

\begin{equation}
\begin{split}\label{lon}
\P \Big( \exists &\text{ $\bv$ with $H\bv=\la\bv$, $\| \bv \|=1$,
$\la \in I$ and } \| \bv \|_\infty \ge
\frac{Q}{N^{1/2}} \Big)  \\
&\leq N^2 \, \P \Big( \exists \text{ $\bv$ with
$H\bv=\la\bv$, $\| \bv \|=1$, $\la \in I$ and } |v_1|^2 \ge
\frac{Q^2}{N} \Big)\\
&\leq C\,  N^2 \P \left( \sum_{\mu_\alpha \in I}
\xi_{\alpha} \leq \frac{4N^2\eta^2}{Q^2}\right) \\
&\leq  C \, N^2  \P \left( \sum_{\mu_\alpha \in I}
\xi_{\alpha} \leq \frac{4N^2\eta^2}{Q^2} \text{ and } \cN_{I} \geq
cN |I| \right) + C\,  N^2   \, \P
\left(\cN_{I} \leq c N |I| \right)
\\&\leq C  \,  N^2 e^{- \wt{c}\sqrt{ N|I|}} + C\,
N^2 e^{- c \, \sqrt{N|I|}} \\
&\leq C e^{-c \sqrt{Q}},
\end{split}
\end{equation}
assuming that $4N^2\eta^2/Q^2 \le cN|I| = cN|I|$, i.e. that
$Q\ge \sqrt{N\eta}$.

Here we used that the deviation from the semicircle law is
subexponentially penalized,
$$
   \P \left(\cN_{I} \leq c N |I| \right) \le e^{-c'\sqrt{N|I|}}
$$
for sufficiently small $c$ and $c'$ if $I$ is away from the edge.
Such a strong subexponential bound does not 
directly follow from the local semicircle law Theorem \ref{lsc}
whose proof was outlined in Section~\ref{sec:refined},
but it can be proved for Wigner matrices with Gaussian decay,
Theorem \ref{wegnerlsc}. 
\qed

\section{Universality for Gaussian convolutions}\label{sec:sine}

\subsection{Strong local ergodicity of the Dyson Brownian Motion}\label{sec:flow}

In this section, we consider the following general question.
Suppose $\mu = e^{-N\cH} /Z $ is a probability measure on the configuration
space $\bR^N$ characterized by some Hamiltonian $\cH :\bR^N\to \bR$, where
$Z= \int e^{-N\cH(\bx)} \rd \bx<\infty$ is the normalization.
We will always assume that $\cH$ is symmetric under permutation of the variables
$\bx =(x_1, x_2, \ldots, x_N)\in \bR^N$.
The typical example to keep in mind is the Hamiltonian of the  general
$\beta$-ensembles \eqref{H}, or the specific GOE $(\beta=1$) or GUE 
$(\beta=2)$ cases.

We  consider time dependent permutation-symmetric
probability density $f_t(\bx)$, $t\ge0$
with respect to the measure $\mu(\rd \bx)=\mu(\bx)\rd\bx$,
i.e. $\int f_t(\bx)\mu (\rd\bx)=1$.
The dynamics is characterized by the forward equation
\be\label{dy}
\partial_{t} f_t =  L f_t, \qquad t\ge 0,
\ee
with a given permutation-symmetric initial data $f_0$.
The generator $L$ is defined via
the Dirichlet form as
\be
D(f): = D_\mu(f) =  -\int  f L f  \rd \mu =  \sum_{j=1}^N \frac{1}{2N}
\int (\partial_j f)^2 \rd \mu,  \quad \pt_j =\pt_{x_j}.
\label{def:dir2}
\ee
Formally, we have $L= \frac{1}{2N}\Delta - \frac{1}{2}(\nabla \cH)\nabla$.
We will ignore the domain questions, we just mention that $D(f)$ is a semibounded quadratic
form, so $L$ can be defined via the Friedrichs extension on $L^2(\rd\mu)$ and
it can be extended to $L^1$ as well. The dynamics is
 well defined for any $f_0\in L^1(\rd\mu)$
initial data. For more details, see
 Appendix A of \cite{ESYY}.

Strictly speaking, we will consider a sequence of Hamiltonians $\cH_N$ and corresponding
dynamics $L_N$ and $f_{t,N}$
parametrized by $N$, but the $N$-dependence will be omitted. All results
will concern the $N\to\infty$ limit.

Alternatively to \eqref{dy}, one could describe the dynamics by a coupled system of stochastic
differential equations \eqref{sde} as mentioned in Section \ref{sec:lrf}, but we
will not use this formalism here.

\bigskip

For any $k\ge 1$ we define the $k$-point
correlation functions (marginals) of the probability measure $f_t\rd\mu$ by
\be
 p^{(k)}_{t,N}(x_1, x_2, \ldots,  x_k) = \int_{\R^{N-k}}
f_t(\bx) \mu(\bx) \rd x_{k+1}\ldots
\rd x_N.
\label{corr}
\ee
The correlation functions of the equilibrium measure  are denoted by
$$
 p^{(k)}_{\mu,N}(x_1, x_2, \ldots,  x_k) = \int_{\R^{N-k}}
\mu(\bx) \rd x_{k+1}\ldots
\rd x_N.
$$

We now list our main assumptions on the initial distribution $f_0$
and on its evolution $f_t$. This formalism is adjusted to generalized
Wigner matrices;  random covariance matrices require some minor modifications
(see \cite{ESYY} for details).
We first define the subdomain
\be
\Sigma_N: = \big\{ \bx\in\bR^N, \; x_1 < x_2 < \ldots < x_N\big\}
\label{def:Sigma}
\ee
of ordered sets of points $\bx$.

\bigskip

{\bf Assumption I.} The Hamiltonian $\cH$ of the equilibrium measure has the
form
\be
  \cH = \cH_N(\bx)  = \beta\Big[ \sum_{j=1}^N U(x_j)
-\frac{1}{N}\sum_{i<j} \log|x_i-x_j| \Big], 
\label{ham}
\ee
where $\beta\ge 1$. The function $U:\bR\to \bR$ is smooth with $U'' \ge 0$,
 and 
$$
  U(x) \ge C  |x|^\delta  \qquad  \text{ for some $\delta> 0$ and $|x|$ large} .
$$

\bigskip

Note that this assumption is automatic for symmetric and hermitian Wigner matrices
with the GOE or GUE being the invariant measure, \eqref{expli2}--\eqref{H}.

Near the $x_{i+1}=x_i$ boundary component of $\Sigma_N$, the generator has the form
$$
   L = \frac{1}{4N} \Big( \pt_u^2 + \frac{\beta}{u}\pt_u\Big) + \mbox{regular operator}
$$
in the relative coordinates $u=\frac{1}{2}(x_{i+1}-x_i)$ when $u\ll 1$.
It is known (Appendix A of \cite{ESYY}) 
that for $\beta\ge1$ the repelling force of the singular diffusion
is sufficiently strong to prevent the particle from  falling into
the origin, i.e., in the original coordinates the trajectories of
the points do not cross. In particular, the ordering of the points 
 will be preserved under the dynamics (for a stochastic proof, see
Lemma 4.3.3 of \cite{AGZ}).
In the sequel we will thus assume
that $f_t$ is a probability measure on $\Sigma_N$.
We continue to use the notation $f$
and $\mu$ for the restricted measure. Note that the correlation
functions $p^{(k)}$ from \eqref{corr} are still defined on $\bR^k$, i.e., their
arguments remain unordered. 

\medskip

It follows from Assumption I that the Hessian matrix of $\cH$ satisfies the
following bound: 
\be
 \big\langle \bv, \nabla^2 \cH(\bx) \bv\big\rangle
\ge  \frac{\beta}{N}
\sum_{i<j} \frac{ (v_i-v_j)^2}{(x_i-x_j)^2}, \qquad
\bv = (v_1, \ldots , v_N)\in\bR^N, \quad \bx\in \Sigma_N.
\label{convex}
\ee

This convexity bound is the key assumption;
our method works for a broad class of  general  Hamiltonians
as long as \eqref{convex} holds.
In particular, an arbitrary many-body potential function $V(\bx)$
can be added to the Hamiltonians \eqref{ham}
as long as $V$ is convex on  $\Sigma_N$.

\bigskip

{\bf Assumption II.} There exists a continuous, compactly supported
density function $\varrho(x)\ge 0$, $\int_\bR \varrho =1$, on the
real line, independent of $N$, such that for any fixed $a,b\in \bR$
\be
\lim_{N\to\infty}   \sup_{t\ge 0}  \Bigg|
\int \frac{1}{N}\sum_{j=1}^N {\bf 1} ( x_j \in [a, b]) f_t(\bx)\rd\mu(\bx)
- \int_a^b \varrho(x) \rd x \Bigg| =0.
\label{assum1}
\ee
\bigskip

In other words, we assume that a limiting density exists; for
Wigner matrices this is the semicircle law.
Let $\gamma_j =\gamma_{j,N}$ denote the location of the $j$-th point
under the limiting density, i.e., $\gamma_j$ is defined by
\be
 N \int_{-\infty}^{\gamma_j} \varrho(x) \rd x = j, \qquad 1\leq j\le N, \quad
\gamma_j\in \mbox{supp} \varrho.
\label{gammaj}
\ee
We will call $\gamma_j$ the {\it classical location} of the $j$-th point.
Note that $\gamma_j$ may not be uniquely defined if the support
of $\varrho$ is not connected
but in this case the next Assumption III
will not be satisfied anyway.

\bigskip

{\bf Assumption III.} There exists an $\fa>0$ such that
\be
 \sup_{t\ge N^{-2\fa}} \int  \frac{1}{N}\sum_{j=1}^N(x_j-\gamma_j)^2
 f_t(\rd \bx)\mu(\rd \bx) \le CN^{-1-2\fa}
\label{assum2}
\ee
with a constant $C$ uniformly in $N$.

\bigskip

Under
Assumption II, the typical spacing between
neighboring points is of order $1/N$ away from the spectral edges, i.e., in
the vicinity of any energy $E$ with $\varrho(E)>0$. Assumption III guarantees
that typically the random points $x_j$ remain in
the $N^{-1/2-\fa}$ vicinity of their classical location.

\bigskip

The final assumption is an upper bound on the local density. 
For any $I\in \R$, let
$\cN_I: = \sum_{i=1}^N {\bf 1}( x_i \in I)$
denote the number of points in $I$.

\bigskip

{\bf Assumption IV.} For any compact subinterval $I_0\subset \{ E\;: \; \varrho(E)>0\}$,
and for  any $\delta>0$,  $\sigma>0$
there are constants $C_n$, $n\in \N$, depending on $I_0$,
$\delta$ and $\sigma$ such that for any interval $I\subset I_0$ with
$|I|\ge N^{-1+\sigma}$ and for any $K\ge 1$, we have
\be
   \sup_{\tau \ge N^{-2\fa}}
    \int {\bf 1}\big\{ \cN_I \ge KN|I| \big\}f_\tau \rd\mu
\le C_{n} K^{-n}, \qquad n=1,2,\ldots,
\label{ass4}
\ee
where $\fa$ is the exponent from Assumption III.
\bigskip

Note that for the symmetric or hermitian Wigner matrices,
Assumption I is automatic, Assumption II is the (global)
semicirle law \eqref{global:sc} and Assumption IV is
the upper bound on the density (Lemma \ref{lm:upper}).
The really serious condition to check is 
\eqref{assum2}.

\medskip

The following main general theorem  asserts
 that the local  statistics of the points $x_j$ in the bulk
with respect to the time evolved distribution $f_t$ coincide with the local statistics
with respect to the equilibrium
measure $\mu$ as long as $t\gg N^{-2\fa}$.

\begin{theorem}\label{thm:main} \cite[Theorem 2.1]{ESYY} Suppose that the Hamiltonian given in
\eqref{ham}  satisfies Assumption I and Assumptions II, III, and IV
hold for the solution $f_t$ of the forward equation
\eqref{dy} with exponent $\fa$.
 Assume that at time $t_0=N^{-2\fa}$
the density $f_{t_0}$ satisfies a bounded entropy
condition, i.e.,
\be
S_{\mu}(f_{t_0}):=\int f_{t_0}\log f_{t_0}\rd\mu \le CN^m
\label{entro}
\ee
 with some fixed exponent $m$. 
 Let $E\in \bR$ and $b>0$
such that $\min\{\varrho(x)\; : \; x\in[E-b,E+b]\}>0$. Then for any $\delta>0$, 
for any integer $k\ge 1$ and for any compactly supported continuous test function
$O:\bR^k\to \bR$, we have, with the notation  $\tau := N^{-2\fa+\delta}$,
\be
\begin{split}
\lim_{N\to \infty} \sup_{t\ge \tau } \;
\int_{E-b}^{E+b}\frac{\rd E'}{2b}
\int_{\R^k} &  \rd\alpha_1
\ldots \rd\alpha_k \; O(\alpha_1,\ldots,\alpha_k) \\
&\times \frac{1}{\varrho(E)^k} \Big ( p_{t,N}^{(k)}  - p_{\mu, N} ^{(k)} \Big )
\Big (E'+\frac{\alpha_1}{N\varrho(E)},
\ldots, E'+\frac{\alpha_k}{ N\varrho(E)}\Big) =0.
\label{abstrthm}
\end{split}
\ee
\end{theorem}
We remark that the limit can be effectively controlled, in fact, in \cite{ESYY} we
obtain that before the $N\to\infty$ limit the left hand side of \eqref{abstrthm}
is bounded by $CN^{2\e'}\big[ b^{-1} N^{-\frac{1}{3}(1+2\fa)} + b^{-1/2} N^{-\delta/2}\big]$.

In many applications, the local equilibrium
statistics can be explicitly computed and in  the $b\to 0$ limit
they become independent of $E$, in particular this is the case for
the classical matrix ensembles. The simplest explicit formula
is for the GUE case, when the correlation functions are
given by the sine kernel \eqref{sineres}.

\subsection{The local relaxation flow}\label{sec:locrelax}

The main idea behind the proof of Theorem \ref{thm:main} is
to analyze the relaxation to equilibrium of the dynamics \eqref{dy}.
The equilibrium is given by an essentially convex Hamiltonian $\cH$
so the Bakry-Emery method \cite{BE} applies. This method was first used
in the context of the Dyson Brownian motion in Section 5.1 of \cite{ERSY};
the presentation here follows \cite{ESYY}.

To explain the idea, assume, temporarily, that the potential
$U$ in \eqref{ham} is uniformly convex, i.e.
$$
   U''(x) \ge U_0>0.
$$
This is certainly the case for the Gaussian ensembles when $U(x) = \frac{1}{4} x^2$.
Then we have the following lower bound on the Hessian of  $\cH$
\be
  \mbox{Hess} \, \cH \ge\beta U_0
\label{conv}
\ee
on the set $\Sigma_N$ \eqref{def:Sigma} since the logarithmic potential is convex.
It is essential to stress at this point  that \eqref{conv} holds
only in the {\it open} set $\Sigma_N$, since the second
derivatives of the logarithmic
interactions have a delta function singularity (with the ``wrong'' sign)
on the boundary. It requires a separate technical argument to
show that for $\beta\ge 1$ the points sufficiently repel each
other so that the Dyson Brownian motion never leaves the open set $\Sigma_N$
and thus the Bakry-Emery method applies. 
See a remark after Theorem \ref{thm:BE}.

 We devote the next pedagogical section
to recall the Bakry-Emery criterion in a general setup
on $\bR^N$.

\subsubsection{Bakry-Emery method}\label{sec:BEM}

Let the probability measure $\mu$ on $\bR^N$ be given by a strictly convex Hamiltonian $\cH$:
\be
  \rd\mu(\bx) = \frac{e^{-\cH(\bx)}}{Z}\rd \bx, \qquad 
 \nabla^2\cH(\bx)=\mbox{Hess} \, \cH(\bx) \ge K >0
\label{convBE}
\ee
with some constant $K$,
and let $L$ be the generator of the dynamics associated with the Dirichlet form
$$
   D(f)= D_\mu(f) = -\int fLf \rd \mu := \frac{1}{2}\sum_j 
  \int (\pt_j f)^2 \rd \mu, \qquad \pt_j = \pt_{x_j}
$$
(note that in this presentation we neglect the prefactor $1/N$ originally present in 
\eqref{def:dir2}).
Formally we have $L= \frac{1}{2}\Delta -\frac{1}{2}(\nabla\cH)\nabla$.
The operator $L$ is symmetric with respect to the measure $\rd\mu$, i.e.
\be
   \int fLg \rd\mu =\int (Lf) g\rd\mu = -\frac{1}{2}\int \nabla f\cdot \nabla g \rd\mu.
\label{symm}
\ee

We define the relative entropy of any probability density $f$ with $\int f\rd\mu=1$
by
$$
   S_\mu(f) = S(f) = \int f(\log f)\rd\mu.
$$
Both the Dirichlet form and the entropy are non-negative, they are zero only for $f\equiv 1$ and
they measure the distance of $f$
from equilibrium $f=1$. The entropy can be used to
control the total variation norm directly via
the {\it entropy inequality}
\be
   \int |f-1|\rd \mu \le \sqrt{ 2 S_\mu(f)}.
\label{entroneq}
\ee

Let $f_t$ be the solution to  the evolution equation
\be
  \pt_t f_t = Lf_t,  \qquad t>0,
\label{dybe}
\ee
with a given initial condition $f_0$ and consider the
evolution of the entropy $S(f_t)$ and  the Dirichlet form
$D(\sqrt{f_t})$. Simple calculation shows
\be
  \pt_t S(f_t) = \int (Lf_t) \log f_t \rd \mu + \int f_t \frac{Lf_t}{f_t} \rd \mu =
 -\frac{1}{2} \int \frac{(\nabla f_t)^2}{f_t}\rd\mu = -4 D(\sqrt{f_t}),
\label{derS}
\ee
where we used that $\int Lf_t \rd\mu = 0$ by \eqref{symm}.
Similarly, we can compute the evolution of the Dirichlet form. Let $h:=\sqrt{f}$ for
simplicity, then
$$ 
   \pt_t h_t = \frac{1}{2h_t} \pt h_t^2 =  \frac{1}{2h_t} L h_t^2 
 =Lh_t + \frac{1}{2h_t} (\nabla h_t)^2.
$$
In the last step we used that $Lh^2=  (\nabla h)^2 +2 hLh$ that can be seen 
either directly from $L= \frac{1}{2}\Delta -\frac{1}{2}(\nabla\cH)\nabla$ or from the 
following identity for any test function $g$:
$$
  \int g Lh^2 \rd\mu =  -\frac{1}{2}\int \nabla g \cdot \nabla (h^2)\rd\mu
 = -\int h(\nabla g)(\nabla h) \rd\mu = \int [-\nabla(hg) + g\nabla h]\nabla h  
\rd\mu=\int g \Big[ (\nabla h)^2 + 2hLh\Big] \rd \mu.
$$
We compute (dropping the $t$ subscript for brevity)
\begin{align}\label{eq:BE}
   \pt_t D(\sqrt{f_t}) =& \frac{1}{2}\pt_t \int  (\nabla h)^2 \rd \mu \non\\
  = & \int (\nabla h) (\nabla Lh) \rd \mu  + \frac{1}{2}
  \int (\nabla h) \cdot \nabla\frac{ (\nabla h)^2 }{h} \rd \mu 
  \non\\
  =  &  \int (\nabla h)[\nabla,L] h \rd \mu + \int (\nabla h)L (\nabla h) \rd\mu +
  \frac{1}{2}\int \sum_{ij}  \pt_i h
 \Big[  \frac{2(\pt_j h ) \pt_i\pt_j h}{h}  - \frac{(\pt_j h)^2\pt_i h}{ h^2 }\Big] \rd\mu
 \non\\
  = & -  \frac{1}{2} \int (\nabla h)(\nabla^2\cH)\nabla h \rd \mu
 - \frac{1}{2}\int \sum_{ij} (\pt_i\pt_j h)^2   \rd \mu  
 +  \frac{1}{2}\int \sum_{ij}  
 \Big[  \frac{2(\pt_j h)(\pt_i h ) \pt_{ij} h}{h}  - \frac{(\pt_j h)^2(\pt_i h)^2 }{h^2}
 \Big] \rd\mu \non\\
  = & -  \frac{1}{2} \int (\nabla h)(\nabla^2\cH)\nabla h \rd \mu
 - \frac{1}{2}\int \sum_{ij} \Big( \pt_{ij} h -\frac{(\pt_i h)( \pt_j h)}{h}\Big)^2   \rd \mu,
\end{align}
where we used the commutator
$$
   [\nabla, L] = -\frac{1}{2} (\nabla^2\cH)\nabla.
$$
Therefore, under the convexity condition \eqref{convBE}, we have
\be
    \pt_t D(\sqrt{f_t}) \le -K D(\sqrt{f_t}).
\label{derD}
\ee
Combining \eqref{derS} and \eqref{derD},
\be
 \pt_t D(\sqrt{f_t}) \le  \frac{K}{4} \pt_t S(f_t).
\label{ptd}
\ee
At $t=\infty$ the equilibrium is achieved, $f_\infty =1$, and both the entropy
and the Dirichlet form are zero. After 
 integrating \eqref{ptd} back from $t=\infty$, we get the 
{\it logarithmic Sobolev inequality}
\be
   S(f_t) \le \frac{4}{K}  D(\sqrt{f_t})
\label{lsi1}
\ee
for any $t\ge 0$, in particular for any initial distribution $f=f_0$.
Inserting this back to \eqref{derS}, we have
$$
    \pt_t S(f_t) \le -K S(f_t).
$$
Integrating from time zero, we obtain the {\it exponential relaxation
of the entropy on time scale $t\sim 1/K$}
\be
    S(f_t) \le e^{-tK} S(f_0).
\label{Sdec}
\ee
Finally, we can integrate \eqref{derS} from time $t/2$ to $t$ to get
$$
   S(f_t) - S(f_{t/2}) = -4 \int_{t/2}^t D(\sqrt{f_\tau})\rd\tau.
$$
Using the positivity of the entropy $S(f_{t})\ge 0$ on the left side
and the monotonicity of the Dirichlet form (from \eqref{derD}) 
on the right side, we get
\be
   D(\sqrt{f_t})\le \frac{2}{t} S(f_{t/2}),
\label{DS}
\ee
thus, using \eqref{Sdec}, we obtain {\it exponential relaxation
of the Dirichlet form on time scale $t\sim 1/K$}
$$
 D(\sqrt{f_t}) \le \frac{2}{t} e^{-tK/2} S(f_0).
$$
We summarize the result of this calculation:
\begin{theorem}\label{thm:BE}\cite{BE}
Assuming the convexity bound on the Hamiltonian, $\nabla^2 \cH \ge K$ with 
some positive constant $K$, the measure $\mu= e^{-\cH}/Z$ satisfies the logarithmic Sobolev
inequality
\be
   S(f) \le \frac{4}{K}  D(\sqrt{f}), \qquad \mbox{for any density $f$ with}\;\; \int f\rd\mu=1, 
\label{lsi2}
\ee
and the dynamics \eqref{dybe} relaxes to equilibrium on the time scale $t \sim 1/K$ 
both in the sense of entropy and Dirichlet form:
\be
    S(f_t) \le e^{-tK} S(f_0), \qquad
 D(\sqrt{f_t}) \le \frac{2}{t} e^{-tK/2} S(f_0).  
\label{dec}
\ee
\qed
\end{theorem}

{\bf Technical remark.} In our application, the dynamics will be restricted to 
the subset $\Sigma_N=\{ \bx \; : \; x_1<x_2< \ldots < x_N\}$, and
thus  we need to check that
the boundary term in the integration by parts 
\be
 \int_{\pt \Sigma} \pt_i h\; \pt_{ij}^2 h \; e^{-\cH}\rd \bx=0
\label{bdry}
\ee
(from the third line to the fourth line in \eqref{eq:BE}) vanishes.
The role of $\cH$ will be played by $N\cH_N$ where $\cH_N$ is
defined in \eqref{ham}. 
Although the density function of the
measure $e^{-N\cH_N}$ behaves as $(x_{i+1}-x_i)^\beta$
near the $x_{i+1}=x_i$ component of the boundary, hence
it  vanishes at the boundary,
the function $f_t$ is the solution to a parabolic equation 
with a singular drift, so in principle it may blow up at the
boundary. Further complication is that $h=\sqrt{f}$ and
the derivative of the square root is singular.
Nevertheless, by using parabolic regularity theory and
cutoff functions properly subordinated to the geometry
of the set $\Sigma_N$, we can prove that \eqref{bdry} vanishes.
This is one reason why the restriction $\beta\ge 1$ is necessary.
For the details, see Appendix B of \cite{ESYY}.

\subsubsection{Universality of gap distribution of the Dyson Brownian Motion for Short Time}

Using the convexity bound \eqref{conv} for the Dyson Brownian motion,
the Bakry-Emery method  guarantees that $\mu$ satisfies the logarithmic
Sobolev inequality and the relaxation time to 
equilibrium is of order one. 

The following result is the main theorem  of Section \ref{sec:locrelax}.
It shows that the relaxation time is in fact
much shorter than order one at least locally and for observables that
depend only on the eigenvalue {\it differences}. 

\begin{theorem}[Universality of the Dyson Brownian Motion for Short Time]\label{thmM}
\cite[Theorem 4.1]{ESYY} \\
Suppose that the Hamiltonian $\cH$ given in \eqref{ham} satisfies
the convexity bound \eqref{convex} with $\beta\ge1$. Let $f_t$ be the solution
of the forward equation \eqref{dy} so that after time $t_0=N^{-2\fa}$ it 
satisfies  $S_{\mu}(f_{t_0}):=\int f_{t_0}(\log f_{t_0} )\rd\mu \le C N^m$ for some $m$
fixed. Set
\be\label{Q}
  Q:= \sup_{t\ge t_0}    \sum_j \int ( x_j-
\gamma_j)^2 f_t \rd \mu ,
\ee
and  assume that $Q\le CN^m$ with some exponent $m$.
Fix $n\ge 1$ and an array of increasing positive  integers, $\bm = (m_1, m_2, \ldots, m_n)\in
\N^n_+$.
Let  $G:\bR^n\to\bR$ be a bounded smooth function with compact 
support and set
\be
  \cG_{i,\bm}(\bx) := 
G\Big( N(x_i-x_{i+m_1}), N(x_{i+m_1}-x_{i+m_2}), \ldots, N(x_{i+m_{n-1}}-x_{i+m_n})\Big).
\label{cG}
\ee
Then for any sufficiently small $\e'>0$, there exist
constants $C, c>0$, depending only on $\e'$ and $G$ such that
 for any $J\subset \{ 1, 2, \ldots , N-m_n\}$ and any $\tau>3t_0=3N^{-2\fa}$,  we have
\be\label{GG}
\Big| \int \frac 1 N \sum_{i\in J} \cG_{i,\bm}(\bx) f_\tau \rd \mu -
\int \frac 1 N \sum_{i\in J} \cG_{i,\bm}(\bx) \rd\mu \Big|
\le C N^{\e'} \sqrt{|J|Q(N\tau)^{-1}}  + Ce^{-cN^{\e'}} .
\ee
\end{theorem}

In Section \ref{sec:int} we explain the intuition behind the proof.
The precise proof will  be given in
Section~\ref{sec:detailproof}.

\subsubsection{Intuitive proof of  Theorem \ref{thmM}}\label{sec:int}

  The key idea is that we can ``speed up''
the convergence to local equilibrium by modifying the dynamics by
adding an auxiliary  potential $W(\bx)$ to the Hamiltonian.
It will have the form
\be
    W(\bx): =    \sum_{j=1}^N
W_j (x_j)   , \qquad W_j (x) := \frac{1}{2R^2} (x_j -\gamma_j)^2,
\label{defW}
\ee
i.e. it is a quadratic confinement on scale $R$ for each eigenvalue
near its classical location, and we define
\be
\wt \cH: = \cH +W.
\label{def:wth}
\ee
The new measure is denoted by
$$
   \rd\om: =\om (\bx)\rd\bx, \quad \om: = e^{-N\wt\cH}/\wt Z 
$$
and it will be called {\it the pseudo equilibrium  measure}
(with a slight abuse of notations we will denote by $\om$ both
the measure and its density function with respect to the Lebesgue measure).
The corresponding generator is denoted by $\wt L$.
We will typically choose $R\ll 1$, so that the additional term $W$  substantially
increases the lower bound \eqref{conv} on the Hessian, hence speeding up the
dynamics from relaxation time on scale $O(1)$ to $O(R^2)$.
 This is the {\bf first step} of the proof and
it will be formulated in Theorem \ref{thm2}  whose proof basically
follows the Bakry-Emery argument from Section \ref{sec:BEM}.

\medskip

In the {\bf second step}, we consider
an arbitrary probability measure of the form $q \om$, with some function $q$, and 
we control  the difference of expectation values 
\be
     \int \bG q\rd\om - \int \bG \rd\om
\label{bgdiff}
\ee
of the  observables
$$
  \bG: =\frac{1}{N}\sum_{i\in J} \cG_{i,\bm}
$$ 
in terms of the entropy and the Dirichlet form of $q$
with respect to $\om$. Eventually this will enable
us to compare the expectations
$$
     \int \bG\, q\rd\om - \int \bG \, q'\rd\om
$$
for any two measures $q\om$ and $q'\om$, in particular
for the measures $f_\tau\mu$ and $\mu$ that will be written in this form
by defining $q= f_\tau\mu/\om$ and $q'=\mu/\om$.

Here we face with an $N$-problem: both the entropy and the
Dirichlet form are  extensive
quantities.
A direct application of the
entropy inequality \eqref{entroneq} to \eqref{bgdiff} would estimate
the observable $\bG$, an order one quantity,  by a quantity of order $O(\sqrt{N})$.
Instead,
we can run the new dynamics up to some time $\tau$ and write
$$
 \int \bG q\rd\om - \int \bG \rd\om =  \int \bG (q-q_\tau)\rd\om 
+  \int \bG (q_\tau-1)\rd\om.
$$
If $\tau$ is larger than the relaxation time of the new dynamics, then
the second term is exponentially small by the entropy inequality, and
this exponential smallness suppresses the $N$-problem.

To estimate  the first term, we want to capitalize
on the fact that $\tau$ is small. By
integrating back the time derivative, $q_\tau -q= \int_0^\tau \pt_t q_t \,\rd t$,
 we could extract a factor proportional
with $\tau$, but after using $\pt q_t = \wt L q_t$ and integrating by parts
 we will have to differentiate the observable that
brings in an additional $N$ factor due to its scaling.
It seems that this method estimates a quantity of order one
by a quantity of order $N$. However, we have proved 
 new estimate, see \eqref{diff} later, that
controls $\int\bG (q-q_\tau)\rd\om$ by $\big(\tau D_\om(\sqrt{q})/N\big)^{1/2}$;
notice the additional $\frac{1}{N}$ factor. The key reason
for this improvement is that the  dynamics relaxes much faster in certain directions, namely
for observables {\it depending only on differences of $x_i$'s}. 
To extract this mechanism, we use that
 the lower bound \eqref{convex} on the Hessian is of order $N$
in the difference variables $v_i-v_j$ and this estimate
can be used to  gain an additional $N$-factor;
this is the content of Theorem \ref{thm3}.
The estimate will have a free parameter $\tau$ that can be optimized.
This parameter stems from the method of the proof: we prove a 
time independent inequality by a dynamical method i.e.,
we run the flow up to some time $\tau$ and we estimate the 
 $q-q_\tau$ and $q_\tau-q_\infty$ differently.

\medskip

Finally, in the {\bf third step},
we  have to compare the original dynamics with the new one
in the sense of entropy and Dirichlet form since $D_\om (\sqrt{f_\tau\mu/\om})$
and $S_\om(f_\tau\mu/\om)$ need to be computed for the estimates in the second step.
 These would be given 
by the standard decay to equilibrium estimates \eqref{dec} if $f_\tau\mu/\om$ were
evolving with respect to the modified dynamics, but $f_\tau$ evolves
by the original dynamics. We thus need to show that 
the error due to the modification of the dynamics by adding $W$
is negligible.

 It turns out 
that the key quantity that determines how much error was made by
the additional potential is the $H^1$ norm of $W$, i.e.
$$
  \Lambda_t: = \int (\nabla W)^2  f_t \rd\mu.
$$
Due to the explicit form of $W$, we have
\be
  \Lambda_t = R^{-4}\sum_i\int  (x_i-\gamma_i)^2 f_t \rd\mu \le C N^{-2\fa}R^{-4}
\label{Las}
\ee
using Assumption III \eqref{assum2}. Given $\fa>0$, we can therefore choose
an $R\ll 1$ so that we still have $\Lambda\ll 1$.  This will complete the proof.

\bigskip

Note that 
the speed of convergence is determined by the {\it second} derivative
of the auxiliary potential, while the modification
in the Dirichlet form and the entropy is determined by the $(\nabla W)^2$.
So one can speed up the dynamics and still compare Dirichlet forms
and entropies of the two equilibrium measures if a strong
apriori bound \eqref{Las} on $\Lambda$ is given. This is one of the 
reasons why the method works.

The other key observation is the effective use of the convexity bound \eqref{convex}
which exploits a crucial property of the dynamics of the Dyson Brownian motion \eqref{sde}.
The logarithmic interaction potential gives rise to a singular  force 
\be
   F(x_i)=-\frac{1}{4}x_i-\frac{1}{2N} \sum_{j\ne i} \frac{1}{x_i-x_j}
\label{Fxi}
\ee
acting on the $i$-th particle. Formally $F(x_i)$ is a mean field
force, and if $x_j$ were distributed according to the semicircle law, then
the bulk of the sum would cancel the $-\frac{1}{4} x_i$ term.
However, the effect of the neighboring particles, $j=i\pm 1$, is huge: they exert a
force of order one on $x_i$. Such a force may move the particle $x_i$ by
a distance of order  $1/N$
within a very short time of order $1/N$. Note that by the order preserving
property of the dynamics, an interval of size $O(1/N)$ is roughly the whole
space available for $x_i$, at least in the bulk.
Thus $x_i$ is likely to relax to its equilibrium within a time scale of order $1/N$
 due to the strong repulsive force from its
neighbors.
Of course this argument is not a proof since
the other particles move as well. However, our observables
involve only eigenvalue differences and in the difference
coordinates the strong drift is persistently present.
This indicates that the eigenvalue differences may relax to
their local equilibrium on a time scale almost of order $1/N$.

For other observables, the relaxation time is necessary longer.
In particular, it can happen
that there is no precise cancellation from the bulk in \eqref{Fxi},
in which case the neighboring particles all feel the same mean field drift
and will move collectively. In fact, if the initial density profile substantially
differs from the semicircle, the relaxation to the semicircle
 may even take order one time
(although such scenario is excluded in our case by \eqref{assum2}).

\subsubsection{Detailed proof of Theorem \ref{thmM}}\label{sec:detailproof}

Every constant in this proof depends on $\e'$ and $G$, and
we will not follow the precise dependence.
Given $\tau >0$, we define $R:= \tau^{1/2} N^{-\e'/2}$.

We now introduce the pseudo equilibrium measure, $\om_N=\om= \psi\mu$,
defined by
\[
\psi:=\frac{Z}{\wt Z}\exp \big(-NW \big),  
\]
where $\wt Z$ is chosen such that $\om$ is a probability measure,
in particular $\omega= e^{-N\wt \cH}/\wt Z$ with
$$
\wt \cH = \cH +W.
$$
The potential $W$ was defined in \eqref{defW} and it
confines
the $j$-th point $x_j$ near its classical location $\gamma_j$.

The local relaxation flow is defined to be  the reversible dynamics w.r.t.
$\omega$.
The dynamics is described by the generator  $\wt L$ defined by
\be\label{Lt}
\int f \wt L  g \rd \omega = - \frac 1 {2N} \sum_j
\int (\partial_j f)( \partial_j g) \rd \omega .
\ee
Explicitly,  $\wt L$ is given by
\be\label{tl}
\wt L = L - \sum_j b_j \partial_j, \quad
b_j = W_j'(x_j)= \frac{x_j -\gamma_j}{R^2}.
\ee
Since the additional potential $W_j$ is uniformly convex with
\be\label{5.9}
\inf_j \inf_{ x \in \bR}     W_j^{\prime \prime}(x)
\ge R^{-2},
\ee
by \eqref{convex} and $\beta \ge 1$ we have
\be\label{convex1}
\Big\langle \bv , \nabla^2 \wt\cH(\bx)\bv\Big\rangle
\ge   \frac{1}{R^2} \,  \|\bv\|^2 + \frac{1}{N}
 \sum_{i<j} \frac{(v_i - v_j)^2}{(x_i-x_j)^2}, \qquad \bv\in\bR^N.
\ee
The $R^{-2}$ in the first term comes from the additional convexity of
the local interaction and it enhances the local
Dirichlet form dissipation.
In particular we have the uniform lower bound
$$
\nabla^2\wt \cH =\mbox{Hess}  (- \log \omega) \ge   R^{-2}.
$$
This guarantees that the relaxation time to equilibrium $\om$
for the $\wt L$ dynamics is bounded above by  $C R^2$.

\bigskip

The first ingredient to prove Theorem \ref{thmM} is the analysis of
the local relaxation flow which  satisfies the logarithmic
Sobolev inequality and the following
dissipation estimate.

\begin{theorem}\label{thm2}
Suppose \eqref{convex1}  holds.
Consider the forward equation
\be
\partial_t q_t=\wt L q_t, \qquad t\ge 0,
\label{dytilde}
\ee
with initial condition $q_0=q$ and
with reversible measure $\omega$. Assume that $q\in L^\infty(\rd\om)$.
Then we have the following estimates
\be\label{0.1}
\partial_t D_{\omega}( \sqrt {q_t}) \le - \frac{1}{R^2}D_{\omega}( \sqrt {q_t}) -
\frac{1}{2N^2}  \int    \sum_{i,j=1}^N
\frac{  ( \pt_i \sqrt{ q_t} - \pt_j\sqrt {q_t} )^2}{(x_i-x_j)^2} \rd \omega ,
\ee
\be\label{0.2}
\frac{1}{2N^2} \int_0^\infty  \rd s  \int    \sum_{ i,j=1}^N
\frac{(\pt_i\sqrt {q_s} - \pt_j\sqrt {q_s} )^2 }{(x_i-x_j)^2}\rd \omega
\le D_{\omega}( \sqrt {q})
\ee
and the logarithmic Sobolev inequality
\be\label{lsi}
S_{\omega}(q)\le C R^2  D_{\omega}( \sqrt {q})
\ee
with a universal constant $C$.
Thus the time to equilibrium is of order $R^2$:
\be\label{Sdecay}
 S_{\omega}(q_t)\le e^{-Ct/R^2} S_\omega(q).
\ee
\end{theorem}

{\it Proof.}  This theorem can be proved   following the standard argument 
presented in Section \ref{sec:BEM}. The key additional input is
the convexity bound \eqref{convex} which gives rise to the second term
on the r.h.s of \eqref{0.1} from the last line of \eqref{eq:BE}.  
In particular, this serves as an upper bound on $\partial_t D_{\omega}( \sqrt {q_t})$,
thus integrating \eqref{0.1} one obtains \eqref{0.2} \qed

\medskip

The estimate \eqref{0.2} on the second  term in \eqref{convex1}
plays a key role in the next theorem.

\begin{theorem}\label{thm3}
Suppose that Assumption I holds and let
 $q\in L^\infty$ be a density, $\int q\rd\om=1$.
Fix $n\ge 1$, $\bm\in \cN_+^n$, 
let  $G:\bR^n\to\bR$ be a bounded smooth function with compact support
and recall the definition of $\cG_{i,\bm}$ from \eqref{cG}.
Then  for any $J\subset \{ 1, 2, \ldots , N-m_n\}$ and any $\tau>0$ we have
\be\label{diff}
\Big| \int \frac 1 N \sum_{i\in J} \cG_{i,\bm}(\bx) q \rd \omega -
\int \frac 1 N \sum_{i\in J} \cG_{i,\bm}(\bx) \rd \omega \Big|
\le C \Big( \frac {|J| D_\omega (\sqrt {q})  \tau}{N^2}  \Big)^{1/2}  + C \sqrt
{S_\om(q)} e^{-c\tau/R^2} .
\ee
\end{theorem}

{\it Proof.} For simplicity, we will consider the case when $\bm =(1,2,\ldots, n)$,
the general case easily follows by appropriately redefining the  function $G$.
Let $q_t$ satisfy
\[
\partial_t q_t = \wt L q_t, \qquad t\ge 0,
\]
with an initial condition $q_0=q$.
We write
\be
   \int \Big[ \frac 1 N \sum_{i\in J} \cG_{i,\bm}\Big] (q-1)\rd\om
  =  \int \Big[ \frac 1 N \sum_{i\in J} \cG_{i,\bm}\Big] (q-q_\tau)\rd\om
 +   \int \Big[ \frac 1 N \sum_{i\in J} \cG_{i,\bm}\Big] (q_\tau-1)\rd\om.
\label{splitt}
\ee
The second term can be estimated by the entropy inequality \eqref{entroneq},
 by the decay of the entropy \eqref{Sdecay} and the boundedness of $G$,
this gives the second term in \eqref{diff}.

To estimate the first term in \eqref{splitt},
we have, by differentiation, by $\pt q_t =\wt L q_t$ and by \eqref{Lt} 
\begin{align}
\int \frac 1 N \sum_i &\cG_{i,\bm}(\bx)q_\tau \rd \omega  -
\int \frac 1 N \sum_i \cG_{i,\bm}(\bx)q_0 \rd \omega \non\\
&= \int_0^\tau  \rd s \int  \frac 1 N \sum_i \sum_{k=1}^n
   \pt_k G\Big( N(x_i-x_{i+1}), \ldots, N(x_{i+n-1}-x_{i+n})\Big)
[\pt_{i+k-1} q_s - \pt_{i+k}q_s]  \rd \omega. \non
\end{align}
{F}rom the Schwarz inequality and $\pt q = 2 \sqrt{q}\pt\sqrt{q}$,
the last term is bounded by
\begin{align}\label{4.1}
2\sum_{k=1}^n & \left [   \int_0^\tau  \rd s \int
\sum_i \Big[\pt_k G \big(N(x_i - x_{i +1}),\ldots, N(x_{i+n-1}-x_{i+n}) \big)\Big] ^2 
(x_{i+k-1}-x_{i+k})^2  \, q_s \rd \omega
\right ]^{1/2} \nonumber \\
& \times \left [ \int_0^\tau  \rd s \int  \frac 1 {N^2 } \sum_i
\frac{1}{(x_{i+k-1}-x_{i+k})^2}  [ \pt_{i+k-1}\sqrt {q_s} -
\pt_{i+k}\sqrt {q_s}]^2  \rd \omega \right ]^{1/2} \nonumber \\
\le &  \; C \Big( \frac{|J|D_\omega(\sqrt {q_0}) \tau}{N^2}\Big)^{1/2},
\end{align}
where we have used \eqref{0.2} and that
$$
  \Big[\pt_k G \Big(N(x_i - x_{i +1}), \ldots, N(x_{i+k-1}-x_{i+k}), \ldots
 N(x_{i+n-1}-x_{i+n}) \Big) \Big]^2
 (x_{i+k-1}-x_{i+k})^2 \le CN^{-2},
$$
since  $G$ is smooth and  compactly
supported.
\qed

\bigskip

As a comparison to Theorem \ref{thm3}, we state the following result which can
be proved in a similar way.

\begin{lemma}\label{thm4}
Let  $G:\bR\to\bR$
 be a bounded smooth function with compact support and let a sequence
$E_i$ be fixed.
Then  we have, for any $\tau>0$,
\be\label{Diff}
\Big|\frac 1 N \sum_{i}   \int G\big( N(x_i-E_i )\big)q \rd \omega -
\frac 1 N \sum_{i}\int  G\big(N (x_i-E_i )\big)   \rd \omega \Big|
\le C\sqrt{ S_\omega (q) \tau}  + C\sqrt{S_\om(q)}
e^{-c\tau/R^2}.
\ee
\end{lemma}

Notice that  by exploiting the local Dirichlet form dissipation
coming from the second term on the r.h.s. of \eqref{0.1},
we have gained the crucial  factor $N^{-1/2}$
in the estimate \eqref{diff} compared with \eqref{Diff}.

\bigskip

The final  ingredient to prove Theorem \ref{thmM} is the following entropy and
Dirichlet form estimates.

\begin{theorem}\label{thm1}
Suppose that  \eqref{convex} holds and recall $\tau = R^2 N^{\e'}\ge 3t_0$ with $t_0=N^{-2\fa}$.
Assume that $S_\mu(f_{t_0}) \le CN^m$ with some fixed $m$.
Let  $g_t: = f_t/\psi$.
Then the entropy and the Dirichlet form satisfy the estimates:
\be\label{1.3}
S_{\omega} (g_{\tau/2}) \le
 C   N R^{-2} Q, \qquad
D_\omega (\sqrt{g_\tau})
\le CN R^{-4}Q.
\ee
\end{theorem}

{\it Proof.}  Recall that  $\pt_t f_t = Lf_t$.
The standard estimate on the entropy of $f_t$ with respect
to the invariant measure  is obtained by differentiating
the entropy twice and using the logarithmic Sobolev inequality
(see Section \ref{sec:BEM}).
The entropy
and the Dirichlet form in \eqref{1.3} are, however, computed
with respect to the measure $\om$. This yields the additional
second  term in the identity \eqref{derS} and we use
 the following identity  \cite{Y} that holds for
any probability density
$\psi_t$:
$$
\partial_t  S_\mu(f_t|\psi_t) = -  \frac{2}{N}  \sum_{j} \int (\partial_j
\sqrt {g_t})^2  \, \psi_t \, \rd\mu
+\int g_t(L-\partial_t)\psi_t \, \rd\mu \ ,
$$
where $g_t: = f_t/\psi_t$ and
$$  
S_\mu(f_t|\psi_t): =   \int f_t \log \frac{f_t}{\psi_t} \rd\mu
$$
is the relative entropy. 

 In our application we set $\psi_t$ to be time independent,
$\psi_t = \psi =\om/\mu$,
hence $S_\mu(f_t|\psi)= S_{\omega} (g_t)$ and
we have, by using \eqref{tl},
$$
\pt_t S_\omega(g_t)
= - \frac{ 2}{N}  \sum_{j} \int (\partial_j
\sqrt {g_t})^2  \, \rd\omega
+\int   \wt L g_t  \, \rd\omega+  \sum_{j} \int  b_j \partial_j
g_t   \, \rd\omega.
$$
Since $\om$ is $\wt L$-invariant, the middle term on the right hand side vanishes, and
from the Schwarz inequality
\be\label{1.1}
\pt_t S_\omega(g_t) \le  -D_{\omega} (\sqrt {g_t})
+   C  N\sum_{j} \int  b_j^2  g_t   \, \rd\omega \le
-D_{\omega} (\sqrt {g_t})
+   C  N\Lambda,
\ee
where we defined
\be\label{bbound}
\Lambda:= Q R^{-4} = \sup_{t\ge 0}  R^{-4}  \sum_j \int ( x_j-
\gamma_j)^2 f_t \rd \mu .
\ee

Together with  \eqref{lsi},  we have
\be\label{1.2new}
\partial_t  S_\omega(g_t) \le  - CR^{-2}  S_\omega(g_t) +
 C  N \Lambda, \qquad t\ge N^{-2\fa}.
\ee
To obtain the first inequality in  \eqref{1.3},
we integrate \eqref{1.2new} from $t_0=N^{-2\fa}$ to $\tau/2$, using that $\tau= R^{2} N^{\e'}$ and
$S_\om(g_{t_0})\le CN^m+N^2Q$ with some finite $m$, depending on $\fa$. This apriori bound 
follows from 
\be
  S_\om(g_{t_0}) = S_\mu(f_{t_0}|\psi) =
 S_\mu(f_{t_0}) - \log Z + \log \wt Z + N \int f_{t_0} W\rd\mu
 \le CN^m + N^2Q,
\label{inentr}
\ee
where we used that $|\log Z|\le CN^m$ and $|\log \wt Z|\le CN^m$, which
can be easily checked.
The second inequality in \eqref{1.3} can be obtained from the first one
by integrating  \eqref{1.1}
from $t=\tau/2$ to $t=\tau$ and using
the monotonicity of the Dirichlet form in time. \qed

\bigskip

Finally, we complete the proof of Theorem \ref{thmM}.
Recall that 
$\tau = R^2 N^{\e'}$ and $t_0=N^{-2\fa}$.
Choose $q_\tau:= g_{\tau}= f_{\tau}/\psi$ as density $q$ in
Theorem \ref{thm3}.
The condition $q_\tau\in L^\infty$ can
be guaranteed by an approximation argument.
 Then  Theorem \ref{thm1}, Theorem \ref{thm3} together with \eqref{inentr} and the 
fact that $\Lambda \tau  = Q\tau^{-1}N^{2\e'}$
directly imply that 
\be\label{diff1}
\Big| \int \frac 1 N \sum_{i\in J} \cG_{i,\bm} f_\tau\rd \mu -
\int \frac 1 N \sum_{i\in J} \cG_{i,\bm}  \rd \omega \Big| \\
\le C N^{\e'}\sqrt{|J|Q (\tau N)^{-1} }
+  Ce^{-cN^{\e'}} ,
\ee
i.e., the local statistics of $f_\tau \mu$ and $\om$
can be compared.
Clearly, equation  \eqref{diff1}  also holds for the special choice
$f_0= 1$ (for which $f_\tau=1$), i.e., local statistics of $\mu$ and
$\om$ can also be compared. This completes the proof of Theorem \ref{thmM}.
\qed

\subsection{From gap distribution to correlation functions: 
Sketch of the proof of Theorem \ref{thm:main}.} 

Our main result Theorem~\ref{thm:main}  will
 follow from Theorem \ref{thmM} and
from the fact that in case $\tau\ge N^{-2\fa+\delta}$, the assumption
\eqref{assum2} guarantees that
$$
N^{\e'}\sqrt{|J|Q(\tau N)^{-1}} \le  N^{\e'-\delta/2} = N^{-\delta/6}\to 0
$$
with the choice $\e'=\delta/3$ and using $|J|\le N$. Therefore the local statistics
of observables involving eigenvalue differences coincide
in the $N\to\infty$ limit.

To complete the proof of  Theorem \ref{thm:main}, we will
have to show that the convergence of the
 observables  $\cG_{i,\bm}$ is sufficient to identify
the correlation functions of the $x_i$'s in the sense prescribed in 
 Theorem \ref{thm:main}.  This is a fairly standard technical argument,
and the details will be given in 
Appendix \ref{sec:corfn}.  Here we just summarize the main points.

Theorem \ref{thmM} detects the joint behavior of eigenvalue differences
on the correct scale $1/N$, due to 
 the factor $N$ in the argument of $\cG$ in \eqref{cG}.
The slight technical problem is that the observable \eqref{cG}, and its
averaged version \eqref{GG}, involve
{\it fixed indices} of eigenvalues, while correlation functions 
involve cumulative statistics.
 
To understand this subtlety, 
consider $n=1$ for simplicity and let $m_1=1$, say.
The observable \eqref{GG} answers to the question: ``What is the
empirical distribution of differences of consecutive eigenvalues?'',
in other words \eqref{GG} directly identifies the {\it gap distribution}
defined in Section \ref{sec:sinegap}. Correlation functions answer to
the question: ``What is the probability that there are two eigenvalues
at a  fixed distance away from each other?'', in other words they 
are not directly sensitive to the possible other eigenvalues in between.
Of course these two questions are closely related and it is easy
to deduce the answers from each other. This is exactly what 
the calculation \eqref{ex} has achieved in one direction,
now we need to go in the other direction: identify correlation
functions from (generalized) gap distributions. 

In fact this
direction is easier and the essence is given by
the following formula:
\begin{align}
\int_{E-b}^{E+b}\frac{\rd E'}{2b} \int_{\R^n} & \rd\alpha_1
\ldots \rd\alpha_n  \; O(\alpha_1,\ldots,\alpha_n)  p_{\tau, N}^{(n)}
\Big (E'+\frac{\alpha_1}{N\varrho(E)},
\ldots, E'+\frac{\alpha_n}{N \varrho(E)}\Big) \label{iddd1} \\
 = & C_{N,n}\int_{E-b}^{E+b}\frac{\rd E'}{2b} \int \sum_{i_1\ne i_2\ne
 \ldots \ne i_n}
  \wt O\big( N(x_{i_1}-E'),  N(x_{i_1}-x_{i_2}),
 \ldots  N(x_{i_{n-1}}-x_{i_n})\big)
 f_\tau\rd\mu, \non \\
= & C_{N,n}\int_{E-b}^{E+b}\frac{\rd E'}{2b} \int \sum_{{\bf m}\in S_n}\sum_{i=1}^N  Y_{i,\bm}
(E',\bx),
\non 
\end{align}
with $C_{N,n}:=N^n(N-n)!/N! =1+O_n(N^{-1})$,
where we let  $S_n$  denote the set of  
increasing positive integers, ${\bf m} = (m_2, m_3, \ldots, m_n) \in \N_+^{n-1}$, 
$m_2< m_3 <\ldots < m_n$, and we
 introduced
\begin{align}
   Y_{i,\bm}(E',\bx) := & \wt O\big(  N(x_i - E'),
N(x_i - x_{i+m_2}) ,\ldots,  N(x_i- x_{i+m_n})   \big) \non\\
\wt O (u_1, u_2, \ldots u_n): = &
O\big( \varrho(E)u_1, \varrho(E)(u_1-u_2), \ldots\big)
\label{Ydef}
\end{align}
We will set $Y_{i,\bm}=0$ if $i+m_n>N$.
The first equality of \eqref{iddd1} is just the definition of the correlation
function after a trivial rescaling. From the second to the third line 
first noticed that by permutational symmetry of $p_{\tau, N}^{(n)}$ 
we can assume that $O$ is symmetric and thus we can restrict
the summation to $i_1<i_2<\ldots < i_n$ upon an overall factor $n!$.
Then we
changed indices $i= i_1$, $i_2= i+m_2$, $i_3=i+m_3, \ldots,$
and performed a resummation over all index differences encoded in $\bm$.
Apart from the first variable $N(x_{i_1}-E')$, the function $ Y_{i,\bm}$
is of the form \eqref{cG}, so Theorem \ref{thmM} will apply. The dependence
on the first variable will be negligble after the $\rd E'$ integration
on a macroscopic interval.

To control the error terms in this argument, especially to show that
even the error terms in the potentially infinite sum over $\bm\in S_n$ 
converge, one needs an apriori bound on the local density. This
is where Assumption IV \eqref{ass4} is used. For the details,
see Appendix \ref{sec:corfn}. \qed

\section{The Green function comparison theorems}\label{sec:4mom}

A simplified version of the Green function comparison theorem was
already stated in Theorem~\ref{thm:4mom}, here we state 
the complete version, Theorem~\ref{comparison}.
It will lead quickly to Theorem~\ref{com} stating that 
the correlation functions  of eigenvalues of two matrix ensembles 
are identical  on scale $1/N$ provided that the first four moments 
of all  matrix elements  of these two ensembles are almost identical.
Here we do not assume that the real and imaginary parts are i.i.d.,
hence the $k$-th moment of $h_{ij}$ is understood as the collection of numbers
$\int \bar h^s h^{k-s}\nu_{ij}(\rd  h)$, $s=0,1,2,\ldots,k$.  
The related Theorem~\ref{thm:TV} from
 \cite{TV} compares the joint distribution of individual eigenvalues ---
which is not covered by our Theorem \ref{comparison} --- 
but it does not address directly the matrix elements of 
Green functions. In Section~\ref{sk:TV} we will sketch some
ideas of the proof of Theorem~\ref{thm:TV} to point out the differences between
the two results.
The key input for both theorems is  the local semicircle law on
 the almost optimal scale $N^{-1+\e}$.
The eigenvalue perturbation  used in  Theorem~\ref{thm:TV}
  requires certain estimates on the  level repulsion; 
the proof of Theorem \ref{comparison} is a straightforward  resolvent perturbation theory.

\begin{theorem}[Green function comparison]\label{comparison}\cite[Theorem 2.3]{EYY}   Suppose that we have  
two generalized $N\times N$ Wigner matrices, $H^{(v)}$ 
and $H^{(w)}$, with matrix elements $h_{ij}$
given by the random variables $N^{-1/2} v_{ij}$ and 
$N^{-1/2} w_{ij}$, respectively, with $v_{ij}$ and $w_{ij}$ satisfying
the uniform subexponential decay condition
$$
   \P \big( |v_{ij}|\ge x )\le C\exp\big(-x^\ttau\big),  \qquad
  \P \big( |w_{ij}|\ge x)\le  C\exp\big(-x^\ttau\big),
$$
with some $C, \ttau>0$.
  Fix a bijective ordering map on the index set of
the independent matrix elements,
\[
\phi: \{(i, j): 1\le i\le  j \le N \} \to \Big\{1, \ldots, \gamma(N)\Big\} , 
\qquad \gamma(N): =\frac{N(N+1)}{2},
\] 
and denote by  $H_\gamma$  the generalized Wigner matrix whose matrix 
elements $h_{ij}$ follow
the $v$-distribution if $\phi(i,j)\le \gamma$ and they follow the $w$-distribution
otherwise; in particular $H^{(v)}= H_0$ and $H^{(w)}= H_{\gamma(N)}$. 
Let $\kappa>0$ be arbitrary
 and suppose that, for any small parameter $\tau>0$ 
and for any  $y \ge N^{-1 + \tau}$,  we have 
the following estimate on the diagonal elements of the resolvent
\be\label{basic}
\P\left(\max_{0 \le \gamma \le \gamma(N)} \max_{1 \le k \le N}  
 \max_{|E|\le 2-\kappa}\left |  \left (\frac 1 {  H_{\gamma}-E- i y} \right )_{k k } 
\right |\le N^{2\tau} \right)\geq 1-CN^{-c\log\log N}
\ee
with some constants $C, c$ depending only on $\tau, \kappa$.
 Moreover, we assume that the first three moments of
 $v_{ij}$ and $w_{ij}$ are the same, i.e.
$$
    \E \ov v_{ij}^s v_{ij}^{u} =  \E \ov w_{ij}^s w_{ij}^{u},
  \qquad 0\le s+u\le 3,
$$
 and the difference between the  fourth moments of 
 $v_{ij}$ and $w_{ij}$ is much less than 1, say
\be\label{4match}
\left|\E \ov v_{ij}^s v_{ij}^{4-s}- \E \ov w_{ij}^s w_{ij}^{4-s}
\right|\leq N^{-\delta}, \qquad s=0,1,2,3,4,
\ee
for some given $\delta>0$.  Let $\e>0$ be arbitrary and choose an 
$\eta$ with $N^{-1-\e}\le \eta\le N^{-1}$.
For any  sequence of positive integers $k_1, \ldots, k_n$, set  complex parameters
$z^m_j = E^m_j \pm i \eta$,   $j = 1, \ldots k_m$, $m = 1, \ldots, n$,
 with $ |E^m_j| \le2-2\kappa $
and with an arbitrary choice of the $\pm$ signs. 
Let  $G^{(v)}(z) =  ( H^{(v)}-z)^{-1}$ denote the resolvent
and let    $F(x_1, \ldots, x_n)$ be a function such that for
any multi-index $\al=(\al_1, \ldots ,\al_n)$ with $1\le |\al|\le 5$
and for any $\e'>0$  sufficiently small, we have
\be\label{lowder}
\max \left\{|\partial^{\al}F(x_1, \ldots, x_n)|: 
\max_j|x_j|\leq N^{\e'}\right\}\leq N^{C_0\e'}
\ee
and
\be\label{highder}
\max\left\{|\partial^\al F(x_1, \ldots, x_n)|:
 \max_j|x_j|\leq N^2\right\}\leq N^{C_0}
\ee
for some constant $C_0$.

Then, there is a constant $C_1$,
depending on $\ttau$, $\sum_m k_m$ and $C_0$ such that for any $\eta$ with
$N^{-1-\e}\le \eta\le N^{-1}$
and  for any choices of the signs 
in  the imaginary part of $z^m_j$, we have 
\begin{align}\label{maincomp}
\Bigg|\E F  \left (  \frac{1}{N}\tr  
\left[\prod_{j=1}^{k_1} G^{(v)}(z^1_{j})\right ]  , \ldots, 
 \frac{1}{N} \tr \left [  \prod_{j=1}^{k_n} G^{(v)}(z^n_{j}) \right ]  \right )  &  -
  \E F\left ( G^{(v)} \to  G^{(w)}\right )  \Bigg| \non\\
\le & C_1 N^{-1/2 + C_1 \e}+C_1 N^{-\delta+ C_1 \e},
\end{align}
where the arguments of $F$ in the second term are changed from the Green functions of $H^{(v)}$
to $H^{(w)}$ and all other parameters remain unchanged.
\end{theorem}

{\it Remark 1:}   We formulated Theorem \ref{comparison} for functions of
traces of monomials of the Green function because this is the form
we need in the application. However, the result (and the proof we are going to present)
holds directly for
matrix elements of monomials of Green functions as well, 
for the precise statement, see \cite{EYY}.
We also remark that  Theorem \ref{comparison} holds for generalized
Wigner matrices if $C_{sup}= \sup_{ij} N\sigma_{ij}^2 <\infty$.
 The positive lower
bound on the variances, $C_{inf}>0$ in \eqref{VV} is not necessary for this theorem.

\bigskip 

{\it Remark 2:}  Although we state Theorem \ref{comparison} for 
hermitian and symmetric ensembles, 
similar results  hold for real and complex sample covariance ensembles;  
the modification of the proof is  obvious.

\medskip

The following result is the main corollary of Theorem \ref{comparison} which  will be proven
later in the section. The important statement is Theorem \ref{comparison}, the proof of its corollary
is a fairly straightforward technicality.

\begin{theorem}[Correlation function comparison]\label{com} \cite[Theorem 6.4]{EYY}
Suppose the assumptions of Theorem \ref{comparison} hold. 
Let $p_{v, N}^{(k)}$ and $p_{w, N}^{(k)}$
be the  $k-$point functions of the eigenvalues w.r.t. the probability law of the matrix $H^{(v)}$
and $H^{(w)}$, respectively. 
Then for any $|E| < 2$,  any
$k\ge 1$ and  any compactly supported continuous test function
$O:\bR^k\to \bR$ we have  
\be \label{6.3}
\int_{\R^k}  \rd\alpha_1 
\ldots \rd\alpha_k \; O(\alpha_1,\ldots,\alpha_k) 
   \Big ( p_{v, N}^{(k)}  - p_{w, N} ^{(k)} \Big )
  \Big (E+\frac{\alpha_1}{N}, 
\ldots, E+\frac{\alpha_k}{N }\Big) =0.
\ee
\end{theorem} 

\subsection{Proof of the Green function comparison Theorem \ref{comparison}}

  The basic idea is that we
 estimate the effect of changing matrix elements of the resolvent one by one 
by a resolvent expansion. Since each matrix element has a typical size of $N^{-1/2}$
and the resolvents are essentially bounded thanks to \eqref{basic},
a resolvent  expansion up to the fourth order will identify the change of 
each element with a precision
$O(N^{-5/2})$ (modulo some tiny corrections of order  $N^{O(\tau)}$). 
The expectation values of the terms up to fourth order involve
only the first four moments of the single entry distribution, which can 
be directly compared. The error terms are negligible even when we sum them
up $N^2$ times, the number of comparison steps needed to replace all matrix elements.

\medskip

To start the detailed proof, we first need an estimate of the type \eqref{basic}
for the resolvents of all intermediate matrices.
 {F}rom  
the trivial bound 
\[
 \im\,  \left ( \frac 1 {H-E- i\eta} \right )_{jj} \le  \left ( \frac {y} {  \eta} \right )\,  \im\,
  \left ( \frac 1 {H-E-iy} \right )_{jj} , \qquad \eta\le y,
 \] 
and from  \eqref{basic}
we have the following apriori bound
\be\label{basic3}
\P\left(\max_{0 \le \gamma \le \gamma(N)} \max_{1 \le k \le N}  \max_{|E|\le 2-\kappa}
\sup_{\eta\ge N^{-1-\e}} \left |  \im\, \left (\frac 1 { H_{\gamma}-E\pm i\eta} \right )_{k k } 
\right | \le N^{3\tau+ \e} \right)\geq 1-CN^{-c\log\log N}.
\ee
Note that the supremum over $\eta$ can be included by establishing the estimate first
for a fine grid of $\eta$'s with spacing $N^{-10}$ and then extend the bound for all $\eta$ by
using that the Green functions are Lipschitz continuous in $\eta$
with a Lipschitz constant $\eta^{-2}$.

Let $\lambda_m$ and $u_m$ denote the eigenvalues and eigenvectors
of $H_\gamma$, then by  the definition of the Green function,   we have 
\[
\left | \left ( \frac 1 {H_\gamma-z} \right )_{jk} \right | 
\le  \sum_{m = 1}^N  \frac {| u_{m}(j)| |  u_m(k)| }{| \lambda_m -z |} 
\le \left [ \sum_{m = 1}^N  \frac {| u_{m}(j)|^2  }{| \lambda_m-z |}  \right ]^{1/2} 
\left [ \sum_{m = 1}^N  \frac {| u_{m}(k)|^2  }{| \lambda_m -z  |}  \right ]^{1/2} .
\]
Define a dyadic decomposition
\be
U_n = \{m:  2^{n-1} \eta \le |\lambda_m - E|<  2^{n} \eta \}, \qquad n=1,2,\ldots, n_0:=C\log N,
\label{dya}
\ee
\[ 
   U_0 = \{m:    |\lambda_m - E|< \eta \},\qquad U_\infty:=
\{m:  2^{n_0} \eta \le |\lambda_m - E| \},
\]
and divide the summation over $m$ into $\cup_n  U_n$
\[
\sum_{m = 1}^N  \frac {| u_{m}(j)|^2  }{| \lambda_m-z  |} 
= \sum_n  \sum_{m \in U_n} \frac {| u_{m}(j)|^2  }{| \lambda_m-z |}   
\le C  \sum_n  \sum_{m \in U_n}  \im \, \frac {| u_{m}(j)|^2  }{ \lambda_m -E - i 2^n \eta }  
\le C  \sum_n   \im \, \left ( \frac 1 {H_\gamma -E - i 2^n \eta} \right )_{jj}. 
\]
Using the estimate \eqref{basic} for $n=0,1, \ldots, n_0$ and a trivial bound of $O(1)$
for $n=\infty$, we have proved that
\be\label{basic4}
\P\left(\sup_{0 \le \gamma \le \gamma(N)} \sup_{1 \le k, \ell \le N}  \max_{|E|\le 2-\kappa}
\sup_{\eta\ge N^{-1-\e}}
  \left |  \left (\frac 1 { H_{\gamma}-E\pm i\eta} \right )_{k \ell } 
\right | \le N^{4\tau+ \e} \right)\geq 1-CN^{-c\log\log N}.
\ee

\medskip
Now we turn to the one by one replacement.
For notational simplicity, we will consider the case when the test function $F$ has only $n=1$
variable and $k_1=1$, i.e., we consider the trace of a first order monomial;
the general case follows analogously.  Consider the telescopic sum of 
differences of expectations 
\begin{align}\label{tel}
\E \, F \left ( \frac{1}{N}\tr  \frac 1 {H^{(v)}-z} \right )   - 
 & \E \, F \left  (  \frac{1}{N}\tr  \frac  1 {H^{(w)}-z} 
\right )  \\
= & \sum_{\gamma=1}^{\gamma(N)}\left[  \E \, 
F \left (  \frac{1}{N}\tr \frac 1 { H_\gamma-z} \right ) 
-  \E \, F \left (  \frac{1}{N}\tr \frac  1 { H_{\gamma-1}-z} \right ) \right] . \non
\end{align}
Let $E^{(ij)}$ denote the matrix whose matrix elements are zero everywhere except
at the $(i,j)$ position, where it is 1, i.e.,  $E^{(ij)}_{k\ell}=\delta_{ik}\delta_{j\ell}$.
Fix an $\gamma\ge 1$ and let $(i,j)$ be determined by  $\phi (i, j) = \gamma$.
We will compare $H_{\gamma-1}$ with $H_\gamma$.
Note that these two matrices differ only in the $(i,j)$ and $(j,i)$ matrix elements 
and they can be written as
$$
    H_{\gamma-1} = Q + \frac{1}{\sqrt{N}}V, \qquad V:= v_{ij}E^{(ij)}
+ v_{ji}  E^{(ji)}
$$
$$
    H_\gamma = Q + \frac{1}{\sqrt{N}} W, \qquad W:= w_{ij}E^{(ij)} +
   w_{ji} E^{(ji)},
$$
with a matrix $Q$ that has zero matrix element at the $(i,j)$ and $(j,i)$ positions and
where we set $v_{ji}:= \ov v_{ij}$ for $i<j$ and similarly for $w$.
Define the  Green functions
$$
      R := \frac{1}{Q-z}, \qquad S:= \frac{1}{H_\gamma-z}.
$$

We first claim that the estimate \eqref{basic4} holds for the Green function $R$ as well. 
To see this, we have, from the 
resolvent expansion, 
\[
 R = S  +   N^{-1/2}SV S + \ldots + N^{-9/5} (SV)^9S+
N^{-5} (SV)^{10} R.
\] 
Since $V$ has only at most two nonzero element, when
computing the $(k,\ell)$ matrix element of this matrix identity,
each term is a finite sum involving
matrix elements of $S$ or $R$ and $v_{ij}$, e.g.  $(SVS)_{k\ell} =S_{ki} v_{ij} S_{j\ell}
+ S_{kj} v_{ji} S_{i\ell}$.  Using the bound \eqref{basic4} for the $S$ matrix elements,
the subexponential decay for $v_{ij}$ and 
 the trivial bound $|R_{ij}| \le  \eta^{-1}$, we obtain that 
the estimate \eqref{basic4} holds for $R$. 

\medskip

We can now start proving the main result by comparing the resolvents of
$H^{(\gamma-1)}$ and $H^{(\gamma)}$ with the resolvent $R$ of the reference
matrix $Q$.
By the resolvent expansion, 
\[
   S = R - N^{-1/2} RVR+ N^{-1} (RV)^2R - N^{-3/2} (RV)^3R + N^{-2} (RV)^4R - N^{-5/2} (RV)^5S,
\]
so we can write
\[
\frac{1}{N} \tr S =  \wh R + \xi, \qquad \xi :=  \sum_{m=1}^4 N^{-m/2}\wh R^{(m)}+N^{-5/2}\Omega
\]
with
\[
   \wh R := \frac{1}{N} \tr R, \qquad \wh R^{(m)}: = (-1)^m\frac{1}{N}\tr (RV)^mR, \qquad
 \Om := - \frac{1}{N}\tr (RV)^5S.
\]
For each diagonal element in the computation
of these traces, the contribution to $\wh R$,  $\wh R^{(m)}$ and $\Omega$ is a sum of 
a few terms. E.g.
\[
  \wh R^{(2)} = \frac{1}{N}\sum_k \Big[ R_{ki} v_{ij} R_{jj} v_{ji} R_{ik} + 
 R_{ki} v_{ij} R_{ji} v_{ij} R_{jk}
+ R_{kj} v_{ji} R_{ii} v_{ij} R_{jk} + R_{kj} v_{ji} R_{ij} v_{ji} R_{ik} \Big]
\]
and similar formulas hold for the other terms.
Then we have 
\begin{align}
\E F  \left(\frac{1}{N}\tr \frac 1 { H_\gamma-z} \right )
= & \E F \left ( \wh R + \xi \right )  
\label{temp6.6}\\
= & \E \left [ F(  \wh R) + F'( \wh R ) \xi + F^{\prime \prime} ( \wh R) \xi^2
+ \ldots+ F^{(5)} ( \wh R+\xi') \xi^5  \right ]  \non\\
= & \sum_{m=0}^5 N^{-m/2} \E    A^{(m)},\non
\end{align}
where $\xi'$ is a number between $0$ and $\xi$ 
and it depends on $ \wh R$ and $\xi$;  the $A^{(m)}$'s are defined as 
\[
A^{(0)} = F(   \wh R), \quad 
A^{(1)}=  F'(\wh R)   \wh R^{(1)},\quad 
A^{(2)}=  F''(\wh R)(  \wh R^{(1)})^2 + F'(\wh R)  \wh R^{(2)},\quad
\]
and similarly for $A^{(3)}$ and $A^{(4)}$. Finally, 
$$
  A^{(5)} = F'(\wh R)\Omega + F^{(5)} (\wh R+ \xi') (\wh R^{(1)})^5 +\ldots.
$$
 The expectation values
of the terms $A^{(m)}$, $m\le 4$, 
with respect to $v_{ij}$ are determined by the first 
four moments of $v_{ij}$, for example
$$
  \E \, A^{(2)} = F'(\wh R)
\Big[ \frac{1}{N}\sum_k R_{ki}R_{jj}R_{ik} + \ldots
   \Big] \E \, |v_{ij}|^2  + F''(\wh R) \Big[
  \frac{1}{N^2}\sum_{k,\ell} R_{ki}R_{j\ell}R_{\ell j} R_{ik} +\ldots\Big]
 \E \, |v_{ij}|^2 
$$
$$
\qquad  +  F'(\wh R) \Big[ \frac{1}{N}\sum_k R_{ki}R_{ji}R_{jk} + \ldots
   \Big] \E \, v_{ij}^2  + F''(\wh R) \Big[
  \frac{1}{N^2}\sum_{k,\ell} R_{ki}R_{j\ell}R_{\ell i} R_{jk} +\ldots\Big]
 \E \, v_{ij}^2 .
$$
Note that the coefficients involve up to four derivatives of $F$ and normalized sums
of matrix elements of $R$.  Using the estimate \eqref{basic4} for $R$ and the derivative
bounds \eqref{lowder}  for the typical values of $\wh R$,
 we see that all these coefficients are bounded by $N^{C(\tau +\e)}$
with a very large probability, where $C$ is an explicit constant. 
We use the bound \eqref{highder} for the extreme values of $\wh R$
but this event  has a very small probability by \eqref{basic4}.
Therefore, the coefficients of the moments $\E\, \bar v_{ij}^s v_{ij}^u$, $u+s\le 4$,
in the quantities $A^{(0)}, \ldots , A^{(4)}$ are essentially bounded,
modulo a factor $N^{C(\tau +\e)}$.
Notice that the fourth moment of $v_{ij}$ appears only in the $m=4$
term that already has a prefactor $N^{-2}$   in \eqref{temp6.6}. Therefore, to
compute the $m\le 4$ terms in \eqref{temp6.6} up to a precision $o(N^{-2})$,
it is sufficient to know the first three moments of $v_{ij}$ exactly and 
the fourth moment only with a precision $N^{-\delta}$; if
$\tau$ and $\e$ are chosen such that $C(\tau +\e) < \delta$, then
the discrepancy in the fourth moment is irrelevant.

Finally, we have to estimate the error term $A^{(5)}$.
All terms without $\Omega$ can be dealt with as before;
after estimating the derivatives of $F$ by $N^{C(\tau + \e)}$,
one can perform the expectation with respect to $v_{ij}$ 
that is independent of $\wh R^{(m)}$. For the terms
involving $\Omega$ one can argue similarly, by
appealing to the fact that the matrix elements of $S$
are also essentially bounded by $N^{C(\tau + \e)}$, see \eqref{basic4},
and that $v_{ij}$ has subexponential decay.
Alternatively, one can use H\"older inequality to decouple
$S$ from the rest and use \eqref{basic4} directly,
for example:
\[
\E  | F'(\wh R)  \Omega | = \frac{1}{N}\E  | F'(\wh R) \tr (RV)^5 S | 
\le  \frac{1}{N}   \left [ \E  ( F'(\wh R))^2 \tr S^2 \right ]^{1/2}    \left [ \E  \, 
   \tr (RV)^5 (VR^*)^5    \right ]^{1/2}
\le C N^{-\frac{5}{2}+C(\tau+ \e)}.
\]

Note that exactly the same perturbation expansion holds for
the resolvent of $H_{\gamma-1}$, just  $v_{ij}$ is replaced
with $w_{ij}$ everywhere. By the moment matching condition,
the expectation values $\E A^{(m)}$ of terms for $m\le 3$  in \eqref{temp6.6}
are identical and the $m=4$ term differs by
$N^{-\delta + C(\tau +\e)}$. Choosing $\tau =\e$,  we have
\[
\E \, F \left ( \frac{1}{N}\tr \frac 1 { H_\gamma-z} \right ) 
-  \E \, F \left ( \frac{1}{N}\tr \frac  1 { H_{\gamma-1}-z} \right )
\le  C N^{-5/2 + C \e}+C N^{-2 -\delta + C \e}.
\]
After summing up in \eqref{tel}  we have thus proved that 
\[
\E \, F \left ( \frac{1}{N}\tr  \frac  1 { H^{(v)} - z}\right ) 
-  \E \, F \left ( \frac{1}{N}\tr \frac  1 { H^{(w)} - z} \right )\le 
 C N^{-1/2+ C \e}+C N^{-\delta+ C \e}.
\]
The proof can be easily generalized to functions of several variables. This concludes the proof
of Theorem~\ref{comparison}.
\qed

\subsection{Proof of the correlation function comparison Theorem \ref{com}}

Define an approximate  delta function (times $\pi$) at the scale $\eta$ by 
\[
\theta_\eta(x):  =  \im \, \frac 1 {x - i \eta}.
\] 
We will choose $\eta\sim N^{-1-\e}$, i.e. slightly smaller than
the typical eigenvalue spacing. This means that an observable of the form
$\theta_\eta$ have sufficient resolution to detect individual eigenvalues. Moreover, polynomials
of such observables detect correlation functions. On the other hand, 
$$
   \frac{1}{N} \im \tr G(E+i\eta) = \frac{1}{N}\sum_i \theta_\eta(\la_i-E),
$$
therefore expectation values of such observables are covered by \eqref{maincomp}
of Theorem   \ref{comparison}. The rest of the proof consists of making
this idea precise. There are two technicalities to resolve. First, correlation
functions involve {\it distinct} eigenvalues (see \eqref{6.5} below), while
polynomials of the resolvent include an  overcounting of coinciding eigenvalues.
Thus an exclusion-inclusion formula will be needed. Second, although $\eta$ is much
smaller than the relevant scale $1/N$, it still does not give
pointwise information on the correlation functions. However,
the correlation functions in \eqref{sineres} are identified only as a weak limit,
i.e., tested against a continuous  function $O$. The continuity of $O$ can be used
to show that the difference between the exact correlation functions
and the smeared out ones on the scale $\eta\sim N^{-1-\e}$ is negligible.
This last step requires an a-priori upper bound on the density to ensure
that not too many eigenvalues fall into an irrelevantly small interval;
this bound is given in \eqref{basic}, and  it will eventually be
verified by the local semicircle law.

\bigskip

For notational simplicity, the detailed proof will be given 
 only for the case of three point correlation functions;
the proof is analogous for the general case.
 By definition of the correlation function, for any fixed $E$, $\al_1, \al_2, \al_3$,
\begin{align}\label{6.5}
    \E^\bw & \frac 1 {N(N-1)(N-2)} \sum_{i \not = j \not = k}
  \theta_\eta\Big(\lambda_i - E-\frac{\alpha_1}{N} \Big) 
 \theta_\eta\Big(\lambda_j-  E-\frac{\alpha_2}{N}\Big) 
 \theta_\eta\Big(\lambda_k- E-\frac{\alpha_3}{N}\Big) 
 \nonumber \\
 & = \int \rd x_1 \rd x_2  \rd x_3 
p_{w, N}^{(3)}(x_1, x_2, x_3)  \theta_\eta(x_1 - E_1)  \theta_\eta(x_2- E_2) 
 \theta_\eta(x_3- E_3),   \qquad E_j :=  E+\frac{\alpha_j}{N},
\end{align}
where $\E^\bw$ indicates expectation w.r.t. the $\bw$ variables.
By the exclusion-inclusion principle,
\be\label{6.6}
\E^\bw \frac 1 {N(N-1)(N-2)} \sum_{i \not = j \not = k}   \theta_\eta(x_1 - E_1)  \theta_\eta(x_2- E_2) 
 \theta_\eta(x_3- E_3)  = \E^\bw  A_1 +  \E^\bw  A_2 +  \E^\bw A_3, 
\ee
where 
\[
A_1: = \frac 1 {N(N-1)(N-2)}  \prod_{j=1}^3 
\left [  \frac 1 {N} \sum_{i }  \theta_\eta(\lambda_i - E_j) \right ],
\]
\[
A_3 :=   \frac 2 {N(N-1)(N-2)}   \sum_{i  }  \theta_\eta(\lambda_i - E_1)  \theta_\eta(\lambda_i - E_2) 
  \theta_\eta(\lambda_i - E_3) + \ldots 
\]
and
\[
A_2 :=  B_1+ B_2 + B_3, \quad\mbox{with}
\quad 
B_3=  -  \frac 1 {N(N-1)(N-2)}   \sum_{i  } 
  \theta_\eta(\lambda_i - E_1)  \theta_\eta(\lambda_i - E_2)  \sum_k  \theta_\eta(\lambda_k - E_3),
\]
and similarly, $B_1$ consists of terms with $j=k$, while  $B_2$ consists of terms with $i=k$.

Notice that, modulo a trivial change in the prefactor,
$\E^\bw A_1$ can be approximated by  
\[
\E^\bw  F \left ( \frac{1}{N}  \im \, \tr\frac 1 { H^{(v)}-z_1}, \ldots, 
\frac{1}{N} \im \, \tr \frac 1 { H^{(v)}-z_3} \right ),
\]
where the function $F$ is chosen to be
 $F(x_1, x_2, x_3) := x_1 x_2 x_3$ if $\max_j |x_j| \le N^\e$ and it is smoothly
cutoff to go to zero in the regime $\max_j |x_j|\ge N^{2\e}$.
The difference between the expectation of $F$ and $A_1$ is negligible, since it 
comes from  the regime where $ N^\e\le \max_j \frac{1}{N}  | \im \, \tr (H^{(v)}-z_j)^{-1}| 
\le N^2$, which has an exponentially small probability by \eqref{basic4}
(the upper bound on the Green function always holds since $\eta\ge N^{-2}$).  
Here the arguments of $F$ are imaginary parts of the trace of the Green function, but
this type of function is allowed when applying Theorem \ref{comparison},
since 
$$
   \im \tr G(z) = \frac{1}{2}\big[ \tr G(z) - \tr G(\bar z)\big].
$$
We remark that the main assumption  \eqref{basic} for Theorem \ref{comparison}  is satisfied by 
using  one of the local semicircle theorems
(e.g. Theorem \ref{lsc} with the choice of $M \sim N$, or Theorem~\ref{45-1}).

Similarly, we can approximate  $\E^\bw B_3$  by  
\[
\E^\bw  G \left ( \frac{1}{N^2}  \tr \left \{ \im \, \frac 1 { H^{(v)}-z_1}
 \im \, \frac 1 { H^{(v)}-z_2}  \right \}, \; 
\frac{1}{N} \im \, \tr \frac 1 { H^{(v)}-z_3} \right ), 
\]
where $G(x_1, x_2) = x_1 x_2$ with an appropriate cutoff for large arguments.
There are  similar expressions for  $B_1, B_2$ and also for $A_3$, the latter
involving the trace of the product of three resolvents. 
By Theorem \ref{comparison}, these expectations w.r.t. $\bw$ in the approximations of $\E^\bw A_i$ 
can be replaced by expectations w.r.t. $\bv$ with only negligible errors provided that 
$\eta \ge N^{-1-\e}$.  We have thus proved that 
\begin{align}\label{6.8}
\lim_{N \to \infty}   \int \rd x_1 \rd x_2  \rd x_3 
\big[ p_{w, N}^{(3)}(x_1, x_2, x_3) -  p_{v, N}^{(3)}(x_1, x_2, x_3)\big]
  \theta_\eta(x_1 - E_1)  \theta_\eta(x_2- E_2) 
 \theta_\eta(x_3- E_3) = 0.
\end{align}

Set $\eta = N^{-1-\e}$ for the rest of the proof.
We now show that the validity of \eqref{6.8} for any choice of $E$, $\al_1, \al_2, \al_3$
(recall $E_j = E + \al_j/N$)
implies that the rescaled correlation functions,
$p_{w, N}^{(3)}(E+\beta_1/N,\ldots, E+ \beta_3/N)$
and $p_{v, N}^{(3)} (E+\beta_1/N,\ldots, E+ \beta_3/N)$, as functions
of the variables $\beta_1, \beta_2, \beta_3$, have the same weak limit.

Let $O$ be a smooth, compactly supported test function and let 
\[
O_\eta(\beta_1, \beta_2, \beta_3): =  \frac{1}{(\pi N)^3} \int_{\R^3}  \rd\alpha_1 \rd\alpha_2
 \rd\alpha_3  O(\alpha_1,\alpha_2,\alpha_3) \theta_\eta\left(\frac{\beta_1-\alpha_1}{N}\right )
\ldots \theta_\eta\left(\frac{\beta_3-\alpha_3}{N}\right ) 
\]
be its smoothing on scale $N\eta$. 
Then  we can write
\begin{align}\label{6.8.1}
  \int_{\R^3}  \rd\beta_1 \rd\beta_2
 \rd\beta_3  & \; O(\beta_1,\beta_2,\beta_3) 
p_{w, N}^{(3)}\left( E+ \frac{\beta_1}{N}, \ldots, E+ \frac{\beta_3}{N}\right)  \nonumber \\
 = &\int_{\R^3}  \rd \beta_1 \rd \beta_2
 \rd\beta_3   \; O_\eta (\beta_1,\beta_2,\beta_3) 
p_{w, N}^{(3)}\left( E+ \frac{\beta_1}{N}, \ldots, E+ \frac{\beta_3}{N}\right)
\nonumber  \\& +  \int_{\R^3}  \rd \beta_1 \rd \beta_2
 \rd\beta_3   \; (O-O_\eta) (\beta_1,\beta_2,\beta_3) 
p_{w, N}^{(3)}\left( E+ \frac{\beta_1}{N}, \ldots, E+ \frac{\beta_3}{N}\right) .
\end{align}
The first term on the right side, after the change of variables $x_j = E + \beta_j /N$, is  equal to 
\begin{align}\label{6.8.2}
&  \int_{\R^3}  \rd\alpha_1 \rd\alpha_2
 \rd\alpha_3   \; O(\alpha_1,\alpha_2,\alpha_3)  \int_{\R^3}\rd x_1 \rd x_2
 \rd x_3  
p_{w, N}^{(3)}(x_1, x_2, x_3)  \theta_\eta(x_1 - E_1)  \theta_\eta(x_2- E_2) 
 \theta_\eta(x_3- E_3)   ,
\end{align}
i.e., it can be written as an integral of expressions of the form \eqref{6.8}
for which limits with $p_{w,N}$ and $p_{v,N}$ coincide.

Finally, the second term on the right hand side of \eqref{6.8.1} is negligible. To see this,
notice that for any  test function $Q$, we have
\begin{align}
 \int_{\R^3}  \rd \beta_1 \rd \beta_2
 \rd\beta_3 &   \; Q (\beta_1,\beta_2,\beta_3) 
p_{w, N}^{(3)}\left( E+ \frac{\beta_1}{N}, \ldots, E+ \frac{\beta_3}{N}\right)
\non\\
& =  N^3 \int_{\R^3}  \rd x_1 \rd x_2
 \rd x_3    \; Q\big( N (x_1-E) , N (x_2-E),  N (x_3-E)\big) 
p_{w, N}^{(3)}(x_1 , x_2, x_3)  \non \\
& =   \Big(1-\frac{1}{N}\Big)\Big(1-\frac{2}{N}\Big)
 \E^\bw \sum_{i\ne j\ne k} Q\big( N (\lambda_i-E), N (\lambda_j-E), N (\lambda_k-E ) \big). 
\label{idQ}
\end{align}
If the test function $Q$ were supported on a ball of size $N^{\e'}$, $\e'>0$, then
this last term were bounded by 
$$
\| Q\|_\infty  \E^\bw \cN_{CN^{-1+\e'}}^3(E) 
\le C\| Q\|_\infty N^{4\e'}.
$$
Here
$\cN_\tau(E)$ denotes the number of eigenvalues in the interval $[E-\tau, E+\tau]$
and in the estimate we used the local semicircle law
 on intervals of size $\tau \ge N^{-1+\e'}$. 

Set now $Q:= O-O_\eta$. {F}rom the definition of $O_\eta$, it is
 easy to see that the function 
\[
Q_1 (\beta_1,\beta_2,\beta_3) = O (\beta_1,\beta_2,\beta_3) -
O_\eta  (\beta_1,\beta_2,\beta_3) \prod_{j=1} ^3  1 ( |\beta_j|\le N^{ \e'})
\]
satisfies the bound $\| Q_1\|_\infty \le \|Q\|_\infty=
 \| O-O_\eta\|_\infty\le C  {N\eta} = CN^{-\e}$. 
So choosing $\e' < \e/4$, the contribution of $Q_1$ is negligible.
Finally, $Q_2 = Q - Q_1$ is given by 
\[
Q_2 (\beta_1,\beta_2,\beta_3) =  -O_\eta  (\beta_1,\beta_2,\beta_3) 
 \left [1-  \prod_{j=1} ^3 1 ( |\beta_j|\le N^{ \e'}) \right ]
\]
and 
\begin{align} 
|Q_2| \le  &  C  \left [  \frac {1} { 1   +  \beta_1^2 } \right ] 
   \left [  \frac {1} { 1   +  \beta_2^2 } \right ] 
   \left [  \frac {1} { 1   +  \beta_3^2 } \right ] 
 \Big \{ 1 ( |\beta_1|\ge N^{ \e'}) +\ldots\Big\}  \non  \\
   \le  & C  \left \{  N^{-\e'} \left [  \frac {N^{\e'} } { N^{2\e'}   +  \beta_1^2 } \right ] 
   \left [  \frac {1} { 1   +  \beta_2^2 } \right ] 
   \left [  \frac {1} { 1   +  \beta_3^2 } \right ]   +\ldots\right \}. 
\end{align} 
Hence the contribution of   $Q_2$ in the last term of \eqref{idQ}  is bounded by 
\be
 C  N^{-3-\e'} \E^\bw \sum_{i,j,k}  \left \{   \left [  
\frac {N^{-1+ \e'} } {   N^{-2+ 2\e'} +  (\lambda_i - E)^2 } \right ] 
 \left [  \frac {N^{-1} } { N^{-2}    +   (\lambda_j - E)^2 } \right ] 
\left [  \frac {N^{-1} } {  N^{-2} +   (\lambda_k- E)^2 } \right ]
+\ldots\right \} .
\label{triple}
\ee
From the local semicircle law,
Theorem \ref{lsc} or Theorem~\ref{45-1}, the  last term is bounded by $N^{-\e'}$ up to some 
logarithmic factor. To see this, note that the Riemann sums for eigenvalues in \eqref{triple}
 can be replaced with an integral because the resolution scale of the functions
involved is at least  $N^{-1}$.
 This completes the proof of
 Theorem   \ref{com}.  \qed

\subsection{Sketch of the proof of Theorem \ref{thm:TV}}\label{sk:TV}

We again  interpolate
between $H$ and $H'$ step by step, by replacing
the distribution of the matrix elements of $H$ from $\nu$ to $\nu'$ one by one
according to a fixed ordering.
 Let $H^{(\tau)}(h)$ be the
matrix where the first $\tau-1$ elements follow $\nu'$ distribution,
the $\tau$-th entry is $h$ and
the remaining ones follow $\nu$. Denote by $\la_i(H^{(\tau)}(h))$ the $i$-th eigenvalue
of $H^{(\tau)}(h)$.  Let
$$
   \cF_\tau(h): = F\Big(N\la_{i_1}(H^{(\tau)}(h)), N\la_{i_2}(H^{(\tau)}(h)) ,\ldots\big)
$$
and we prove that
\be
    \big| \E \cF_\tau(h) -  \E' \cF_\tau (h')\big|\le CN^{-2-c_0}.
\label{enough}
\ee
Since the number of replacement steps is of order $N^2$, this
will show \eqref{FF}. Let $\tau$ represent the $(pq)$
matrix element and we can drop the index $\tau$.

We will prove that 
\be
   \Big|  \frac{\pt^n\cF}{\pt h^n} \Big| \le CN^{O(c_0)+ o(1)}
\label{derb}
\ee
for any $n\le 5$. 
Then, by Taylor expansion,
\be
  \cF(h) = \sum_{n=0}^4 \frac{1}{n!} \frac{\pt^n\cF}{\pt h^n} (0) h^n +
  N^{-5/2 + O(c_0)+ o(1)},
\label{taylor}
\ee
since $|h|\le N^{-5/2 + o(1)}$ with very high probability.
After taking expectations for $h$ with respect to $\nu$ and
$\nu'$, since the first four moments match, the contributions of
the fully expanded terms in \eqref{taylor} coincide and this proves
\eqref{enough}.

To see \eqref{derb}, we assume for simplicity that $F$ has only one
variable and $i_1=i$. Then
$$
   \frac{\pt\cF}{\pt h}(h) = N F'(\lambda)  \frac{\pt\la_i}{\pt h}(h).
$$
By standard first order perturbation theory, with $h=h_{pq}$,
\be
    \frac{\pt\la_i}{\pt h} = 2\re u_i(p)\bar u_i(q),
\label{lader}
\ee
where $\bu_i = (u_i(1), u_i(2), \ldots , u_i(N))$ 
is the eigenfunction belonging to $\la_i$.
Since the eigenvectors are delocalized, 
\be
\| \bu\|_\infty \approx N^{-1/2}
\label{delocev}
\ee
(modulo logarithmic corrections, see \eqref{eq:deloc}), so we obtain
\be
   \Big|  \frac{\pt\la_i}{\pt h}\Big|\lesssim O(N^{-1})
\label{1der}
\ee
and thus
$$
  \Big|  \frac{\pt\cF}{\pt h}\Big|\lesssim N\cdot N^{c_0}\cdot N^{-1} = N^{c_0}.
$$
For higher order derivatives, we have to differentiate the eigenfunctions as well.
This gives rise to the resonances, for example  with  $h=h_{pq}$
$$ 
  \frac{\pt u_i(p)}{\pt h_{pq}} = \sum_{j\ne i} \frac{1}{\la_i-\la_j} 
  u_j(p)\big[ \bar u_j(p) u_i(q) + \bar u_j(q) u_i(p)\big] 
$$
thus
$$
 \Big|  \frac{\pt u_i(p)}{\pt h} \Big|\le
  N^{-1/2} \frac{1}{N}  \sum_{j\ne i} \frac{1}{|\la_i-\la_j|}  \lesssim N^{-1/2+c_0},
$$ 
assuming that the eigenvalues regularly follow the  semicircle law
and no two neighboring eigenvalues get closer than $N^{-1-c_0}$,
see \eqref{lowtail}.
Substituting this bound into the derivative of \eqref{lader}, we have
$$
   \Big| \frac{\pt^2\la_i}{\pt h^2}\Big|\lesssim CN^{-1+c_0}
$$
Combining this bound with \eqref{1der}, and reinstating 
the general case when $F$ has more than one variable,  we have
$$
 \Big|  \frac{\pt^2\cF}{\pt h^2}\Big|\lesssim CN \cdot |\pt_{ii} F||\pt^2\la_i| + 
  + CN^2 |\pt_{ij} F| |\pt \la_j||\pt \la_i| \le CN^{2c_0}.
$$
The argument for the higher derivatives is similar. The key technical
inputs are the delocalization bound on the eigenvectors \eqref{delocev}
that can be obtained from local semicircle law and  the lower tail
estimate \eqref{lowtail}.

\section{Universality for Wigner matrices: putting it together}\label{sec:put}

In this section we put the previous information together to prove 
our main result Theorem \ref{mainthm} below.
We will state our most general result from \cite{EYY2}.
The same result under somewhat more restrictive conditions
was proved in our previous papers, e.g. \cite[Theorem 2.3]{ESY4},
\cite[Theorem 3.1]{ESYY} and \cite[Theorem 2.2]{EYY}.

\medskip

Recall that $p_N(\lambda_1, \lambda_2, \ldots ,\lambda_N)$ denotes
the symmetric joint density  of the eigenvalues of the $N\times N$
Wigner matrix $H$. For simplicity we will use the formalism as if the joint
distribution of the eigenvalues 
were absolutely continuous with respect to the Lebesgue measure, but 
it is not necessary for the proof. Recall the
definition of the $k$-point
correlation functions (marginals) $p_N^{(k)}$ from \eqref{pk}.
We will use the notation $p^{(k)}_{N, GUE}$ and  $p^{(k)}_{N, GOE}$ 
for the correlation functions of the GUE and GOE ensembles.

We consider the rescaled correlation functions about a
fixed energy $E$ under a scaling that guarantees that the
local density is one. The sine-kernel universality for the GUE
ensemble states that
the rescaled correlation functions converge weakly
to the determinant of the sine-kernel, $K(x)= \frac{\sin \pi x}{\pi x}$, i.e.,
\be
   \frac{1}{[\varrho_{sc}(E)]^k} p^{(k)}_{N, GUE}
  \Big( E + \frac{\al_1}{N\varrho_{sc}(E)},\ldots
 E + \frac{\al_k}{N\varrho_{sc}(E)}\Big) \to 
  \det \big( K(\al_\ell - \al_j)\big)_{\ell,j=1}^k
\label{GUEuniv}
\ee
as $N\to\infty$
for any fixed energy $|E|<2$ in the bulk of the spectrum
\cite{MG, D}.
Similar result holds for the GOE case; the sine kernel
being replaced with a similar but somewhat more complicated
universal function, see \cite{M}. Our main result is that
universality \eqref{GUEuniv} holds for  hermitian or symmetric
generalized
Wigner matrices after averaging a bit in the energy $E$:

\begin{theorem}\label{mainthm} \cite[Theorem 2.2]{EYY2}
 Let $H$ be an $N\times N$ symmetric or hermitian generalized
Wigner matrix. In the hermitian
case we assume that the real and imaginary parts are i.i.d.
Suppose that the distribution $\nu$ of the rescaled matrix elements $\sqrt{N}h_{ij}$ 
have subexponential decay \eqref{subexpuj}. Let $k\ge 1$ and 
  $O: \R^k \to \R$ be a continuous, compactly supported function. 
Then for any $ |E| < 2 $, we have 
\begin{equation}\label{mainres}
\begin{split}
\lim_{ b\to 0} \lim_{N \to \infty}  & \frac{1}{2b} \int_{E-b}^{E+b}  \rd v
 \int_{\R^k}  \rd\alpha_1 
\ldots \rd\alpha_k \; O(\alpha_1,\ldots,\alpha_k) 
\\ &\times  \frac{1}{[\varrho_{sc}(v)]^k} \Big ( p_N^{(k)}  - p_{N, \#} ^{(k)} \Big )
  \Big (v+\frac{\alpha_1}{N\varrho_{sc}(v)}, 
\ldots, v+\frac{\alpha_k}{N \varrho_{sc}(v)}\Big) =0,
\end{split} 
\end{equation}
where $\#$ stands for GOE or GUE for the symmetric or hermitian cases, respectively.
\end{theorem}

{\it Proof.} 
For definiteness, we consider the symmetric case, i.e., the
limit will be the Gaussian Orthogonal Ensemble (GOE),
corresponding to the parameter $\beta=1$
in the general formalism. The joint distribution of the 
eigenvalues $\bx= (x_1, x_2, \ldots , x_N)$  
is given by the following measure 
\be\label{H1}
\mu=\mu_N(\rd{\bf x})= 
\frac{e^{-N\cH({\bf x})}}{Z}\rd{\bf x},\qquad \cH({\bf x}) =  
 \sum_{i=1}^N \frac{x_{i}^{2}}{4} -  \frac{1}{N} \sum_{i< j} 
\log |x_{j} - x_{i}|,
\ee
and we assume that the eigenvalues are ordered, i.e.,
$\mu$ is restricted to 
$\Sigma_N = \{ \bx\in \R^N\; : \; x_1 < x_2<  \ldots < x_N\}$.

Let $\wh H$ be a symmetric Wigner matrix with single entry
distribution satisfying the subexponential decay \eqref{subexpuj}.
 We let the matrix  evolve according to the
matrix valued Ornstein-Uhlenbeck process, \eqref{process},
$$
  \rd H_t = \frac{1}{\sqrt{N}}\rd\beta_t - \frac{1}{2}H_t\rd t, \qquad H_0= \wh H,
$$
and recall that the distribution of $H_t$, for each fixed $t>0$ is the same as
\be
   e^{-t/2} \wh H + (1-e^{-t})^{1/2} V,
\label{HT}
\ee
where $V$ is an independent GOE matrix. The distribution $\nu_t(\rd x)= u_t(x) \rd x$
of the matrix elements evolves according to 
 the  Ornstein-Uhlenbeck process  on $\bR$, i.e., 
\be\label{ou}
\partial_{t} u_t =  A u_t, \quad 
   A  = \frac{1}{2}\frac{\pt^2}{\pt x^2}
- \frac{ x}{2} \frac{\pt}{\pt x}.
\ee
Note that the initial distribution $\nu=\nu_0$ may be singular, but for any $t>0$
the distribution $\nu_t$ is absolutely continuous.

The Ornstein-Uhlenbeck  process \eqref{ou} induces \cite{Dy} the Dyson Brownian motion 
 on the eigenvalues with a  generator given by 
\be
 L=   \sum_{i=1}^N \frac{1}{2N}\partial_{i}^{2}  +\sum_{i=1}^N
 \Bigg(- \frac{1}{4} x_{i} +  \frac{1}{2N}\sum_{j\ne i} 
\frac{1}{x_i - x_j}\Bigg) \partial_{i}
\label{L}
\ee
acting on $L^2(\mu)$. The measure $\mu$ is invariant
and reversible with respect to the dynamics generated by $L$.

Denote the distribution of the eigenvalues  at  time $t$
 by $f_t ({\bf x})\mu(\rd {\bf x})$.
 Then $f_t$ satisfies 
\be\label{dy1}
\partial_{t} f_t =  L f_t
\ee
with initial condition $f_0$ given by the eigenvalue density of the
Wigner matrix $\wh H$. With the previous notations, $p_N = f_0\mu_N$,
where  $p_N$ and hence $f_0$ may be singular with respect to
the Lebesgue measure.
Due to $\beta\ge 1$, the eigenvalues do not cross,
i.e., the dynamics \eqref{dy1} is well defined on $\Sigma_N$.
By using a straighforward symmetrization, one can extend the equilibrium measure, 
the density functions and the dynamics 
to the whole $\R^N$.  We will use the formalism of ordered eigenvalues everywhere,
except  in the definition of the correlation functions \eqref{pk}, where
the symmetrized version is easier.  With a small abuse of notations, we will disregard this
difference. 

\bigskip

Theorem \ref{mainthm} was originally proved in \cite{ESY4}
for standard Wigner matrices and under more restrictive conditions
on the single entry distribution.  Here
we present a more streamlined proof, following \cite{EYY}
but for notational simplicity we consider the case of standard Wigner matrices only.
The main part of the proof of Theorem \ref{mainthm} consists of three steps:

\medskip

{\it Step 1.} First we show that there exists an $\e_0>0$ such that
 the correlation functions of {\it any}
Wigner ensemble with a Gaussian convolution of 
variance $t\sim N^{-\e_0}$ coincide with the GOE.
In other words, {\it any} ensemble of the form \eqref{HT}
with $t\ge N^{-\e_0}$ (and with subexponential decay
on the matrix elements of $\wh H$) has a universal local
statistics.

\medskip

{\it Step 2.} Set $t=N^{-\e_0}$. We then show 
 that for any given Wigner matrix $H$ we can 
find another Wigner matrix $\wh H$
such that the first three moments of $H$ and $H_t= e^{-t/2} \wh H + (1-e^{-t})^{1/2} V$
coincide and the fourth moments are close by $O(N^{-\e_0})$.

\medskip

{\it Step 3.} Theorem \ref{com}, which was the corollary
of the Green function comparison theorem, shows that the
local correlation functions of $H$ and $H_t$ from Step 2
coincide. Together with Step 1, this will complete the proof of Theorem \ref{mainthm}. \qed

\medskip

Now we formulate the statements in  Step 1 and Step 2 more precisely,
Step 3 is already completed.

\subsection{Step 1: Universality for Gaussian convolutions}

This step is just an application of Theorem \ref{thm:main}, we formulate it
for our special case:

\begin{theorem}
Suppose that the probability distribution of the initial symmetric Wigner matrix $\wh H$ 
has subexponential decay \eqref{subexpuj} with some exponent $\ttau$ and
let $H_t$ be given by the Gaussian convolution \eqref{HT}. Let $p^{(k)}_{t,N}$
denote the $k$-point correlation function of the eigenvalues of $H_t$.
Then there exists an $\e_0>0$, depending on the parameters in \eqref{subexpuj} such that 
for any $t\ge N^{-\e_0}$ we have
\begin{equation}\label{mainresgoe}
\begin{split}
\lim_{ b\to 0} \lim_{N \to \infty}  & \frac{1}{2b} \int_{E-b}^{E+b}  \rd v
 \int_{\R^k}  \rd\alpha_1 
\ldots \rd\alpha_k \; O(\alpha_1,\ldots,\alpha_k) 
\\ &\times  \frac{1}{[\varrho_{sc}(v)]^k} \Big ( p_{t,N}^{(k)}  - p_{N, GOE} ^{(k)} \Big )
  \Big (v+\frac{\alpha_1}{N\varrho_{sc}(v)}, 
\ldots, v+\frac{\alpha_k}{N \varrho_{sc}(v)}\Big) =0
\end{split} 
\end{equation}
for any continuous, compactly supported test function $O$. 
\end{theorem}

We remark that the threshold exponent $\e_0$ can be given explicitly. 
If we use the local semicircle law from Theorem~\ref{lsc} and
its corollary, Theorem~\ref{prop:lambdagamma}, then $\e_0$ can be chosen as any number smaller than 1/7.
Using the strong local semicircle law, Theorem \ref{45-1}, the exponent $\e_0$ can be chosen
as any number smaller than 1.

\medskip

{\it Proof.} We just
have to check that the assumptions of Theorem  \ref{thm:main} are satisfied. 
First, the Hamiltonian of the equilibrium measure is  \eqref{H1}, and 
it is clearly of the form \eqref{ham}, so Assumption I is automatic.
The entropy assumption  \eqref{entro}
on the initial state $f_0$ may not be satisfied since $f_0$ can even be
singular. However, by the semigroup property of the OU flow, one
can consider the initial condition $f_{t_0}$ for the flow $f_t$, $t\ge t_0$,
for some $t_0\le N^{-\e_0}$  since  the statement of Theorem \ref{thm:main} concerns 
only the time $t \ge N^{-\e_0}$.
Thus it is sufficient to show that the entropy condition is satisfied for
some very small $t_0\ll N^{-\e_0}$.

 To see this,  
let $\nu_t$ denote the single entry distribution of $H_t$ 
and $ \bar \nu_t$  the  probability measure of the matrix $H_t$. 
Let $\bar\nu_{GOE}$ denote the probability measure 
of the GOE ensemble and $\nu_{GOE}$ the probability measure of 
its $ij$-th element which is a Gaussian measure 
with mean zero and variance $1/N$. Since the dynamics of  matrix 
elements are independent (subject to the symmetry condition), 
and the entropy is additive,
we have the identity 
\be\label{entrcomp}
\int  \log \left ( \frac {d \bar \nu_t} { \rd \bar \nu_{GOE}}  \right ) \rd \bar \nu_t 
= \sum_{i\le j} \int  \log \left ( \frac {\rd  \nu_t} 
{ \rd \nu_{GOE} }  \right ) \rd \nu_t \le CN^2 \int  \log \left ( \frac {\rd  \nu_t} 
{ \rd \nu_{GOE}}  \right ) \rd  \nu_t
\ee
since the summation runs over the indices of all independent elements $1\le i\le j\le N$.
Clearly, the process $t \to \nu_t$ is an Ornstein-Uhlenbeck  process and each entropy term 
on the right hand side of \eqref{entrcomp} is bounded by $ C N$ provided that $t \ge t_0:=1/N$
and $ \nu_0$ has a  subexponential decay. Since the entropy of the marginal distribution 
on the eigenvalues is bounded by the entropy of the total  measure
 on the matrix, we have proved that 
\be
\int f_{1/N}\log f_{1/N} \rd\mu\leq CN^3,
\ee
and this verifies \eqref{entro}. 
Therefore, in order to  apply Theorem \ref{thm:main}, we only have to 
verify the Assumptions II, III and IV. 
Clearly,  Assumptions  II and IV follow from 
the local semicircle law, Theorem \ref{lsc} with $\varrho(E)=\varrho_{sc}(E)$
 (note that in the bounded variance case $M\sim N$),
and Assumption III was proven in Theorem \ref{prop:lambdagamma}.
Now we can apply  Theorem \ref{thm:main} and we get \eqref{mainresgoe}
with any $\e_0<\e$, where $\e$ is obtained from Theorem~\ref{prop:lambdagamma},
i.e. $\e_0$ can be any number smaller than $1/7$.
If we use the strong local semicircle law, Theorem~\ref{45-1}, then 
\eqref{rigidity} implies 
$$
    \E \sum_j (\la_j-\gamma_j)^2 \lesssim N^{-1}, 
$$
i.e. Assumption III, \eqref{assum2} holds with any $\fa<1/2$ and thus $\e_0$ can be
any number smaller than 1.
 \qed

\subsection{Step 2: Matching Lemma}

For any real random variable $\xi$, denote by $m_k(\xi) =\E \xi^k$
its $k$-th moment. By Schwarz inequality, the sequence of moments,
$m_1, m_2, \ldots $ are not arbitrary numbers, for example $|m_1|^k\le m_k$
and $m_2^2\le m_4$, etc, but there are more subtle relations. For example, if $m_1=0$, then
\be
   m_4 m_2 - m_3^2 \ge m_2^3
\label{34}
\ee
which can be obtained by
$$
  m_3^2= \big[ \E \xi^3\big]^2 =  \big[ \E \xi(\xi^2-1)\big]^2
  \le \big[ \E \xi^2\big]\big[ \E (\xi^2-1)^2\big] = m_2 (m_4 - 2m_2^2 +1)
$$
and noticing that  \eqref{34} is scale invariant, so it is
sufficient to prove it for $m_2=1$.
In fact, it is easy to see that \eqref{34} saturates if and only of
the support of $\xi$ consists of exactly two points (apart from the trivial
case when $\xi\equiv 0$).

This restriction shows that given a sequence of four admissible moments,
$m_1=0$, $m_2=1$, $m_3, m_4$,
there may not exist a Gaussian divisible random variable $\xi$ 
with these moments; e.g. the moment sequence $(m_1, m_2, m_3, m_4)= (0,1,0,1)$
uniquely characterizes the standard Bernoulli variable ($\xi=\pm 1$ with 1/2-1/2
probability). However, if we allow a bit room in the fourth moment, then
one can match any four admissible moments with a small Gaussian convolution.
This is the content of the next lemma 
which completes Step~2.

\begin{lemma}\label{fmam}\cite[Lemma 6.5]{EYY}
 Let  $m_3$ and $m_4$ be two real numbers  such that 
\be\label{m4m3cc}
m_4-m_3^2-1\ge 0,\,\,\, m_4\leq C_2
\ee
for some positive constant  $C_2$. Let $\xi^G$ be a Gaussian random variable 
 with mean $0$ and variance $1$. 
Then for any sufficient small $\gamma>0$ (depending on  $C_2$),
 there exists a real  random 
variable $\xi_\gamma$ with subexponential decay 
and independent of $\xi^G$, such that the first four moments of 
\be\label{defxi'}
\xi'=(1-\gamma)^{1/2}\xi_\gamma+\gamma^{1/2}\xi^G
\ee 
are $m_1(\xi')=0$, $m_2(\xi')=1$, $m_3(\xi')=m_3$ and $m_4(\xi')$, and 
\be\label{m4m4}
|m_4(\xi')-m_4|\leq C\gamma
\ee 
for some $C$ depending on $C_2$.
\end{lemma}

{\it Proof.} It is easy to see by an
explicit construction that:

\medskip

{\bf Claim:} For any given numbers $m_3, m_4$, with $m_4-m_3^2-1\ge 0$
there is a random variable $X$ with  first four moments
$0, 1, m_3, m_4$ and with subexponential decay. \qed

\medskip

For any real random variable
$\zeta$, independent of $\xi^G$, and with the first 4 moments being $0$, $1$, $m_3(\zeta)$
 and $m_4(\zeta)<\infty$, the first 4 moments of 
\be
\zeta'=(1-\gamma)^{1/2}\zeta+\gamma^{1/2}\xi^G
\ee
are $0$, $1$, 
\be\label{relm3}
m_3(\zeta')=(1-\gamma)^{3/2}m_3(\zeta)
\ee
 and 
 \be\label{relm4}
m_4(\zeta')=(1-\gamma)^{2}m_4(\zeta) +6\gamma-3\gamma^2.
 \ee

Using the Claim, we obtain that for any $\gamma>0$
there exists a real random variable $\xi_\gamma$ such that the 
first four moments are $0$, $1$, 
\be\label{m3xi}
m_3(\xi_\gamma)=(1-\gamma)^{-3/2} m_3
\ee
 and
 $$
 m_4(\xi_\gamma)=m_3(\xi_\gamma)^2+(m_4-m^2_3).
 $$ 
With $m_4\leq C_2$, we have $m_3^2\leq C_2^{3/2}$, thus   
 \be
 |m_4(\xi_\gamma)-m_4 |\leq C\gamma
 \ee
for some $C$ depending on $C_2$. 
Hence with \eqref{relm3} and \eqref{relm4}, we obtain that $\xi'
=(1-\gamma)^{1/2}\xi_\gamma+\gamma^{1/2}\xi^G$
 satisfies $m_3(\xi')=m_3$ and \eqref{m4m4}.
This completes the proof of Lemma \ref{fmam}.
\qed

\appendix

\section{Large Deviation Estimates: proof of Lemma~\ref{generalHWT}}\label{sec:lde}

 The estimates in Lemma~\ref{generalHWT} are weaker
than the corresponding results of Hanson and Wright \cite{HW}, used
in \cite{ESY3, ESYY}, but they require only independent, not necessarily
identically distributed random variables with subexponential decay,
moreover the proofs are much simpler. Thus the Gaussian decay requirement
of Hanson and Wright is relaxed to subexponential, but the tail
probability estimate is weaker.

\medskip

{\it Proof of \eqref{resgenHWTD}.} 
Without loss of generality, we may assume that $\sigma=1$. 
The assumption \eqref{assumdelta2} implies that the $k-$th moment of $a_i$ is bounded by: 
\be\label{Eaik}
\E|a_i|^k\leq (Ck)^{\al k}
\ee
for some $C>0$. 
\par First,  for $p\in \N$, we estimate
\be
\E\left|\sum_{i=1}^Na_iA_i\right|^p.
\ee 
With the Marcinkiewicz--Zygmund inequality, for $p\geq 2$,  we have  
\be\label{iiEaA}
\E\left|\sum_{i}a_iA_i\right|^p\leq (Cp)^{p/2}\E\left(\sum_{i}|a_iA_i|^2\right)^{p/2}
\ee
(for the estimate of the constant, see e.g. Exercise 2.2.30 of \cite{Str}).
Inserting \eqref{Eaik} into \eqref{iiEaA}, we have
\be\label{temp9.6}
\E\left|\sum_{i}a_iA_i\right|^p\leq (Cp^{\frac{1}{2}+\al})^p\left(\sum_{i}|A_i|^2\right)^{p/2},
\ee
which implies \eqref{resgenHWTD}
 by choosing $p=\log N$ and applying a high moment Markov inequality. 
\qed
\bigskip

{\it Proof of  \eqref{diaglde}}.
 Notice that the random variables $|a_i|^2-1$ ($1\leq i\leq N$) are independent random variables
 with mean $0$ and variance less than some constants $C$. Furthermore, the $k$-th moment of
 $|a_i|^2-1$ is bounded as 
\be
\E(|a_i|^2-1)^k\leq (Ck)^{2\alpha k} .
\ee 
Then following the proof of \eqref{resgenHWTD} with $|a_i|^2-1$ replacing $a_i$, 
we obtain  \eqref{diaglde}.

\medskip

{\it Proof of \eqref{resgenHWTO}}. 
For  any $p\in \N$, $p\ge 2$, we estimate
\be
\E\left|\sum_{i>j}\overline a_i\xi_i\right|^p\equiv 
\E\left|\sum_{i>j}\overline a_iB_{ij}a_j\right|^p
\ee 
where $\xi_i:=\sum_{j<i}B_{ij}a_j$. Note that $a_i$ and $\xi_i$ are independent for any fixed $i$. 
 By the definition, 
\be
X_n\equiv \sum_{i=1}^n \overline a_i\xi_i
\ee
is martingale. Using  the Burkholder inequality, we have that
\be\label{iiEaA2}
\E\left|\sum_{i}\overline a_i\xi_i\right|^p\leq
 (Cp)^{3p/2}\E\Big(\sum_{i}|\overline a_i\xi_i|^2\Big)^{p/2}
\ee
(for the constant, see Section VII.3 of \cite{Shy}).
By the generalized Minkowski inequality, by the independence of $a_i$ and $\xi_i$
and using \eqref{Eaik}, we have
\[
 \left [ \E\Big(\sum_{i}|\overline a_i\xi_i|^2\Big)^{p/2}\right]^{2/p}
 \le  \sum_{i}  \bigg [ \, \E |\overline a_i\xi_i|^{p} \, \bigg ]^{2/p} 
 =  \sum_{i}  \bigg [ \, \E(|\overline a_i|^p)\E(|\xi_i|^p) \, \bigg ]^{2/p}
\le
(C p)^{ 2 \alpha} \sum_{i}  \bigg [ \, \E(|\xi_i|^p) \, \bigg ]^{2/p}.
\]
Using \eqref{temp9.6}, we have
$$
    \E(|\xi_i|^p) \le (Cp^{\frac{1}{2}+\al})^p \Big(\sum_{j} |B_{ij}|^2\Big)^{p/2}.
$$
Combining this with \eqref{iiEaA2} we obtain
\be\label{iiEaA3}
\E\left|\sum_{i}\overline a_i\xi_i\right|^p\leq
(Cp)^{2p(1+\alpha)}\Big( \sum_{i} \sum_j |B_{ij}|^2 \Big)^{p/2}.
\ee
 Then choosing $p=\log N$ and applying Markov inequality, we obtain \eqref{resgenHWTO} . 
\qed

\section{Proof of Theorem \ref{thm:main}}\label{sec:corfn}

Recalling the notations around \eqref{Ydef},
we start with the identity \eqref{iddd1}
\begin{align}
\int_{E-b}^{E+b}\frac{\rd E'}{2b} \int_{\R^n} \rd\alpha_1
\ldots \rd\alpha_n  & \; O(\alpha_1,\ldots,\alpha_n)  p_{\tau, N}^{(n)}
\Big (E'+\frac{\alpha_1}{N\varrho(E)},
\ldots, E'+\frac{\alpha_n}{N \varrho(E)}\Big) \label{iddd} \\
 = & \int_{E-b}^{E+b}\frac{\rd E' }{2b}\sum_{{\bf m}\in S_n}\sum_{i=1}^N  Y_{i,\bm}
(E',\bx) \non
\end{align}
We have to show that
\be
   \lim_{N\to\infty}
  \Bigg| \int_{E-b}^{E+b}\frac{\rd E'}{2b} \int \sum_{{\bf m}\in S_n}\sum_{i=1}^N  Y_{i,\bm}
(E',\bx) (f_\tau-1)\rd \mu \Bigg|=0.
\label{goal}
\ee
Let $M$ be an $N$-dependent parameter chosen at the end of the proof. Let 
$$
  S_n(M): = \{ \bm \in S_n \; , \; m_n \le M\}, \quad S_n^c(M):= S_n\setminus S_n(M),
$$
and note that $|S_n(M)|\le M^{n-1}$. 
To prove \eqref{goal}, it is sufficient to show that
\be
   \lim_{N\to\infty}
  \Bigg| \int_{E-b}^{E+b}\frac{\rd E'}{2b} \int \sum_{{\bf m}\in S_n(M)}\sum_{i=1}^N  Y_{i,\bm}
(E',\bx) (f_\tau-1)\rd \mu\Bigg|=0.
\label{goal1}
\ee
and that
\be
   \lim_{N\to\infty}  \sum_{{\bf m}\in S_n^c(M)}
  \Bigg| \int_{E-b}^{E+b}\frac{\rd E'}{2b} \int \sum_{i=1}^N  Y_{i,\bm}
(E',\bx) f_\tau\rd \mu \Bigg|=0
\label{goal2}
\ee
holds for any $\tau>N^{-2\e+\delta}$ (note that $\tau =\infty$ 
corresponds to the equilibrium, $f_\infty =1$).
\bigskip

\noindent
{\it Step 1: Small $\bm$ case; proof of \eqref{goal1}.}

\bigskip
 After performing the $\rd E'$ integration, we will eventually
apply Theorem \ref{thmM} to the function
$$
G\big( u_1, u_2, \ldots \big)
 : = \int_{\R} \wt O\big( y,
 u_1, u_2 ,\ldots,   \big) \rd y ,
$$
i.e., to the quantity
\be
 \int_\R \rd E'\; Y_{i,\bm}(E',\bx)= 
 \frac{1}{N} G\Big( N(x_i-x_{i+m_2}), \ldots \Big) 
\label{OO}
\ee
for each fixed $i$ and ${\bf m}$.

For any $E$ and  $0<\xi<b$ define sets of integers
$J=J_{E,b,\xi}$ and $J^\pm= J^\pm_{E,b,\xi}$  by
$$
  J : = \big\{ i\; : \; \gamma_i \in [E-b, E+b]\big\},\quad
  J^\pm : = \big\{ i\; : \; \gamma_i \in [E-(b\pm\xi), E+b\pm\xi]\big\},
$$
where $\gamma_i$ was defined in \eqref{gammaj}.  Clearly $J^-\subset J \subset J^+$.
With these notations, we have
\be
   \int_{E-b}^{E+b} \frac{\rd E'}{2b} \sum_{i=1}^N
 Y_{i,\bm}(E',\bx )=  \int_{E-b}^{E+b} \frac{\rd E'}{2b} \sum_{i\in J^+} Y_{i,\bm}(E',\bx )
 + \Omega^+_{J,\bm}(\bx).
\label{uppe}
\ee
 The error term $\Omega^+_{J,\bm}$, defined by \eqref{uppe}
indirectly, comes from  those $i\not\in J^+$ indices,
for which $x_i \in [E-b, E+b] + O(N^{-1})$ since 
$Y_{i,\bm}(E',\bx)=0$ unless $|x_i-E'|\le C/N$, the constant
depending on the support of $O$. Thus
$$
   |\Omega^+_{J,\bm}(\bx)| \le  CN^{-1}b^{-1}\# \{ \; i \; : \; |x_i-\gamma_i|\ge \xi/2 \}
$$
for any sufficiently large $N$ assuming $\xi\gg 1/N$
and using that $O$ is a bounded function. The additional $N^{-1}$ factor
comes from the $\rd E'$ integration. 
 Taking the expectation with respect to the
measure $f_\tau\rd\mu$, we get
\be
    \int |\Omega^+_{J,\bm}(\bx)| f_\tau\rd\mu \le Cb^{-1}\xi^{-2}N^{-1} 
 \int \sum_i (x_i-\gamma_i)^2 f_\tau\rd\mu 
   = Cb^{-1}\xi^{-2} N^{-1-2\fa}
\label{exp}
\ee
using Assumption III \eqref{assum2}.
We can also estimate
\begin{align}
    \int_{E-b}^{E+b} \frac{\rd E'}{2b} \sum_{i\in J^+} Y_{i,\bm}(E',\bx)
  \le & \int_{E-b}^{E+b} \frac{\rd E'}{2b} \sum_{i\in J^-} Y_{i,\bm}(E',\bx) + 
CN^{-1} |J^+\setminus J^-|  \non \\
   = & \int_\R \frac{\rd E'}{2b} \sum_{i\in J^-}Y_{i,\bm}(E',\bx)  + C(Nb)^{-1}|J^+\setminus J^-|+
  \Xi^+_{J,\bm}(\bx) \label{fol} \\
 \le & \int_\R \frac{\rd E'}{2b} \sum_{i\in J}Y_{i,\bm}(E',\bx)+
C(Nb)^{-1}|J^+\setminus J^-|+C(Nb)^{-1}|J\setminus J^-|+
  \Xi^+_{J,\bm}(\bx), \non
\end{align}
where the error term $\Xi^+_{J,\bm}$, defined by \eqref{fol}, 
comes from indices $i\in J^-$ such that $x_i \not \in [E-b, E+b]+O(1/N)$.
It satisfies the same bound \eqref{exp} as $\Omega^+_{J,\bm}$.
By the continuity  of $\varrho$, the density of $\gamma_i$'s is
bounded by $CN$, thus $|J^+\setminus J^-|\le CN\xi$ and
$|J\setminus J^-|\le CN\xi$.
 Therefore, 
summing up the formula \eqref{OO} for $i\in J$,
 we obtain from \eqref{uppe} and \eqref{fol}
$$
    \int_{E-b}^{E+b} \frac{\rd E'}{2b} \int \sum_{i=1}^N
Y_{i,\bm}(E',\bx)  f_\tau\rd\mu\le \frac{1}{2b} \int \frac{1}{N}\sum_{i\in J}
 G \Big( N(x_i-x_{i+m_2}), \ldots \Big)
    f_\tau\rd\mu + Cb^{-1}\xi + Cb^{-1}\xi^{-2} N^{-1-2\fa}
$$
for each $\bm\in S_n$.
A similar lower bound can be obtained analogously, and after choosing
$\xi= N^{-1/3}$, we obtain
\be
  \Bigg|  \int_{E-b}^{E+b} \frac{\rd E'}{2b} \int \sum_{i=1}^N
 Y_{i,\bm}(E',\bx) f_\tau\rd\mu-  \int \frac{1}{N}
\sum_{i\in J} G \Big( N(x_i-x_{i+m_2}), \ldots \Big)
    f_\tau\rd\mu \Bigg|\le CN^{-1/3}
\label{seccc}
\ee
for each $\bm\in S_n$, where $C$ depends on $b$. It is possible to
optimize the choice of $\xi$, depending on $b$ and $\fa$, and
this would yield the effective bound mentioned after Theorem~\ref{thm:main},
but in this presentation we will not pursue the effective bound, see \cite{ESYY}
for more details.

 Adding up \eqref{seccc}
for all $\bm\in S_n(M)$, we get
\be
  \Bigg|  \int_{E-b}^{E+b} \!\frac{\rd E'}{2b} \int \!\!\!\sum_{\bm\in S_n(M)}\sum_{i=1}^N
 Y_{i,\bm}(E',\bx) f_\tau\rd\mu-  \int \!\!\sum_{\bm\in S_n(M)} \frac{1}{N}
\sum_{i\in J} G \Big( N(x_i-x_{i+m_2}), \ldots \Big)
    f_\tau\rd\mu \Bigg|\le CM^{n-1}N^{-1/3},
\label{seccc1}
\ee
and the same estimate holds for the equilibrium, i.e.,
if we set $\tau=\infty$ in \eqref{seccc1}.
Subtracting these two formulas and applying \eqref{GG} from
 Theorem \ref{thmM}
to each summand on the second term in \eqref{seccc} we conclude 
that
\be
    \Bigg|  \int_{E-b}^{E+b} \frac{\rd E'}{2b} \int \sum_{\bm\in S_n(M)}\sum_{i=1}^N
 Y_{i,\bm}(E',\bx) (f_\tau-1)\rd\mu  \Bigg|\le  CM^{n-1}(N^{-1/3}+ N^{-\delta/6}).
\label{ssd}
\ee
Choosing 
\be
    M : =N^{\min \{ 1/3, \delta/6\}/n},
\label{Mdef}
\ee
we obtain that \eqref{ssd} vanishes as $N\to\infty$, and this proves \eqref{goal1}.

\bigskip
\noindent
{\it Step 2. Large $\bm$ case; proof of \eqref{goal2}.}
\bigskip

For a fixed $y\in \R$, $\ell >0$, let
$$
   \chi(y,\ell) : =\sum_{i=1}^N {\bf 1}\Big\{ x_i\in \big[ y- \frac{\ell}{N},
 y +\frac{\ell}{N}\big] \Big\}
$$
denote the number of points in the interval $[y-\ell/N, y+\ell/N]$.
 Note that for a fixed $\bm=(m_2, \ldots , m_n)$, we have
\be
   \sum_{i=1}^N |Y_{i,\bm} (E',\bx)| \le C\cdot\chi(E',\ell)
 \cdot {\bf 1}\Big(\chi(E',\ell)\ge m_n\Big) \le C\sum_{m=m_n}^\infty m \cdot
 {\bf 1}\Big(\chi(E',\ell)\ge m\Big),
\label{Ofi}
\ee
where $\ell$ denotes the maximum of $|u_1|+\ldots + |u_n|$
in the support of  $\wt O(u_1, \ldots , u_n)$.

 Since the summation over
all increasing sequences
$\bm = (m_2, \ldots, m_n)\in \N_+^{n-1}$ with a fixed $m_n$
contains at most $m_n^{n-2}$ terms,
we have
\be
  \sum_{\bm \in S_n^c(M)} \Bigg| \int_{E-b}^{E+b}  \frac{\rd E'}{2b} \; 
\int \sum_{i=1}^N|Y_{i,\bm} (E',\bx)|  f_\tau \rd\mu \Bigg|
 \le C \int_{E-b}^{E+b}  \frac{\rd E'}{2b} \;\sum_{m=M}^\infty m^{n-1}
  \int  {\bf 1}\Big(\chi(E',\ell)\ge m\Big) f_\tau\rd\mu.
\label{toc}
\ee
Now we use  Assumption IV for the interval
$I = [E' - N^{-1+\sigma}, E' + N^{-1+\sigma}]$ with 
$\sigma:=\frac{1}{2n}\min \{ 1/3, \delta/6\}$.
 Clearly $\cN_I\ge \chi(E',\ell)$ for
sufficiently large $N$, thus we get from  \eqref{ass4} that
$$
   \sum_{m=M}^\infty m^{n-1}
  \int  {\bf 1}\Big(\chi(E',\ell)\ge m\Big) f_\tau\rd\mu 
 \le C_a \sum_{m=M}^\infty m^{n-1} \Big(\frac{m}{N^\sigma}\Big)^{-a} 
$$
holds for any $a\in \N$. By the choice of $\sigma$,
we get that $\sqrt{m}\ge N^\sigma$ for any $m\ge M$
(see \eqref{Mdef}), and thus choosing $a=2n+2$, we get
$$
   \sum_{m=M}^\infty m^{n-1}
  \int  {\bf 1}\Big(\chi(E',\ell)\ge m\Big) f_\tau\rd\mu 
 \le \frac{C_a}{M} \to 0
$$
as $N\to\infty$.
Inserting this into \eqref{toc},
this completes the proof of \eqref{goal2} and the
proof of Theorem \ref{thm:main}. \qed

\end{document}